\documentclass[12pt,preprint]{elsarticle}
\usepackage[active]{srcltx}    
\usepackage{listings}
\usepackage{hyperref}
\usepackage{graphics}
\usepackage{epsfig}
\usepackage{amsmath}
\usepackage{mathtools}
\usepackage[normalem]{ulem}
\usepackage{graphicx}
\usepackage{url} 
\usepackage{color} 
\usepackage{footnote}
\usepackage[absolute]{textpos}
\makeatletter
\def\ps@pprintTitle{%
 \let\@oddhead\@empty
 \let\@evenhead\@empty
 \def\@oddfoot{}%
 \let\@evenfoot\@oddfoot}
\makeatother

\renewcommand\sout{\bgroup \color{blue}\ULdepth=-.5ex \ULset}

\newcommand{\fti}{\langle N_B\rangle+\langle N_{\bar{B}}\rangle}
\newcommand{\ftis}{S}

\newcommand{\zsqminus}{z^2-\langle N_B\rangle\,\langle N_{\bar{B}}\rangle}
\newcommand{\zsqminuss}{Q}

\newcommand{\ntimesn}{\langle N_B\rangle\,\langle N_{\bar{B}}\rangle}
\newcommand{\ntimesns}{P}

\newcommand{\nb}{\langle N_B\rangle_{GC}}
\newcommand{\nbb}{\langle N_{\bar{B}}\rangle_{GC}}
\newcommand{\nbc}{\langle N_B\rangle}
\newcommand{\nbbc}{\langle N_{\bar{B}}\rangle}
\newcommand{\nplusn}{\langle N_B\rangle+\langle N_{\bar{B}}\rangle}
\newcommand{\nplusns}{S}
\newcommand{\hk}{r}
\newcommand{\pb}{\alpha_B}
\newcommand{\pbb}{\alpha_{\bar{B}}}
\newcommand{\snn}{\ensuremath{ \sqrt{s_{\rm NN} } } }

\RequirePackage{lineno}
\begin{document}

\begin{textblock}{5}(10,1)
\noindent\small{CERN-TH-2020-116}
\end{textblock}

\begin{frontmatter}
\title{Relativistic nuclear collisions: Establishing a non-critical baseline for fluctuation measurements}

\author[EMMI,HU]{P.~Braun-Munzinger }
%\ead{p.braun-munzinger@gsi.de}
\author[GSI]{B.~Friman}
%\ead{b.friman@gsi.de}
\author[WU,GSIR]{K.~Redlich}
%\ead{krzysztof.redlich@uwr.edu.pl}
\author[GSI,NNRC]{A.~Rustamov}
%\ead{a.rustamov@cern.ch}
\author[HU]{J.~Stachel}
%\ead{stachel@physi.uni-heidelberg.de}

\address[EMMI]{Extreme Matter Institute EMMI, GSI, Darmstadt, Germany}
\address[HU]{Physikalisches Institut, Universit\"{a}t Heidelberg, Heidelberg, Germany}
\address[GSI]{GSI Helmholtzzentrum f{\"u}r Schwerionenforschung, Darmstadt, Germany}
\address[WU]{Institute of Theoretical Physics, University of Wroclaw, Wroclaw, Poland}
\address[GSIR]{Theoretical Physics Department, CERN, CH-1211 Gen\`eve 23, Switzerland}
\address[NNRC]{National Nuclear Research Center, Baku, Azerbaijan}

\begin{abstract}
We study the influence of  global baryon number conservation   on the non-critical baseline  of   net baryon  cumulants in heavy-ion collisions  in a given acceptance, accounting for the asymmetry between the mean-numbers of baryons  and antibaryons. We derive the probability distribution of net baryon number in a restricted phase space  from the canonical partition function that incorporates exact conservation of baryon number in the full system. Furthermore, we provide  tools to compute cumulants  of any order  from the generating function of uncorrelated baryons constrained by exact baryon number conservation. The results are applied to quantify the non-critical baseline for  cumulants of   net proton number fluctuations obtained in heavy-ion collisions by the STAR collaboration  at different RHIC energies and by the ALICE collaboration at the LHC. Furthermore, volume fluctuations are added by a Monte Carlo procedure based on the centrality dependence of charged particle production as measured experimentally. Compared to the predictions based on the hadron resonance gas model or Skellam distribution  a clear suppression of fluctuations is observed due to exact baryon-number conservation. The suppression increases with the order of the cumulant and towards lower collision energies. Predictions for net proton cumulants up to the eight order  in heavy-ion collisions are given for experimentally accessible collision energies.   

\end{abstract}
\end{frontmatter}

\section{Introduction}
\label{sec:Intro}
One of the  goals of current experimental and theoretical studies of strongly interacting matter and the quark-gluon plasma is  to  unravel  the  phase  structure of QCD and the connections to  the restoration of  chiral symmetry at finite temperature and density \cite{BraunMunzinger:2009zz,Andronic:2017pug}.

At finite  quark masses the chiral $SU(2)_L\times SU(2)_R$ symmetry is explicitly broken in the QCD vacuum.  Moreover, 
it is well established through lattice QCD (LQCD) investigations~\cite{Aoki:2006we} that chiral symmetry is restored in a crossover  transition at
vanishing net baryon density and a (pseudo-critical) temperature of $T_{pc}\simeq 156.5$ MeV~\cite{Bazavov:2018mes}. In ref.~\cite{Pisarski} it was argued, based on analytical arguments, that in the limit of massless u and d quarks the chiral crossover becomes a genuine  second-order  phase  transition  belonging  to the O(4) universality class in three dimensions. Indeed, recent numerical LQCD calculations provide strong indications for a second-order chiral transition at a critical temperature  $T_c\simeq 132$ MeV,  \cite{Ejiri:2009ac,Sarkar:2019jwa,Ding:2019prx}.

At present, systematic LQCD studies of the properties of QCD matter are possible only at small net baryon densities. Consequently, first-principle results on the nature of the chiral transition at high baryon densities are not yet available. However, studies of strongly-interacting matter in effective models of QCD suggest that, at sufficiently  large baryon chemical potential  $\mu_B$, QCD matter exhibits a first order chiral phase transition \cite{Asakawa:1989bq,Rajagopal:1992qz,Halasz:1998qr,Stephanov:1998dy,Stephanov:2006dn,Schaefer:2006ds,Sasaki:2007db,Ding:2015ona,Buballa:2018hux}.  The endpoint of such a first-order transition line in the $(T,\mu_B)$-plane is the conjectured  chiral critical endpoint (CP) \cite{Asakawa:1989bq}.  At such a CP, the system would exhibit a  2nd order phase transition belonging  to  the Z(2) universality class.

A dedicated beam energy scan program at RHIC  has been established to explore the large net baryon density region of the QCD phase diagram  and to search for the  CP in collisions of heavy ions at relativistic energies \cite{Aggarwal:2010cw,Bzdak:2019pkr}.  By varying the beam energy, the properties of strongly interacting matter are studied in a broad range of net baryon densities. Important related research objectives are pursued by the ALICE collaboration at the LHC~\cite{ Rustamov:2017lio,Arslandok:2020mda,Acharya:2019izy}, by the NA61 collaboration at the CERN SPS~\cite{Mackowiak-Pawlowska:2020glz}, and by the HADES collaboration at GSI~\cite{Adamczewski-Musch:2020slf}.
At the LHC, symmetry is observed  between  yields of produced  matter and antimatter~\cite{Andronic:2017pug}. Consequently, in the mid-rapidity region of central heavy ion collisions, QCD matter emerges with (nearly) vanishing chemical potentials. This corresponds to a particularly interesting part of the thermal parameter space, where unique first principle predictions on the properties of the equation of state of QCD matter and fluctuation observables are available from LQCD calculations \cite{Akiba:2015jwa,Bazavov:2020bjn,Borsanyi:2018grb}.

Fluctuations of conserved charges have been identified as promising observables for identifying and probing chiral criticality in QCD  matter  \cite{Bazavov:2020bjn,Hatta:2002sj,Hatta:2003wn,Ejiri:2004bh,Allton:2005gk,Ejiri:2005wq,Karsch:2005ps,Sasaki:2006ww,Skokov:2010uh,Bluhm:2020mpc}. These are experimentally accessible and
are studied theoretically in  LQCD, as well as in effective models. Fluctuations of the  net baryon number are a particulary interesting probe, owing to their  direct connection to critical behavior at the chiral phase boundary \cite{Sarkar:2019jwa,Bazavov:2020bjn,Hatta:2002sj,Hatta:2003wn,Karsch:2010ck,Friman:2011pf}. Moreover, in heavy ion collisions, as well as in LQCD calculations,  there is a well established baseline for hadron  yields \cite{Andronic:2017pug} and for the  non-critical behavior of the   net baryon number   fluctuations \cite{Sarkar:2019jwa,Allton:2005gk,Karsch:2010ck,Bazavov:2012jq,Ratti:2010kj}. This  is provided by the thermodynamic potential of the hadron resonance gas (HRG)~\cite{Andronic:2017pug} and the resulting Skellam probability distribution for fluctuations of the net baryon number \cite{BraunMunzinger:2011dn,Braun-Munzinger:2014lba}.

First  data  on the  variance,  skewness  and  kurtosis for fluctuations of the net proton number~\footnote{The net proton number  is used as a  proxy for the net baryon number.} in  heavy-ion  collisions have been obtained by the STAR Collaboration at RHIC  \cite{Adamczyk:2013dal,Adam:2020unf}.  Recently,  high statistics results on the mean and variance of the net proton  distributions  have been   obtained  in  nucleus-nucleus collisions at the LHC by the ALICE Collaboration  \cite{Rustamov:2017lio,Arslandok:2020mda,Acharya:2019izy}. The STAR  data  on higher-order cumulants and their ratios  exhibit  deviations from the HRG baseline, with a possible  non-monotonic dependence on the collision energy for the kurtosis times variance of the net proton number.  It has been conjectured that these data could provide a first indication  for the  existence  of a CP in the QCD phase diagram at finite baryon density \cite{Adam:2020unf}. For a recent summary of the experimental situation see ~\cite{Rustamov:2020ekv}.
Nevertheless, in order to conclusively establish the structure of the QCD phase diagram at large net baryon densities, further in-depth studies of all available fluctuation observables are needed~\cite{ Bazavov:2020bjn,Bollweg:2020yum,Almasi:2017bhq}.

In heavy ion collisions, cumulants of net proton number in a given experimental acceptance  are influenced by a number of non-critical effects, which also give rise to deviations from the HRG baseline.
Two  such effects are currently discussed: volume fluctuations \cite{Skokov:2012ds,Braun-Munzinger:2016yjz,Sugiura:2019toh}, which are linked to event-by-event fluctuations of the number of participating nucleons and constraints imposed by the conservation of the net baryon 
number in full phase space~ \cite{Braun-Munzinger:2016yjz,Braun-Munzinger:2018yru,Bzdak:2012an,Braun-Munzinger:2019yxj,Begun:2004gs,Vovchenko:2020tsr,Barej:2020ymr}.

In this work, we present a detailed study of how global baryon number  conservation influences the  non-critical background of the  net baryon  number  cumulants in a given experimental acceptance. To this end we introduce  the partition function for  baryons in a finite system, where the net baryon number is conserved. Given the probability distribution of the net baryon number in a particular acceptance \cite{Bzdak:2012an},  we derive general expressions for cumulants of any order, allowing for different acceptances for protons and antiprotons.
Moreover, we quantify the  effect of baryon number conservation for the acceptance employed by the STAR collaboration  for extracting the net-proton-number cumulants in heavy ion collisions at a few representative energies. These results are obtained by either using proton and antiproton distributions measured over the full rapidity range or by extrapolation of data over a restricted rapidity range using the concept of limiting fragmentation~\cite{Benecke:1969sh}. In particular, except for LHC energies with  $\snn \ge 2.76$ TeV, the resulting acceptance probabilities for protons and antiprotons differ substantially. This observation motivated the present study of net baryon fluctuations, where we allow for different acceptances for protons and antiprotons.~\footnote{Bzdak {\em et al.} \cite{Bzdak:2012an} derived the probability distribution for fluctuations of the net baryon number and the corresponding cumulant generating function, constrained by global conservation of the baryon number, with different acceptances for baryons and antibaryons. However, analytic and numerical results for cumulants were given only for the special case with equal acceptances.}

We find that the constraints imposed by baryon number conservation imply a significant reduction of the  net proton fluctuations. In addition, we include the effect of volume fluctuations, using the method developed in~\cite{Braun-Munzinger:2016yjz}.  Finally, we compare results for chemical freeze-out parameters obtained by analysis of particle multiplicities with those determined from higher cumulants.  We demonstrate that, especially at lower beam energies, great care must be exercised if physically meaningful results are to be obtained from chemical analysis of fluctuation observables.

We argue that our results provide a robust non-critical baseline which is relevant for the interpretation of experimental results on event-by-event fluctuations of the net baryon number in nuclear collisions. In the derivation we employ a theoretical framework which, in conjuction with experimental input, accounts for the leading order effects in the fugacity expansion. Terms of higher order in this expansion, which are not included in our scheme, are for cumulants of order $n<6$ expected to be subleading in the relevant temperature range.

The paper is organized in the following way. We start, in Sect.~\ref{sec:FlucCE}, by constructing the canonical partition function for baryons and antibaryons, thereby implementing exact baryon number conservation. As necessary input from experiment we use here the mean number of protons and antiprotons within the acceptance where  fluctuations are measured experimentally. To obtain the fraction of (anti-)protons in the acceptance over the total number of (anti-)baryons produced in the collision we use, at {c.m.} energies of $\sqrt{s_{NN}}< 80$ GeV, experimental data plus, when needed, a limiting fragmentation prescription, detailed in Sect.~\ref{sec:Simulations}. The measured data are supplemented by using the statistical hadronization model, described in detail in \cite{Andronic:2017pug} and references therein. At the highest  energies, we model the rapidity dependence of antibaryon production following the one measured for charged particles over a wide rapidity range (10 units) and neglect baryons from the fragmentation regions. Volume fluctuations are added by a Monte Carlo procedure based on the centrality dependence of charged particle production as measured experimentally. Analytical expressions for canonical cumulants of baryons at any order are obtained as functions of the total baryon and antibaryon numbers and the (anti-)proton acceptances.
 All results are cross-checked with numerical and Monte Carlo calculation. The final results for the 1st through 4th as well as 6th cumulants are then compared as a non-critical baseline to the available experimental data. Numerical values for the fifth- and eighth-order cumulants are presented as well. A short concluding section summarizes the results obtained and puts them into context. Technical details are presented in three appendices.

\section{Fluctuations in the Canonical Ensemble}
\label{sec:FlucCE}
In this section we present the theoretical framework for computing fluctuations in a system where the net baryon number $B$ is conserved.  Our starting point is the canonical (C) partition function for a system in a finite volume $V$ at temperature $T$ ~\cite{BraunMunzinger:2003zd,Hagedorn:1984uy, Gorenstein:2013gra}~\footnote{This form of the partition function corresponds to the leading term in a fugacity expansion. We neglect terms of higher order in this expansion, since lattice results indicate that they are subleading in the relevant temperature range~\cite{Vovchenko:2017xad}. Moreover, we note that the approach presented here is not restricted to thermal statistical models. Indeed it is more general and applies to Poisson distributed baryon and antibaryon multiplicities under the constraint of baryon number conservation. In particular, the form of the partition function applies also to a non-uniform system, where the composition depends on the rapidity, if the baryons and antibaryons in each rapidity interval are Poisson distributed. This follows from the fact that the sum of two (or more) Poisson distributed variables is also Poisson distributed.} 

\begin{eqnarray}\label{eq:canonical-partition} 
Z_{B}(V,T)&=&\sum_{N_{B}=0}^{\infty}\sum_{N_{\bar{B}}=0}^{\infty}\frac{(\lambda_{B}\,z_B)^{N_{B}}}{N_{B}!}\frac{(\lambda_{\bar{B}}\,z_{\bar{B}})^{N_{\bar{B}}}}{N_{\bar{B}}!}\delta({N_{B}-N_{\bar{B}}-B}) \nonumber\\
&=&\int_0^{2\pi}\frac{d\phi}{2\pi}\,e^{-iB\phi}\exp\left({\lambda_B\,z_B\,e^{i\phi}+\lambda_{\bar B}\,z_{\bar{B}}\,e^{-i\phi}}\right)
\nonumber\\
&=&\left(\frac{\lambda_{B}\,z_B}{\lambda_{\bar{B}}\,z_{\bar{B}}}\right)^{\frac{B}{2}}\,I_{B}(2\,z\,\sqrt{\lambda_{B}\,\lambda_{\bar{B}}})
\label{eq:zCE}
\end{eqnarray}
where $I_{B}$ denotes the modified Bessel function of the first kind and $z=\sqrt{z_Bz_{\bar B}}$. The single particle partition functions for baryons $z_{B}$ and antibaryons $z_{\bar{B}}$ involve integrations over position and momentum space~\footnote{For one particle species in a homogeneous system~\cite{Cleymans:2004iu}, $z_B=z_{\bar{B}}=(V/(2\,\pi)^3)\int d^3p\, e^{-\sqrt{p^2+m^2}/T}$. In general, e.g., when strange hadrons are included in the partition sum, $z_B\neq z_{\bar{B}}$.} and are directly related to the mean number of baryons and antibaryons in the grand canonical ensemble $\nb=e^{\mu_B/T}\,z_B$ and $\nbb=e^{-\mu_B/T}\,z_{\bar{B}}$.  It follows that $z=\sqrt{\nb\,\nbb}$. 

The auxiliary parameters  $\lambda_{B,\bar{B}}$ are introduced for the calculation of the mean number of baryons and antibaryons. In the final results they are set equal to unity. The resulting mean multiplicities in the canonical ensemble are~\footnote{Expectation values without a subscript, $\langle \dots \rangle$, refer to the canonical ensemble (\ref{eq:canonical-partition}).}
\begin{eqnarray}\label{NCM}
\langle N_B\rangle&=& \lambda_B{\frac{\partial\ln Z_B} {\partial\lambda_B}}|_{\lambda_B,\lambda_{\bar B}=1}=z\,\frac{I_{B-1}(2\, z)}{I_{B}(2\, z)},\\\label{NCP}
\langle N_{\bar{B}}\rangle&=&\lambda_{\bar B}{\frac{\partial\ln Z_B} {\partial\lambda_{\bar B}}}|_{\lambda_B,\lambda_{\bar B}=1}=z\,\frac{I_{B+1}(2\, z)}{I_{B}(2\, z)}.
\end{eqnarray}

\noindent
The conservation of the net baryon number in the canonical ensemble,
\begin{equation}\label{eq:net-baryon-number}
\langle N_B\rangle-\langle N_{\bar{B}}\rangle=B,
\end{equation}
is fulfilled by (\ref{NCM}) and (\ref{NCP}). This follows from the recurrence relation for Bessel functions, $2\,\nu\, I_\nu(x)=x\,\big[I_{\nu-1}(x)-I_{\nu+1}(x)\big]$.

Consider now the fluctuations in a subsystem.  In an experiment this is defined by the acceptance, which generally corresponds to cuts in momentum space. Consequently, the single particle partition functions are split into two parts, $z_A$ for baryons in the acceptance and $z_R$ for those outside of the acceptance, with analogous expressions for antibaryons. In order to obtain the partition function for the subsystem, we rewrite (\ref{eq:canonical-partition}) using the binomial theorem for $\big(z_B\big)^{N_B}=\big(z_A+z_R\big)^{N_B}$ and $\big(z_{\bar{B}}\big)^{N_{\bar{B}}}=\big(z_{\bar{A}}+z_{\bar{R}}\big)^{N_{\bar{B}}}$ 
\begin{eqnarray}\label{eq:canonical-split}
Z_{B}(V,T)&=&\sum_{B_A=-\infty}^\infty\,\Bigg\{\sum_{N_{\bar{A}}=0}^\infty\,\frac{\big(\lambda_{A}\,z_A\big)^{N_{\bar{A}}+B_A}}{\big(N_{\bar{A}}+B_A\big)!}\,\frac{\big(\lambda_{\bar{A}}\,z_{\bar{A}}\big)^{N_{\bar{A}}}}{\big(N_{\bar{A}}\big)!}\Bigg\}\nonumber\\
&\times&\Bigg\{\sum_{N_{\bar{R}}=B_A-B}^\infty\,\frac{\big(z_R\big)^{N_{\bar{R}}+B-B_A}}{\big(N_{\bar{R}}+B-B_A\big)!}\,\frac{\big(z_{\bar{R}}\big)^{N_{\bar{R}}}}{\big(N_{\bar{R}}\big)!}\Bigg\}\\
&=&\sum_{{B_A}=-\infty}^\infty\,\left(\frac{\lambda_{A}\,z_A}{\lambda_{\bar{A}}\,z_{\bar{A}}}\right)^{B_A/2}\,I_{B_A}\big(2\,\sqrt{\lambda_{A}\,\lambda_{\bar{A}}\,z_A\,z_{\bar{A}}}\big)\nonumber\\
&\times&\left(\frac{z_R}{z_{\bar{R}}}\right)^{(B-B_A)/2}\,I_{B-B_A}\big(2\,\sqrt{z_R\,z_{\bar{R}}}\big).\nonumber
\end{eqnarray}

In (\ref{eq:canonical-split}) $N_{\bar{A}}$ and $N_{\bar{R}}$ are the antibaryon numbers in the acceptance and in the complimentary subsystem, respectively, while $B_A$ is the (fluctuating) net baryon number in the acceptance. Moreover, we have introduced the parameters $\lambda_{A}$ and $\lambda_{\bar{A}}$ to facilitate the calculation of the mean baryon and antibaryon numbers in the acceptance.  Using Graf's addition formula~\cite{Watson:1966} and properties of the Bessel functions we find, 
\begin{eqnarray}\label{eq:NCM-subsystem}
\langle N_B\rangle_{A}&=& \lambda_{A}{\frac{\partial\ln Z_B} {\partial\lambda_{A}}}|_{\lambda_{ A},\lambda_{\bar { A}}=1}=\pb\,z\,\frac{I_{B-1}(2\, z)}{I_{B}(2\, z)}=\pb\,\langle N_B\rangle,\\\label{NCP-subsystem}
\langle N_{\bar{B}}\rangle_{A}&=&\lambda_{\bar A}{\frac{\partial\ln Z_B} {\partial\lambda_{\bar A}}}|_{\lambda_A,\lambda_{\bar A}=1}=\pbb\,z\,\frac{I_{B+1}(2\, z)}{I_{B}(2\, z)}=\pbb\,\langle N_{\bar{B}}\rangle,
\end{eqnarray}
where we have introduced the acceptance parameters $\pb=z_A/z_B$ and $\pbb=z_{\bar{A}}/z_{\bar{B}}$. The ratios of the baryon and antibaryon single particle partition functions in the acceptance to the corresponding total ones are equal to the probabilities for observing a baryon and an antibaryon, respectively.  In a given experiment, these probabilities are determined by the experimental acceptance for the various particle species.

Using (\ref{eq:canonical-split}) we deduce the normalized probability distribution  for the number of baryons, $N_A$, and antibaryons, $N_{\bar{A}}$,
in the acceptance, 
\begin{eqnarray}\label{eq:probability-acceptance-1}
P_A(N_{A},N_{\bar{A}})&=&\frac{\big(\pb\,z\big)^{N_{A}}}{N_{A}!}\,\frac{\big(\pbb\,z\big)^{N_{\bar{A}}}}{N_{\bar{A}}!}\,\left(\frac{1-\pb}{1-\pbb}\right)^{(B-N_{A}+N_{\bar{A}})/2}\\
&\times&\,\frac{I_{B-N_{A}+N_{\bar{A}}}\big(2\,z\, {{ \sqrt{  (1-\pb)(1-\pbb)}}}\big)}{I_B\big(2\,z\big)}.\nonumber
\end{eqnarray}
Moreover, by summing (\ref{eq:probability-acceptance-1}) over $N_{A}$, keeping $B_A=N_{A}-N_{\bar{A}}$ fixed, one recovers the probability distribution for the net baryon number in the acceptance~\cite{Bzdak:2012an},
\begin{eqnarray}\label{eq:probability-acceptance-2}
P_A(B_A)&=&\left(\frac{\pb}{\pbb}\right)^{B_A/2}\,\left(\frac{1-\pb}{1-\pbb}\right)^{(B-B_A)/2}\\
&\times&\frac{I_{B_A}\big(2\,z\,\sqrt{\pb\,\pbb}\big)\, I_{B-B_A}\big(2\,z\,\sqrt{(1-\pb)\,(1-\pbb)}\big)}{I_B\big(2\,z\big)}.\nonumber
\end{eqnarray}

In the framework presented here, we take only correlations that arise from the conservation of the net baryon number into account. Consequently the resulting probability distributions depend only on the baryon number and are independent of any other quantum numbers, like, e.g., isospin. Thus, the probability distribution for the net proton number in the acceptance is obtained from (\ref{eq:probability-acceptance-2}) by simply replacing the binomial  probabilities $\pb$ and $\pbb$ by  $\alpha_p=\langle N_p\rangle_{A}/\langle N_B\rangle$ and $\alpha_{\bar p}=\langle N_{\bar p}\rangle_A/\langle N_{\bar B}\rangle$,  respectively,    where $\langle N_p\rangle_A$ and  $\langle N_{\bar p}\rangle_A$  are  the mean-number of protons and antiprotons  in the acceptance window.

The moments of the net baryon number in the acceptance are given by
\begin{eqnarray} 
\mu_n=\langle (N_B-N_{\bar{B}})^n\rangle_A&=&\sum_{N_{A},N_{\bar{A}}}\,(N_{A}-N_{\bar{A}})^n\,P_A(N_{A},N_{\bar{A}})\\
&=&\sum_{B_A}\,(B_A)^n\, P_A(B_A).\nonumber
\end{eqnarray}
The corresponding cumulants are polynomials in the moments,
\begin{eqnarray}
\kappa_1&=&\mu_1,\nonumber\\
\kappa_2&=&\mu_2-(\mu_1)^2,\\
\kappa_3&=&\mu_3-3\,\mu_2\,\mu_1+2\,(\mu_1)^3,\nonumber\\
\kappa_4&=&\mu_4-4\,\mu_3\,\mu_1-3\,(\mu_2)^2+12\,\mu_2\,(\mu_1)^2-6\,(\mu_1)^4.\nonumber
\end{eqnarray}  
In general the cumulants can be expressed in terms of moments
by making explicit use of  Bell polynomials~\cite{Comtet:1974}
\begin{equation}
\kappa_n=\sum_{k=1}^n\,(-1)^{k-1}\,(k-1)!\,B_{n,k}(\mu_1,\mu_2,\dots,\mu_{n-k+1}).\
\end{equation}

In Appendix A we derive analytical expressions for the canonical cumulants at any order. Here we give explicit expressions for the first four,
\begin{eqnarray}
\kappa_1&=&B\,k_+^{(1)}+S\,k_-^{(1)}=\nbc\,c^{(1)}_B-\nbbc\,c^{(1)}_{\bar{B}},\nonumber\\
\kappa_2&=&B\,k_+^{(2)}+S \,k_-^{(2)}
+4\,Q\,\big[k_-^{(1)}\big]^2,\nonumber\\
\kappa_3&=&B\,k_+^{(3)}+S\,k_-^{(3)}
+12\,Q \,k_-^{(1)}\,k_-^{(2)}
+8\,\Big(Q-W\Big)\,\big[k_-^{(1)}\big]^3,\\
\kappa_4&=&B\,k_+^{(4)}+S\,k_-^{(4)}+4\,Q\,\Big[3\,\big[k_-^{(2)}\big]^2
+4\,k_-^{(1)}\,k_-^{(3)}\Big]
\nonumber\\
&+&48\, \Big(Q-W\Big)\,\big[k_-^{(1)}\big]^2\,k_-^{(2)}+16\,\Big(W\,\big(S-1\big)-2Q^2+Q\Big)
\big[k_-^{(1)}\big]^4,\nonumber
\end{eqnarray}
where
\begin{eqnarray}\label{eq:SPQW}
S&=&\nplusn,\nonumber\\
P&=&\ntimesn,\\
\zsqminuss&=&z^2-\ntimesns,\nonumber\\
W&=&Q\,S-\ntimesns.\nonumber
\end{eqnarray}
Moreover, 
\begin{equation}
    k_{\pm}^{(n)}=\frac{1}{2}(c_B^{(n)}\pm c_{\bar{B}}^{(n)}),
\end{equation}
where
\begin{eqnarray}
c^{(n)}_B&=&\delta_{n,1}+(-1)^{1+n}\, {\rm Li}_{1-n}(1-1/\pb),\\
c^{(n)}_{\bar{B}}&=&\delta_{n,1}+ {\rm Li}_{1-n}(1-1/\pbb),\nonumber\\
\end{eqnarray}
and ${\rm Li}_{n}(x)$ is the polylogarithm. One finds, e.g.,  $c^{(1)}_B=\pb$, $c^{(2)}_B=\pb\,(1-\pb)$, $c^{(1)}_{\bar{B}}=\pbb$ and $c^{(2)}_{\bar{B}}=-\pbb\,(1-\pbb)$. Further details are found in Appendix A.
    
In the high-energy limit, the net baryon number is small compared to either the baryon or antibaryon number. In Appendix B we derive a compact expression for the cumulant of any order in this limit,
\begin{equation}
\kappa_n^{(he)}=S\,r^{(n)}+B\,k_+^{(n)}+\frac12\,\big[r^{(n)}-k_-^{(n)}\big],
\end{equation}
with     
\begin{equation}
r^{(n)}=B_n\left(k_-^{(1)},k_-^{(2)},\dots,k_-^{(n)}\right),    
\end{equation}
where $B_n\left(x_1,x_2\dots,x_n\right)$ is a complete Bell polynomial.

Finally, in the low-energy limit,  the cumulants reduce to those of a binomial distribution,
\begin{equation}
\kappa_n^{(le)}=B\,c^{(n)}_B, 
\end{equation}
as shown in Appendix C.

\section{Simulations}
\label{sec:Simulations}
The goal of this section is to simulate the experimental conditions, paving the way for a quantitative understanding of the measured net baryon fluctuations. We recall that, in experiments, the net baryon number fluctuates due to the finite acceptance. The relevant experimental acceptance is typically defined by imposing selection criteria on transverse momentum and rapidity of baryons and antibaryons. To proceed further, we rewrite Eq.~(\ref{eq:canonical-partition}) as
\begin{eqnarray}\label{eq:canonical-partition2}
Z_{B}(V,T)&=&\sum_{N_{\bar{B}}=0}^{\infty} \frac{(z_B)^{N_{\bar{B}}+B}\,(z_{\bar{B}})^{N_{\bar{B}}}}{(N_{\bar{B}}+B)!N_{\bar{B}}!}=\left(\frac{z_B}{z_{\bar{B}}}\right)^{B/2}I_{B}(2z),
\end{eqnarray}
where we explicitly set both $\lambda_{B}$ and $\lambda_{\bar{B}}$ to unity.
\begin{figure}[!htb]
    \centering
    \includegraphics[width=1.\linewidth,clip=true]{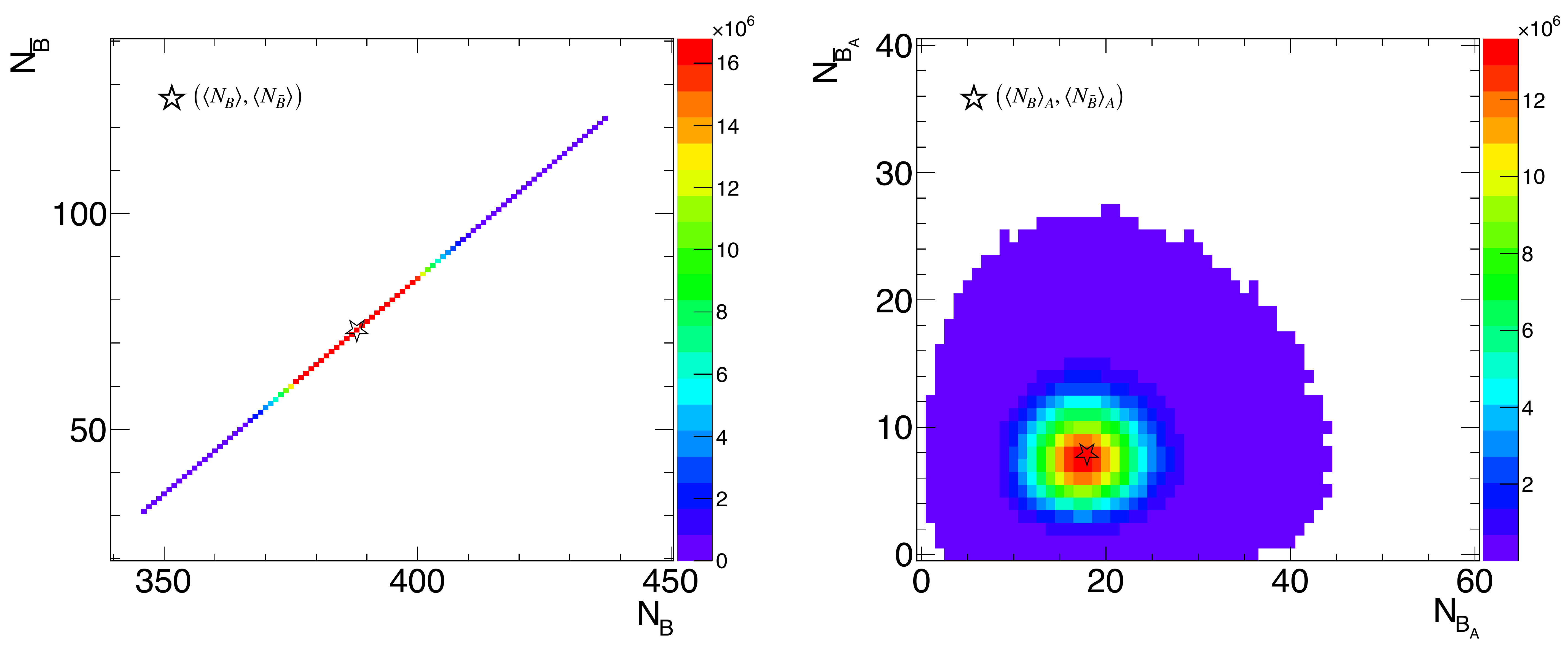}
    \caption{Left panel: Density plot of the number of baryons and antibaryons in  full phase space generated in a Monte Carlo simulation based on the canonical ensemble, using Eq.~(\ref{prob}). While the number of baryons and antibaryons fluctuates from event-to-event, the correlation imposed by baryon number conservation eliminates fluctuations of the net baryon number $N_B-N_{\bar{B}}$. Right panel: The correlation between the baryon and antibaryon numbers measured in a limited acceptance window is largely lifted. The stars mark the mean numbers.}
\label{fig:MonteCarlo-canonical}
\end{figure} 
From Eq.~(\ref{eq:canonical-partition2}) it follows that, for a given value of net baryon number $B$ of the full system, the underlying normalized canonical probability distribution for $N_{\bar{B}}$ is given by\footnote{Equation (\ref{prob}) is also obtained by taking the limit $\alpha_B, \alpha_{\bar{B}}\to 1$ in (\ref{eq:probability-acceptance-1}) and integrating over $N_A$. Note that in this limit $N_{\bar{A}}=N_{\bar{B}}$.}  
\begin{eqnarray}\label{prob}
P_B( N_{\bar{B}})=\frac{1}{I_{B}(2z)}\frac{z^{B}z^{2N_{\bar{B}}}}{(N_{\bar{B}}+B)!N_{\bar{B}}!}.
\end{eqnarray}

We first randomly generate the number of antibaryons $N_{\bar{B}}$ and baryons $N_{\bar{B}}+B$ using Eq.~\ref{prob}. 
Next, in order to obtain non-zero net baryon fluctuations, we apply an acceptance folding procedure based on empirical baryon and antibaryon rapidity distributions. 

This is illustrated in Fig.~\ref{fig:MonteCarlo-canonical}, where we show the simulated number of baryons and antibaryons, both in full phase space (left panel) and in a limited acceptance (right panel). Here we used the empirical rapidity distributions of baryons (protons) and antibaryons (antiprotons) for Au-Au collisions at $\sqrt{s_{NN}}=$ 62.4 GeV, discussed in section~\ref{sec:rapidity-distributions}. 

\begin{figure}[!htb]
\centering
\includegraphics[width=.495\linewidth,clip=true]{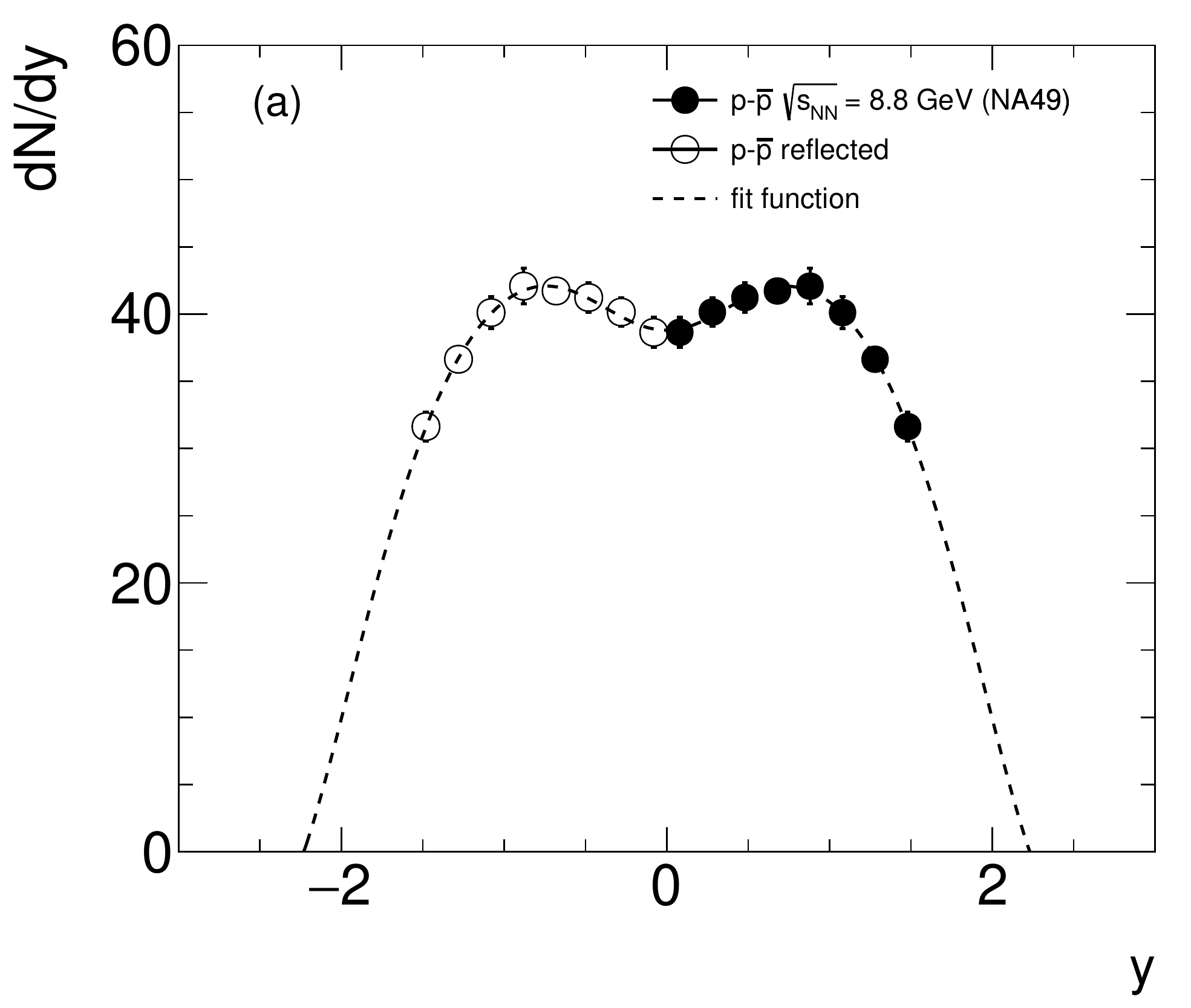}
\includegraphics[width=.495\linewidth,clip=true]{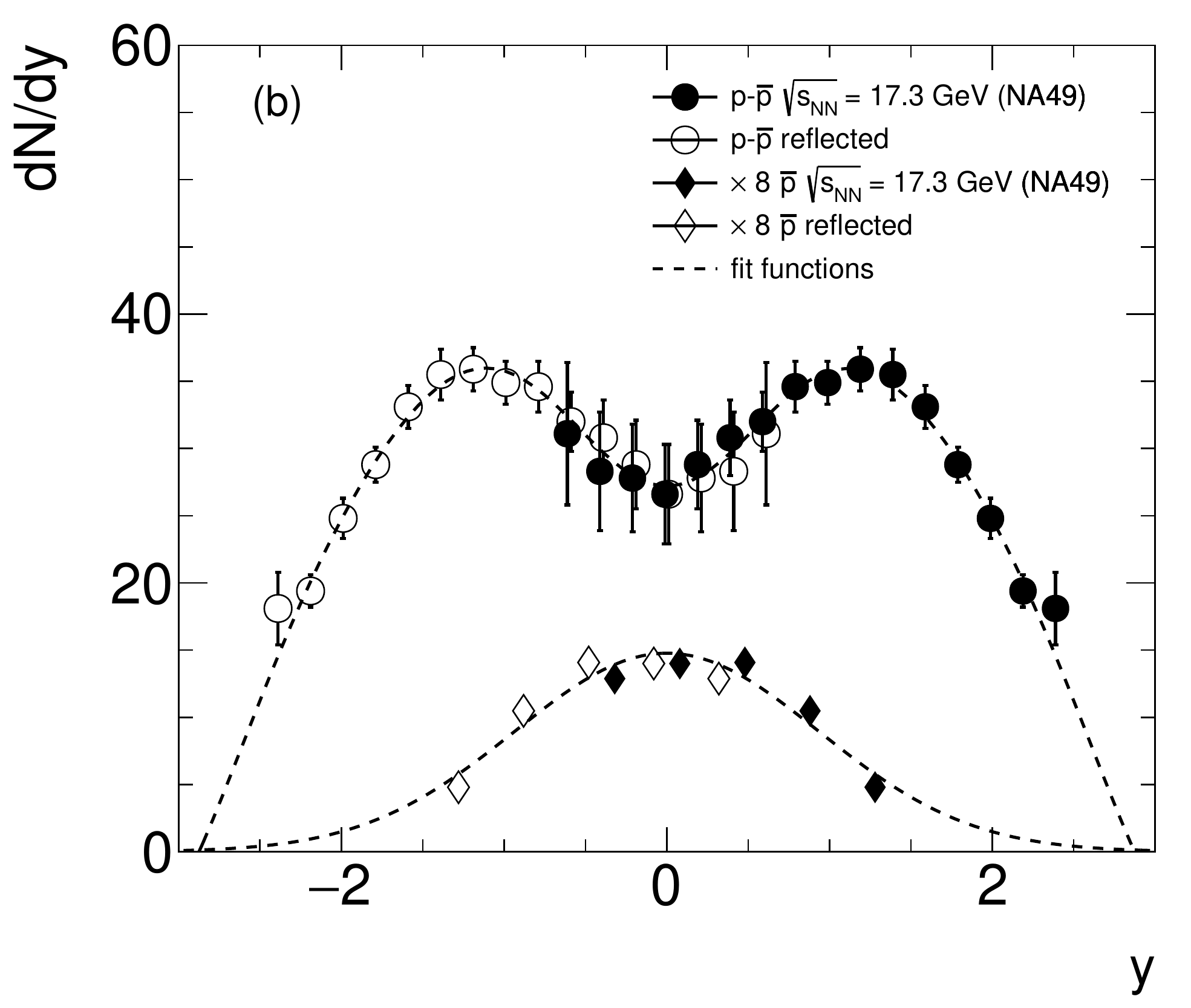}
\includegraphics[width=.495\linewidth,clip=true]{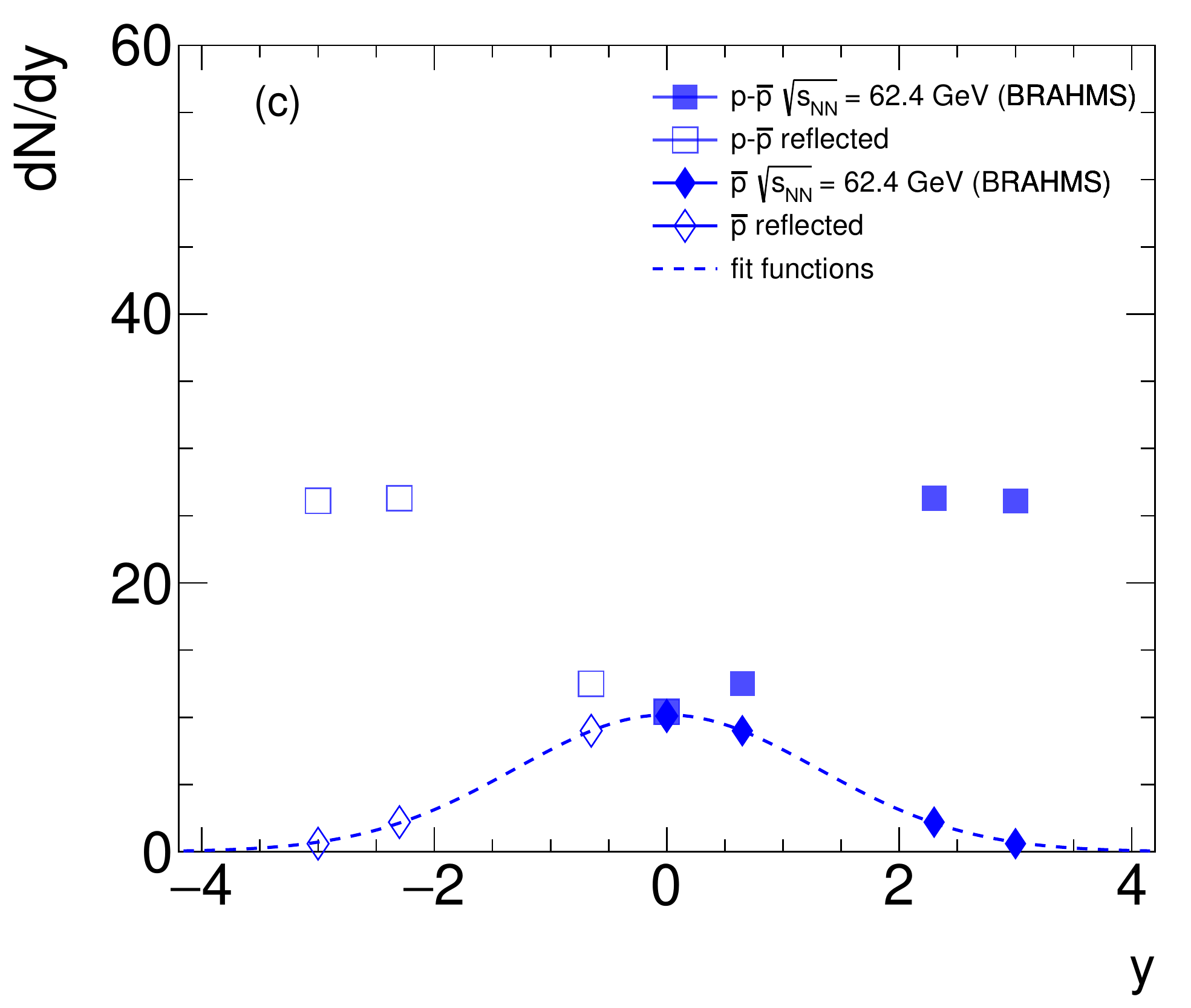}
\caption{Panel (a): Rapidity distributions of protons (black solid circles) and their reflected values (black open circles) as measured by the NA49 collaboration in central Pb--Pb collisions at beam momentum of 40$A$ GeV/$c$ ($\sqrt{s_{NN}}$= 8.8 GeV). Lines represent a parametrization of the data with a polynomial plus a Gaussian function.
Panel (b): Similar to the  panel (a) for beam momentum of 158$A$ GeV/$c$ ($\sqrt{s_{NN}}= 17.3$ GeV) , where also antiproton data (black diamonds) and their reflected values (open diamonds) are presented. Antiproton distributions  are fitted with a single Gaussian function. For a better visibility the antiproton distribution is scaled by a factor of 8. Panel (c). Si\-mi\-lar to panel (b) for BRAHMS data at $\sqrt{s_{NN}}= 62.4$ GeV. Only measurements in four rapidity bins are available in this case. For references to the original data see text.}
\label{fig:rapdist}
\end{figure}

As illustrated in the left panel of Fig.~\ref{fig:MonteCarlo-canonical}, the net baryon number does not fluctuate in full phase space. These results were obtained by counting all particles generated with the probability distribution Eq.~(\ref{prob}). On the other hand, as shown in the right panel of Fig.~\ref{fig:MonteCarlo-canonical}, the net baryon number in a limited acceptance is not conserved but rather fluctuates from event to event\footnote{Since the protons and antiprotons constitute only a part of the baryons and antibaryons in the system, the net number of protons is not conserved but rather fluctuates even in full acceptance.}.

As noted in Sect.~\ref{sec:FlucCE} above, in the absence of isospin correlations, the probability distribution for net protons is obtained from (\ref{eq:probability-acceptance-2}), by replacing the probabilities $\alpha_B$ and $\alpha_{\bar{B}}$ by the corresponding ones for protons and anti\-protons, $\alpha_p$ and $\alpha_{\bar{p}}$. The latter are needed for further analysis at all relevant energies. They are determined empirically below by using the mean number of protons and antiprotons obtained from the SPS, RHIC and LHC  experiments (see section~\ref{sec:rapidity-distributions}). 

For the special case of a binomial acceptance folding we can use Eq.~(\ref{eq:probability-acceptance-1}) to generate the number of baryons and antibaryons in a given acceptance, defined by the parameters $\pb$ and $\pbb$. However, in the presence of, e.g., short-range correlations there would be additional correlations between produced baryons and antibaryons in momentum space, which are not captured by Eq.~(\ref{eq:probability-acceptance-1}). On the other hand, a simulation procedure starting from the canonical probability distribution for the full system (Eq.~(\ref{prob})) and imposing the acceptance cuts on the generated particles allows us to consider more general cases, where the acceptance folding is not binomial. For example, in~\cite{Braun-Munzinger:2019yxj} some of us explored the effect of particle production combined with short-range correlations in rapidity space. In the present work we consider only long range correlations, consistent with the recent results of the ALICE collaboration~\cite{Acharya:2019izy}. Thus, in the acceptance folding procedure we consider only binomial losses. If data at lower energies would require the presence of shorter range correlations, this can easily be incorporated into the simulations.  The simulation module is based on $p_T$ integrated rapidity distributions of baryons, antibaryons, protons and antiprotons. 

\subsection{Rapidity distributions and their parametrization}
\label{sec:rapidity-distributions}
In order to determine the acceptances for protons and antiprotons covered by the STAR measurements, the essential ingredients of our simulations are rapidity distributions of protons and antiprotons for the full phase space. Since the STAR measurements are at mid-rapidity only, we will resort to other data available from RHIC and the SPS with a larger phase space coverage. In Fig.~\ref{fig:rapdist} experimental rapidity distributions are presented as obtained for net protons and antiprotons by the NA49~\cite{Appelshauser:1998yb,Anticic:2010mp} and BRAHMS~\cite{Arsene:2009aa} collaborations at $\sqrt{s_{NN}}=$ 8.8, 17.3 and 62.4 GeV for Pb-Pb and Au-Au collisions, respectively. 

At 8.8 GeV the amount of antibaryons is negligibly small, hence at this energy we assume proton and net proton rapidity distributions to be identical. The precision and large acceptance coverage of the NA49 measurements allows a fit of the measured rapidity distributions with analytic functions, also shown in the figure. 

\begin{figure}[!htb]
\centering
\includegraphics[width=.495\linewidth,clip=true]{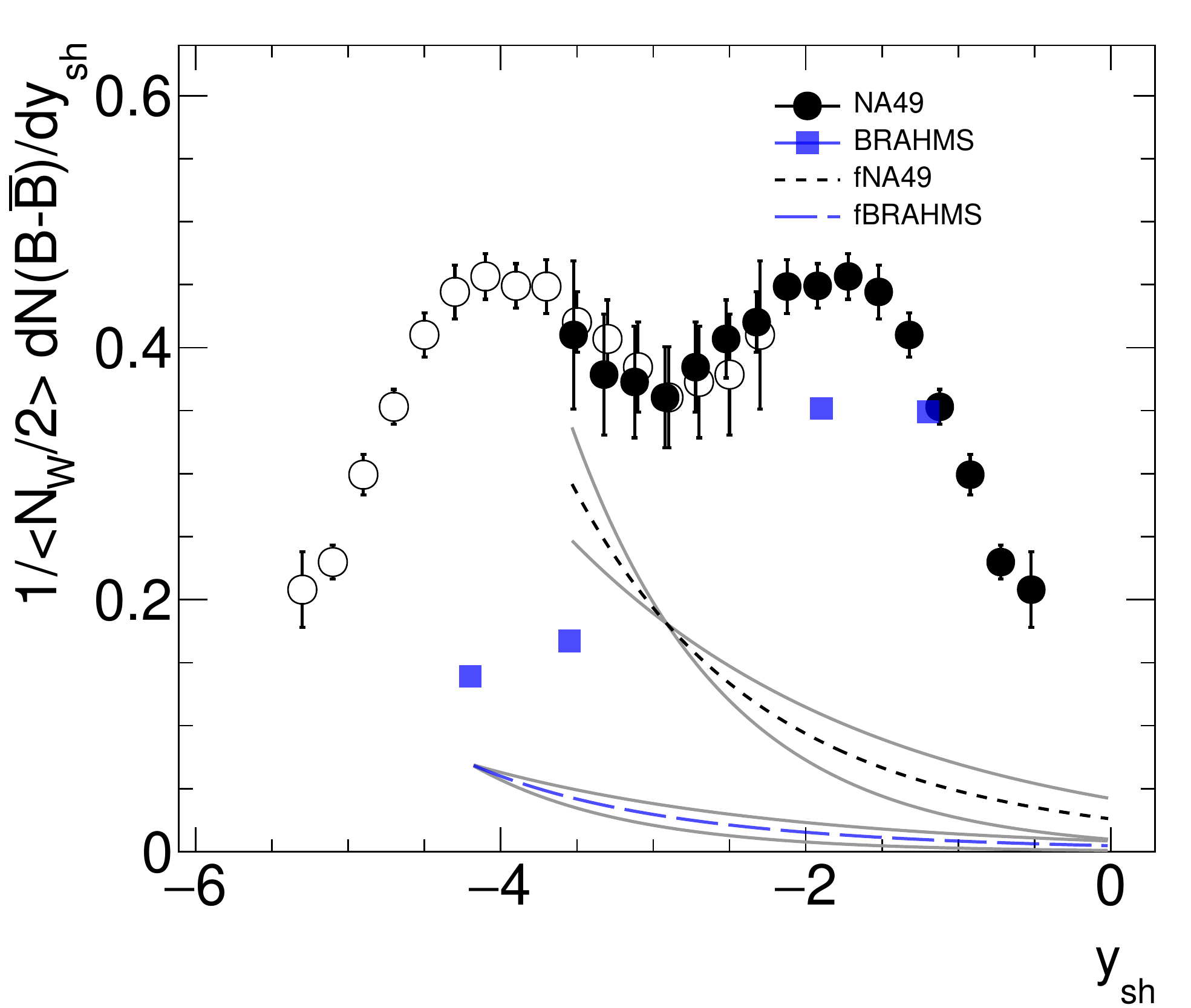}
\includegraphics[width=.495\linewidth,clip=true]{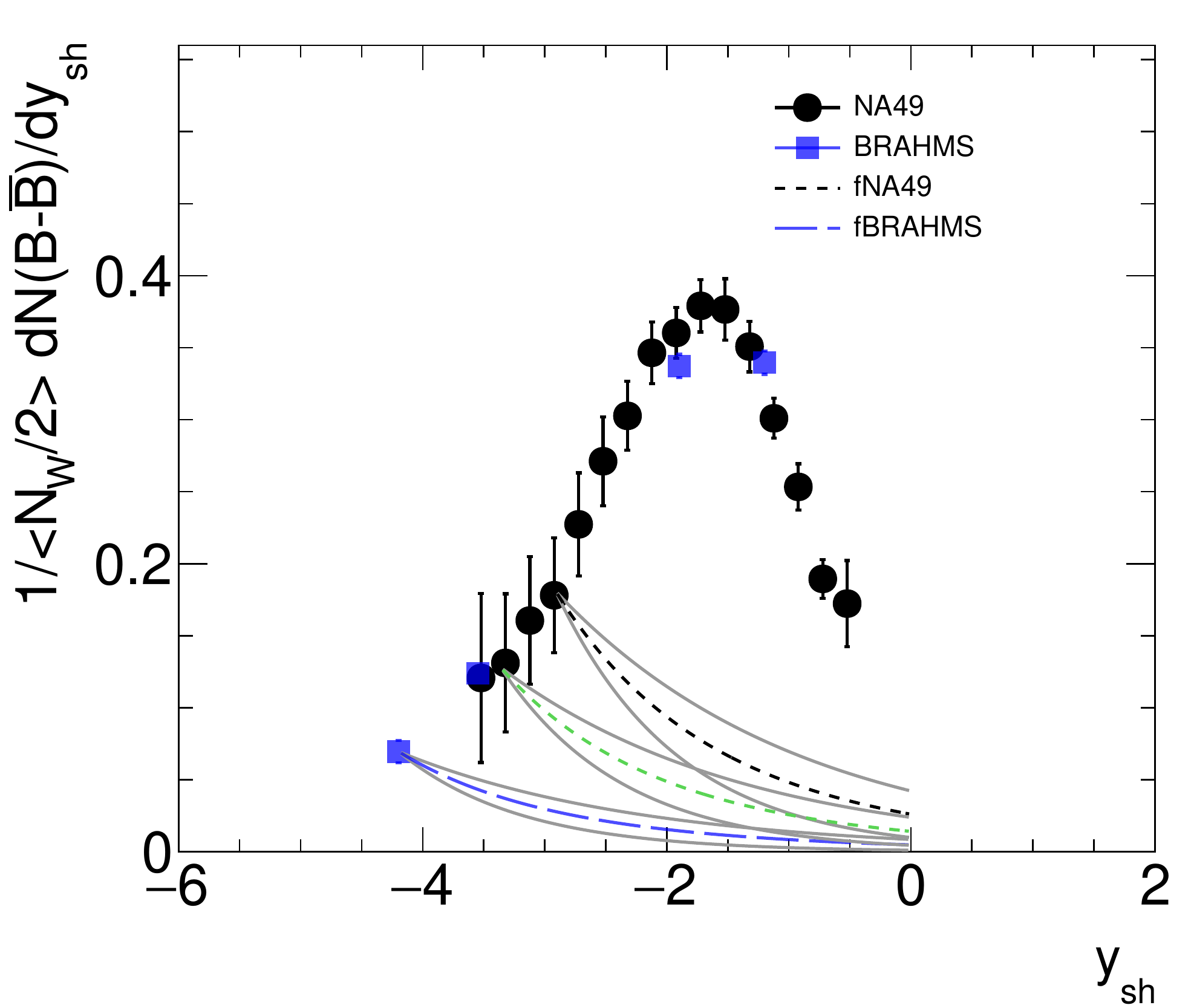}
\caption{Left panel: Normalized net baryon rapidity densities as measured by the NA49~\cite{Appelshauser:1998yb}
 collaboration (black solid circles) and their reflected values (black open circles) in central Pb--Pb collisions at beam momentum of 158$A$ GeV/$c$ ($\sqrt{s_{NN}}= 17.3$ GeV), plotted in the beam rapidity frame. Here, the shifted rapidity is defined as $y_{sh}=y-y_{b} $ with $y_{b}$ the beam rapidity in the nucleon-nucleon center-of-mass frame. The blue solid squares represent the corresponding BRAHMS data~\cite{Arsene:2009aa} at $\sqrt{s_{NN}}= 62.4$ GeV. The black dashed and blue long dashed lines, calculated with Eq.~\ref{eq:targetCont}, represent the contributions from the 'target' regions for NA49 and BRAHMS data, respectively. Right panel: Normalized net baryon rapidity densities after subtracting the corresponding target contributions. In addition the 'target' contribution for $\sqrt{s_{NN}} = 27$ GeV is shown as green dashed line.}
\label{fig:beam-rapdist}
\end{figure} 

The antiproton distributions measured by both, the NA49 and BRAHMS collaborations, can be well parametrized by a Gaussian distribution. The corresponding values are added to the net proton rapidity distributions in order to recover, together with the fit to the net proton distributions, a full rapidity distributions of protons. For BRAHMS data, however, this approach is not directly applicable, because the underlying rapidity distribution of net protons is not complete enough, with only 4 measured rapidity bins as presented in panel (c) of Fig.~\ref{fig:rapdist}. To address this problem we present the net baryon rapidity densities, also given by NA49 and BRAHMS, in the beam rapidity frame, 
$y_{sh}=y-y_{b} $ 
with $y_{b}$ the beam rapidity in the nucleon-nucleon center-of-mass frame. This is inspired by the phenomenologically well established concept of limiting fragmentation \cite{Benecke,Stasto}. In Fig.~\ref{fig:beam-rapdist} we present the net baryon rapidity densities at $\sqrt{s_{NN}}=$ 17.3 and 62.4 GeV. In order to make the data comparable the distributions are normalized to the mean number of wounded nucleons for either experiment. For this purpose we scaled the net-baryon distribution of NA49 such that its integral yields the mean number of wounded nucleons $\langle N_{W}\rangle = 352$ in the most central 5\% of the events. The BRAHMS net baryon data for the most central 10\% are already normalized to 314 wounded nucleons in~\cite{Arsene:2009aa}. 

\begin{figure}[!htb]
    \centering
   \includegraphics[width=.495\linewidth,clip=true]{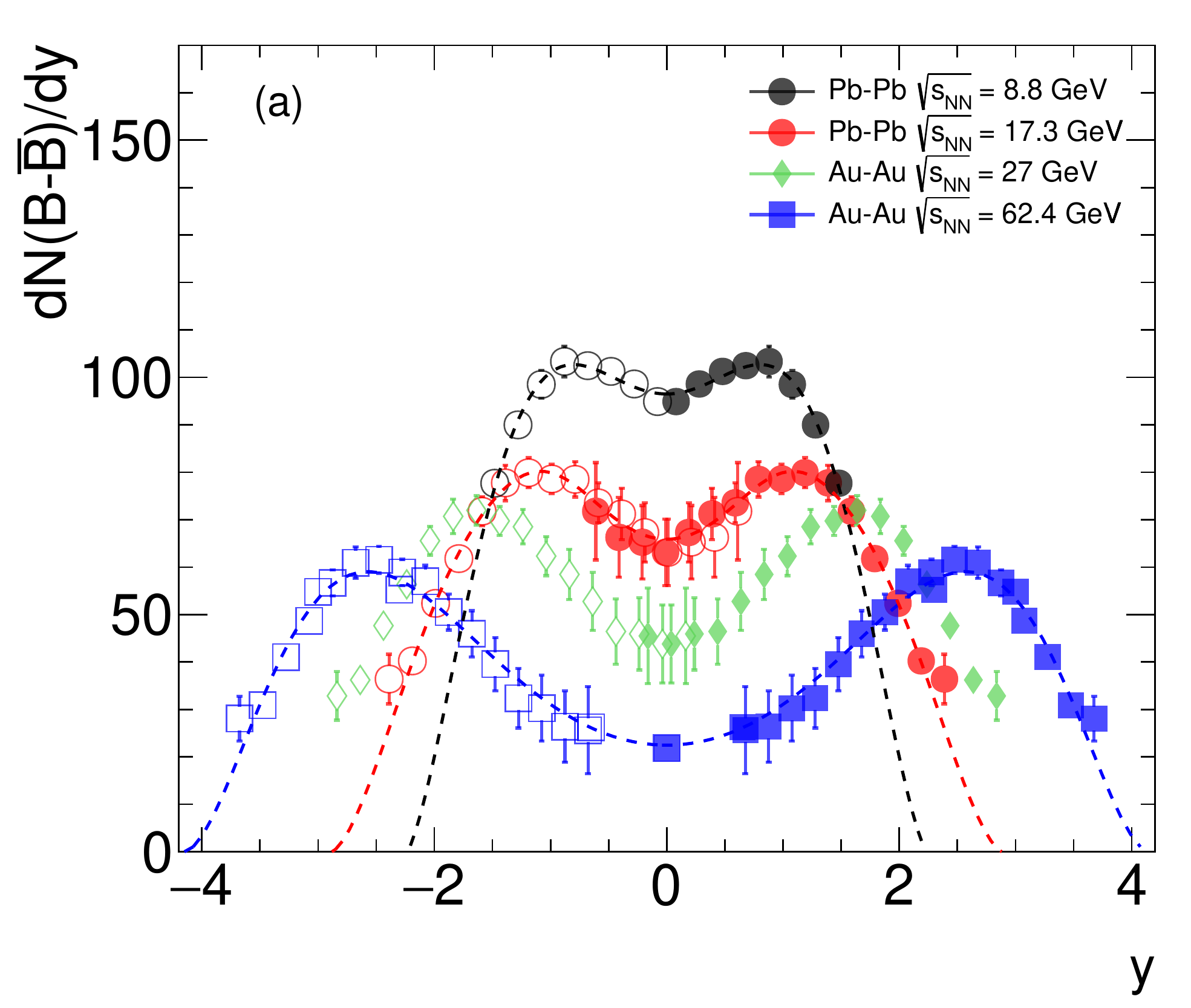}
   \includegraphics[width=.495\linewidth,clip=true]{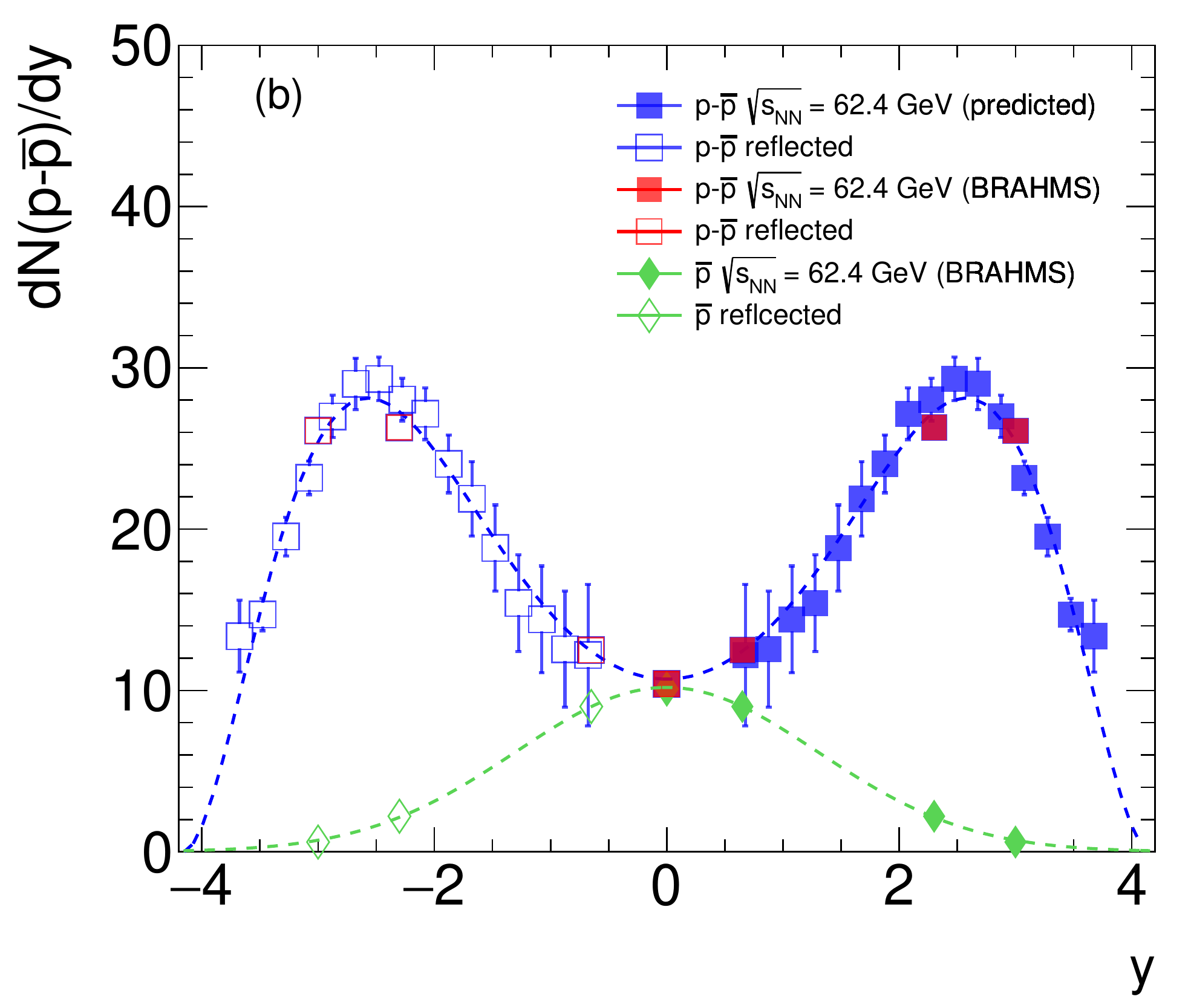}
    \caption{(a): Rapidity distributions of net baryons at $\sqrt{s_{NN}} =$ 8.8 and 17.3 GeV (measured distributions from NA49) and 27 and 62.4 GeV (constructed using the limiting fragmentation concept described in the text). (b): Constructed (blue symbols) and BRAHMS measured (red symbols) rapidity distributions of net-protons at  $\sqrt{s_{NN}} = 62.4$ GeV.}
    \label{fig:rapdistBRAHMS}
\end{figure} 

A central aspect of limiting fragmentation is the possibility to compare data at different collision energies. To this end it is useful to separate the net baryon distribution into a beam and target component, noting that at mid-rapidity these components are equal and at beam rapidity the target component should be negligible. We follow here a procedure also used by \cite{Arsene:2009aa} where the components from the target region are parametrized as the average of two exponential functions based on \cite{frag1:1984, frag2:1989}

\begin{figure}[!htb]
    \centering
    \includegraphics[width=.495\linewidth,clip=true]{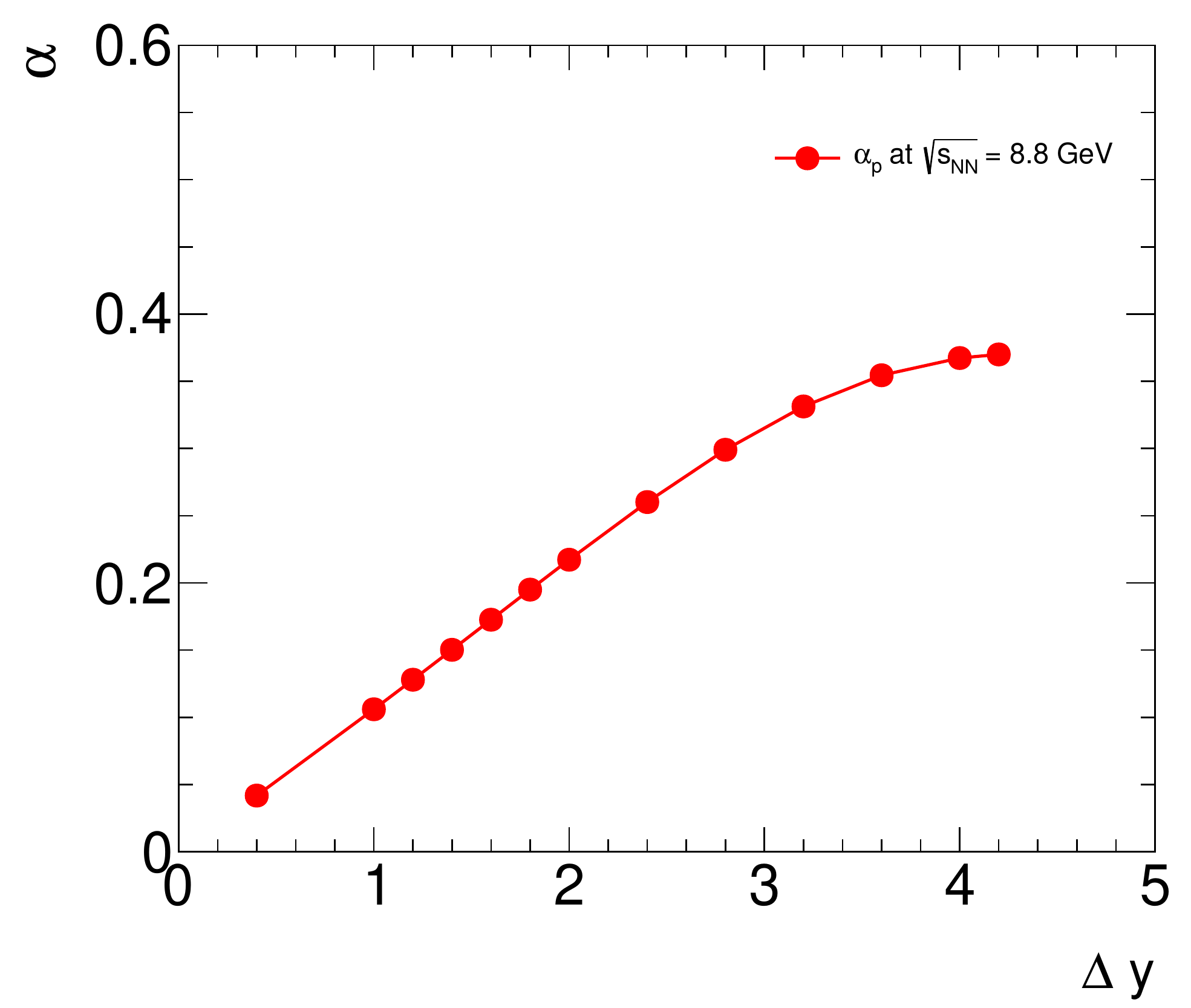}
    \includegraphics[width=.495\linewidth,clip=true]{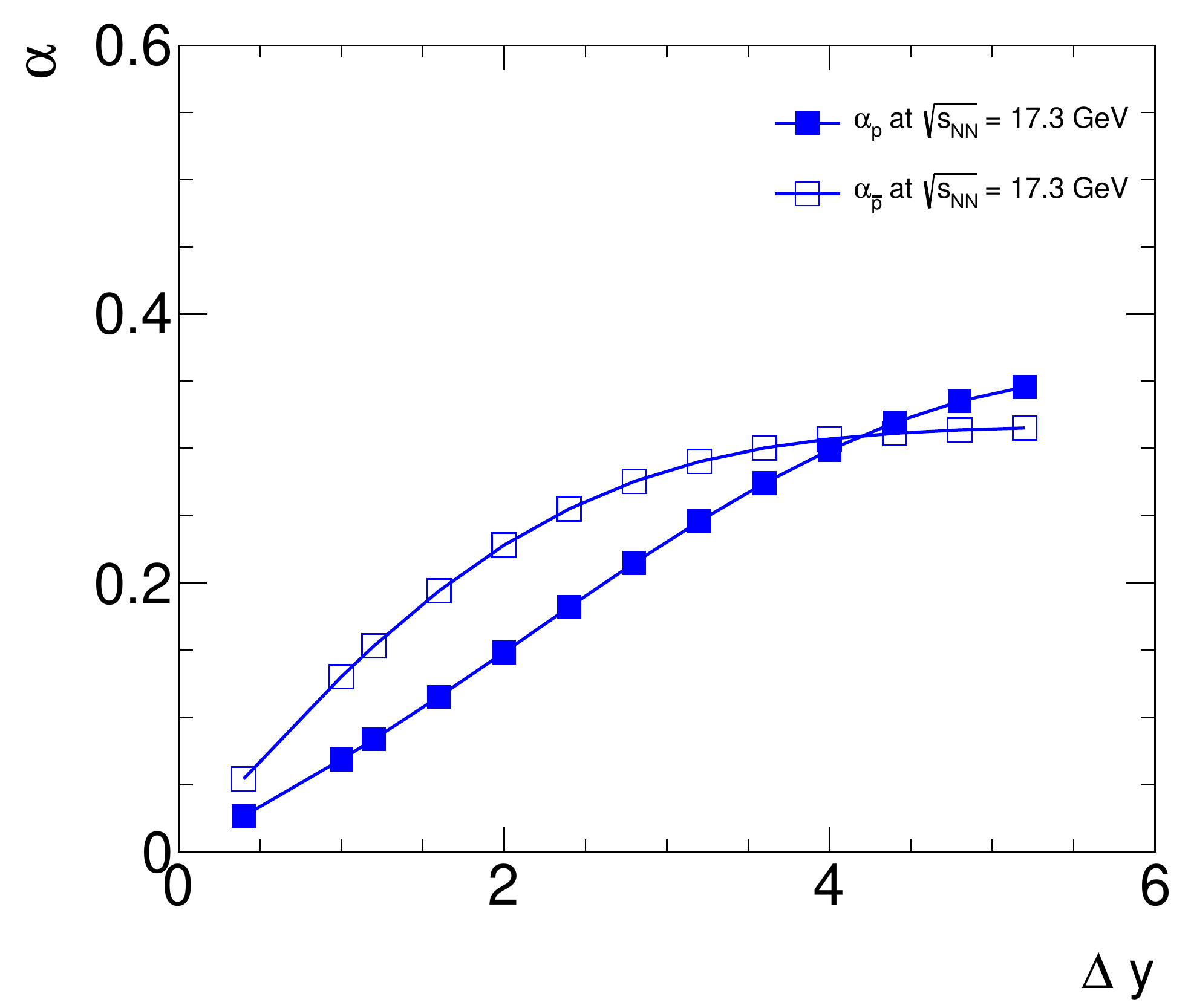}
    \includegraphics[width=.495\linewidth,clip=true]{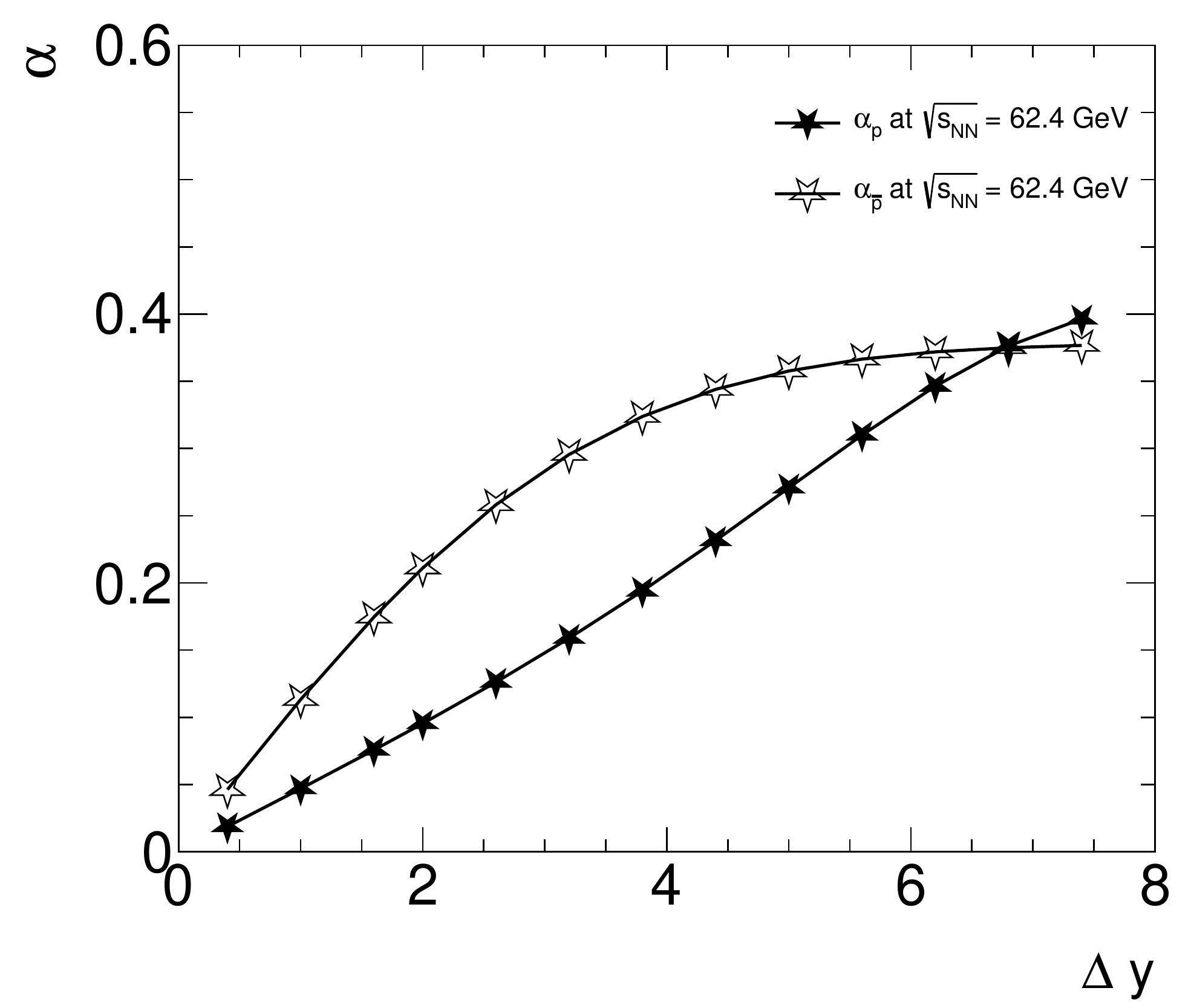}
    \includegraphics[width=.495\linewidth,clip=true]{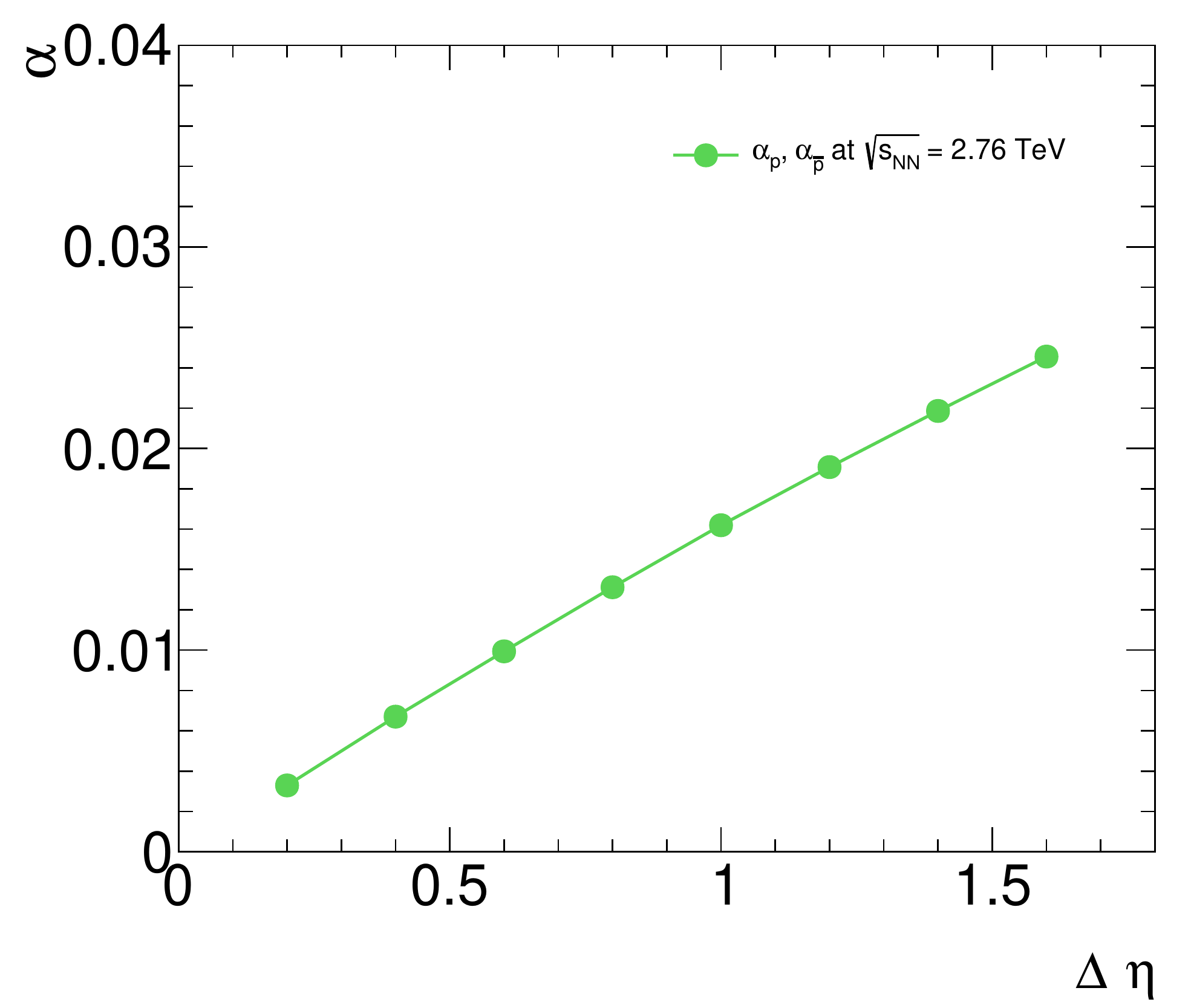}
    \caption{Ratio of accepted protons (antiprotons) to the number of baryons (antibaryons) in the full phase space for the energy region from low SPS  to full LHC energy.}
    \label{fig:alpha-values}
\end{figure} 

\begin{equation}
f_y(y_{sh}) =C_{sh}\frac{\exp(-y_{sh}) + \exp(-y_{sh}/2)}{2},
\label{eq:targetCont}
\end{equation}
where the normalization factor $C_{sh}$ is fixed from symmetry considerations to yield half of the measured net baryon densities at mid-rapidity. The functions $f_y(y_{sh})$ for the BRAHMS and NA49 data are illustrated in Fig.~\ref{fig:beam-rapdist} with the black (dashed) and blue (long dashed) lines, respectively. In the right panel of Fig.~\ref{fig:beam-rapdist} the NA49 and BRAHMS data are presented after subtracting the parametrized target contributions. The four measured BRAHMS data points, presented by the solid blue squares, closely follow the NA49 data represented with the solid black circles, thereby establishing a common baseline for the net baryon distribution in the beam fragmentation region. In a similar way the green dashed line in the right panel of Fig.~\ref{fig:beam-rapdist} represents the target contribution for the intermediate energy $\sqrt{s_{NN}} = 27$ GeV. 

Using this common baseline for the beam fragmentation region for the not so well measured forward distributions and adding back the corresponding target distributions we retrieve nearly complete rapidity distributions of net-baryons at $\sqrt{s_{NN}} = 62.4$ and 27 GeV.
The results, after boosting back to the nucleon-nucleon center of mass system, are presented in the left panel of Fig.~\ref{fig:rapdistBRAHMS}, where blue and green symbols are our constructed rapidity densities for $\sqrt{s_{NN}} = 62.4$ and 27 GeV, respectively. The remaining small gap to beam rapidity is filled by a fit, where we took care that the fit functions, defined as Gaussian plus polynomial distributions and shown as dashed lines in the figure, 
vanish at beam rapidity. In the right panel of Fig.~\ref{fig:rapdistBRAHMS} the net proton distribution at $\sqrt{s_{NN}} = 62.4$ is presented using the conversion factor between baryons and protons quoted in \cite{Arsene:2009aa}. The four red symbols correspond to the actual BRAHMS data points. The entire proton rapidity distribution is recovered by adding the contributions from antiprotons, as shown above and fitted by a Gaussian to cover the entire rapidity range. The measured antiproton points are represented by the green diamonds in the right panel of Fig.~\ref{fig:rapdistBRAHMS}.

\subsection{Comparison of net proton cumulants to current experimental data}
\label{sec:STAR data}

The net proton and net baryon distributions derived in the previous subsection for the four collision energies will be used to determine the acceptance for the net proton cumulants measured by STAR. To this end, we first compute the number of baryons $\langle N_B\rangle$ and antibaryons $\langle N_{\bar{B}}\rangle$ in full phase space. Since at $\sqrt{s_{NN}}$ = 8.8 GeV, only an upper limit on the number of produced antibaryons is known (less than 1 \% of the number of baryons), we set $\langle N_{\bar{B}}\rangle$ = 2 and $\langle  N_B\rangle = \langle N_{W} \rangle + 2$. We note, however, that the contribution of the antiprotons to the cumulants is very small at this energy. Thus, they are to a very good approximation given by the binomial cumulants, discussed in Appendix C. In the simulation we therefore use $\langle N_{\bar{B}}\rangle=\langle N_{\bar{p}}\rangle=0$ and $\langle N_{W} \rangle$ = 351 at $\sqrt{s_{NN}}=8.8$ GeV. 

For $\sqrt{s_{NN}}$ = 17.3 GeV we estimate the total number of antibaryons as

\begin{equation}
    \langle N_{\bar{B}}\rangle = \langle \bar{p} + \bar{n} + \bar{\Lambda} + \bar{\Sigma}^{0} + \bar{\Sigma}^{+} + \bar{\Sigma}^{-} + \bar{\Xi}^{+} + \bar{\Xi^{0}} + \bar{\Omega}\rangle,
\end{equation}
where  $\langle \bar{p} \rangle $ is obtained by integrating the antiproton rapidity distribution presented in the panel (b) of Fig.~\ref{fig:rapdist}, $\langle \bar{\Lambda} + \bar{\Sigma}^{0} \rangle $  and $\langle \bar{\Xi}^{+} \rangle $ are obtained from~\cite{Alt:2008qm}\footnote{The numbers provided in this reference are for the 10 \% most central collisions. We transformed these numbers to the  5 \% most central collisions by assuming that, in central collisions,  mean multiplicities are proportional to the mean number of wounded nucleons.}, and $\langle \bar{\Sigma}^{+} \rangle $  and $\langle \bar{\Omega}\rangle $ multiplicities are calculated using their ratios to antiprotons as obtained from the HRG model~\cite{Andronic:2017pug}. The multiplicities  $\langle \bar{n} \rangle $, $\langle \bar{\Xi}^{0} \rangle $,  $\langle \bar{\Sigma}^{-} \rangle $ are estimated using isospin symmetry. For $\sqrt{s_{NN}}$ = 62.4 GeV the number of baryons and antibaryons are fixed using two conditions: (i) $\langle N_{B} - N_{\bar{B}}\rangle $ = $\langle N_{W} \rangle $ and (ii) $\langle N_{\bar{B}} \rangle /\langle N_{B}\rangle$ = $\langle N_{\bar{p}}\rangle / \langle N_{p}\rangle $.  The resulting numbers for the three energies are presented in Table~\ref{tab:inputValues}, where we also provide the corresponding numbers of protons, antiprotons. 
For $\sqrt{s_{NN}}$ = 27 GeV the values are computed using a different procedure, discussed below. 

Results presented for ALICE energy are obtained with $\langle N_{\bar{B}}\rangle = 749$, $\langle N_B\rangle$ = B + $\langle N_{\bar{B}}\rangle$ and $\Delta y $ = 1. We neglect baryon transport to mid-rapidity, and thus set $B$ = 0. In order to explore the sensitivity to this assumption, we also considered  the other extreme with $B$ = $\langle N_{W}\rangle \approx$ 383. For the 6th cumulant we find that the resulting difference  is less than 6\%.  For lower cumulants the differences are negligible.

\begin{table}
\centering
 \begin{tabular}{||c | c | c | c | c | c| c||} 
 \hline
 $\sqrt{s_{NN}}$ & $\langle N_{B}\rangle$ & $\langle N_{\bar{B}}\rangle$ & $\langle N_{p}\rangle$ & $\langle N_{\bar{p}} \rangle$ & $z$ & $\gamma_{p}$ ($\gamma_{\bar{p}}$) \\ [0.5ex] 
 [GeV] &  &  &  &  &  & \\ [0.5ex] 
 \hline\hline
 8.8 & 353 & 2 & 130 & 0.51 & 26.61 & 0.9 ($-$) \\ 
 \hline
 17.3 & 368 & 16 & 154.6 & 4.36 & 76.83 & 0.8 (1.2)\\
 \hline
 27 & $373 \, (377)$ & $30 \, (34)$ & $-$ & $-$ & $105.91 \, (113.35)$ & $-$\\
 \hline
 62.4 & 384 & 70 & 181.5 & 33.23 & 164.13 & 0.9 (0.8) \\[0.5ex]
 \hline
\end{tabular}
\caption{\label{tab:inputValues} Estimated values of $\langle N_{B}\rangle$, $\langle N_{\bar{B}}\rangle$ and the corresponding values of the single–particle partition function $z$. For reference, we also provide the corresponding numbers of protons and antiprotons in full phase space. For $\sqrt{s_{NN}}$ = 27 GeV we provide lower and upper bounds (in brackets), see the text below. In the last column we show the values of $\gamma_{p}$  ($\gamma_{\bar{p}}$) as calculated using Eq.~(\ref{eq:normconditions}).}
\end{table}

We identify the mean multiplicities presented in Table~\ref{tab:inputValues}  with the particle numbers in the canonical ensemble. Thus, for each collision energy, we solve Eq.~(\ref{NCM}) for the single-particle partition function $z$. The resulting $z$ values are also presented in Table~\ref{tab:inputValues}. Alternatively one could use~Eq.(\ref{NCP}) to determine $z$. Owing to the exact conservation of the net baryon number, reflected in the identity (\ref{eq:net-baryon-number}), the two procedures are completely equivalent.

In the following we are comparing the results of our simulations based on $10^9$ events with the published STAR Au--Au  and ALICE Pb--Pb data. Before proceeding with the comparison a comment is in order. By applying the rapidity cut of $|y|<0.5$ we can effectively introduce the STAR rapidity coverage used for protons and antiprotons. However, it is quite intricate to account for all other effects present in the STAR data, such as cuts on transverse momenta, contributions from weakly decaying hadrons etc. In order to deal with this difficulty we use the following normalization conditions 

\begin{eqnarray}
\gamma_p\,\int_{-0.5}^{0.5}\left[\frac{dn_{p}}{dy}\right]dy=\left<p\right>,\,\,\ \gamma_{\bar{p}}\,\int_{-0.5}^{0.5}\left[\frac{dn_{\bar{p}}}{dy}\right]dy=\left<\bar{p}\right>, 
\label{eq:normconditions}
\end{eqnarray}
where $[dn_{p}/dy]$ and $[dn_{\bar{p}}/dy]$ are the rapidity distributions for protons and antiprotons presented in Figs.~\ref{fig:rapdist} and \ref{fig:rapdistBRAHMS}, while $\langle p\rangle$ and $\langle \bar{p}\rangle$ are the mean numbers of protons and antiprotons in the STAR acceptance. The so obtained values of $\gamma_{p}$ ($\gamma_{\bar{p}}$) are presented in the last column of Table~\ref{tab:inputValues}.  Next we proceed and compute for the four collision energies where we have constructed complete rapidity distributions the acceptance values and the effect of the finite acceptance on the net proton cumulants.

We use the STAR data measured with phase space coverage $|y|<0.5$ and $0.4<p_{T}<2$ GeV/$c$ directly for $\sqrt{s_{NN}}$ = 64.2 and 27 GeV and interpolate between nearby STAR measurements for 17.3 and 8.8 GeV to obtain the mean numbers of protons $\left<p\right>$ and antiprotons $\left<\bar{p}\right>$. The values of $\gamma_p$ and $\gamma_{\bar{p}}$ resulting from (\ref{eq:normconditions}) are then used to compute the acceptance parameters for protons and antiprotons

\begin{eqnarray}
\alpha_p = \frac{\gamma_{p}\int_{y_{min}}^{y_{max}}\left[\frac{dn_{p}}{dy}\right]dy}{\left < N_{B}\right >},\,\,\ \alpha_{\bar{p}} = \frac{\gamma_{\bar{p}}\int_{y_{min}}^{y_{max}}\left[\frac{dn_{\bar{p}}}{dy}\right]dy}{\left <N_{\bar{B}}\right>},
\label{eq:alphadef}
\end{eqnarray}
with $|y_{max} - y_{min}| = \Delta y$.

In Fig.~\ref{fig:alpha-values} we present the acceptance probabilities for protons and antiprotons as functions of the rapidity coverage $\Delta y$ from our simulations using (\ref{eq:alphadef}) for selected energies from low SPS to full LHC energy. 

\begin{figure}[!htb]
    \centering
    \includegraphics[width=.495\linewidth,clip=true]{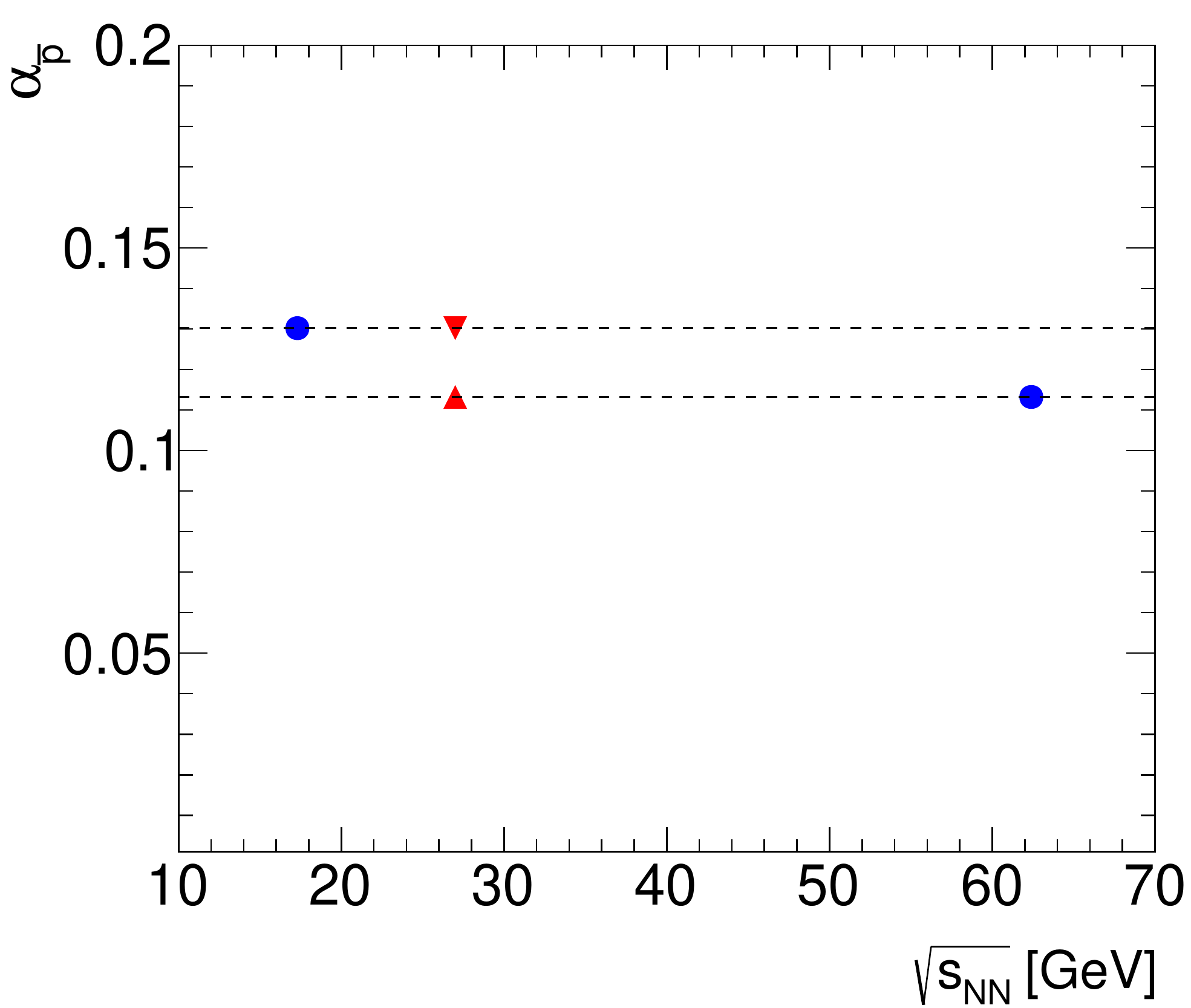}
    \includegraphics[width=.495\linewidth,clip=true]{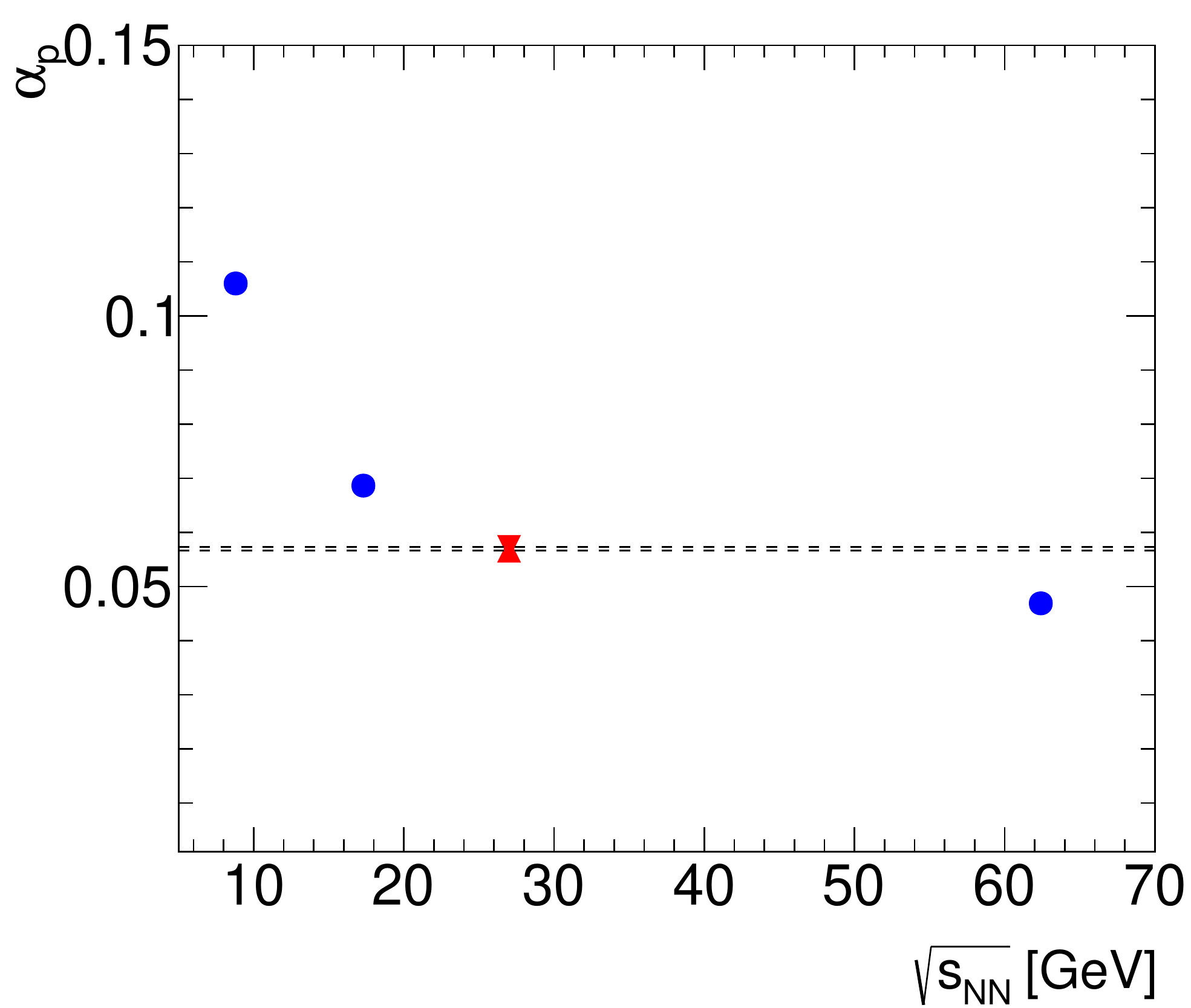}
    \caption{Collision energy dependence of $\alpha_{\bar{p}}$ (left panel) and $\alpha_{p}$ (right panel) for $\Delta y = 1$. The red triangles indicate the estimated upper and lower bounds for $\sqrt{s_{NN}} = 27$ GeV.}
    \label{fig:acc-27}
\end{figure} 

For $\sqrt{s_{NN}} =$ 27 GeV data we were able to construct only the net baryon rapidity distribution as presented in the left panel of Fig.~\ref{fig:rapdistBRAHMS}. The antiproton rapidity distribution is not measured at this energy and therefore we cannot determine the number of baryons and antibaryons separately from measured data alone. In order to obtain baryon and antibaryon numbers in full phase space we first inspect the energy dependence of $\alpha_{\bar{p}}$ for $\Delta y = 1$, illustrated in the left panel of Fig.~\ref{fig:acc-27}. Assuming a monotonic decrease of $\alpha_{\bar{p}}$ with increasing collision energy we obtain upper and lower bounds of $\alpha_{\bar{p}}$ indicated by the red triangles. Combining this with the measured mean multiplicities of antiprotons at $\sqrt{s_{NN}} = 27$ GeV we obtain the corresponding lower and upper limits for the mean number of antibaryons in full phase space. The number of baryons in full phase space is then computed as the sum of the number of wounded nucleons $\langle N_W\rangle$ at 27 GeV and the number of antibaryons. The resulting values are presented in Table~\ref{tab:inputValues}. In the right panel of Fig.~\ref{fig:acc-27} we show the corresponding  energy dependence of $\alpha_{p}$. The point at $\sqrt{s_{NN}} = 27$  GeV, estimated as just described, nicely follows the smooth trend established by our values at other energies indicated by the filled blue circles. All cumulants at $\sqrt{s_{NN}} = 27$ GeV presented below are calculated as averages of the values obtained using the lower and upper limits of $\langle N_{B}\rangle$, $\langle N_{\bar{B}}\rangle$, $\alpha_{p}$ and $\alpha_{\bar{p}}$ while the differences are attributed to the  systematic uncertainties.  

Also shown in Fig.~\ref{fig:alpha-values} are the $\alpha$ values for ALICE at $\sqrt{s_{NN}}$ = 2.76 TeV, where acceptances are introduced in the pseudo-rapidity and momentum space, from $|\eta|<0.1$ to $|\eta|<0.8$ and for $0.6 < p < 1.5$ GeV/$c$~\cite{Acharya:2019izy}. Due to the large increase in beam rapidity they exhibit a significant reduction in the acceptance probabilities compared to those obtained for the RHIC energies (see also the acceptance discussion in the context of Fig.~\ref{fig:kn-to-k2} below). 

\begin{figure}[!htb]
    \centering
    \includegraphics[width=.495\linewidth,clip=true]{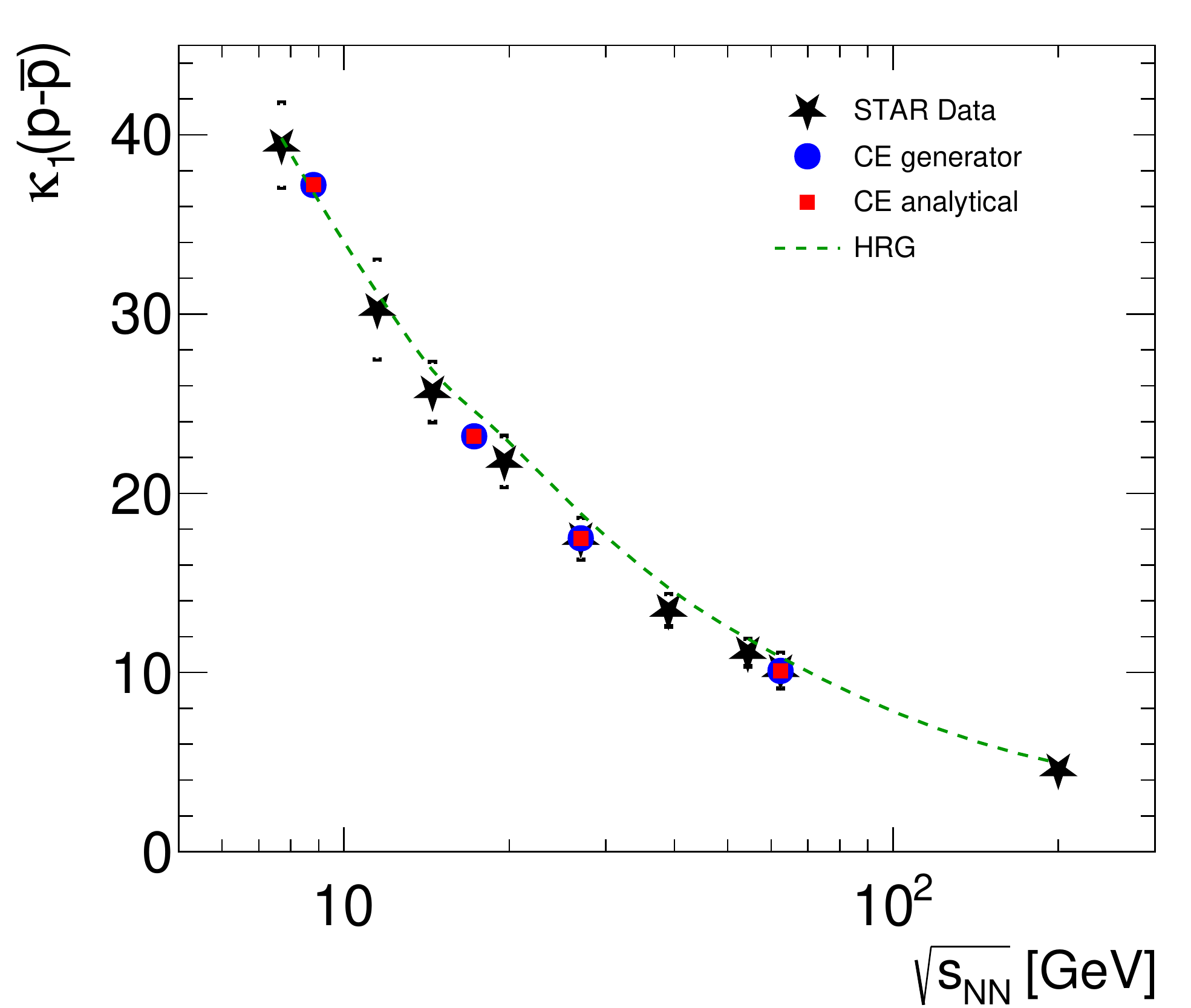}
    \includegraphics[width=.495\linewidth,clip=true]{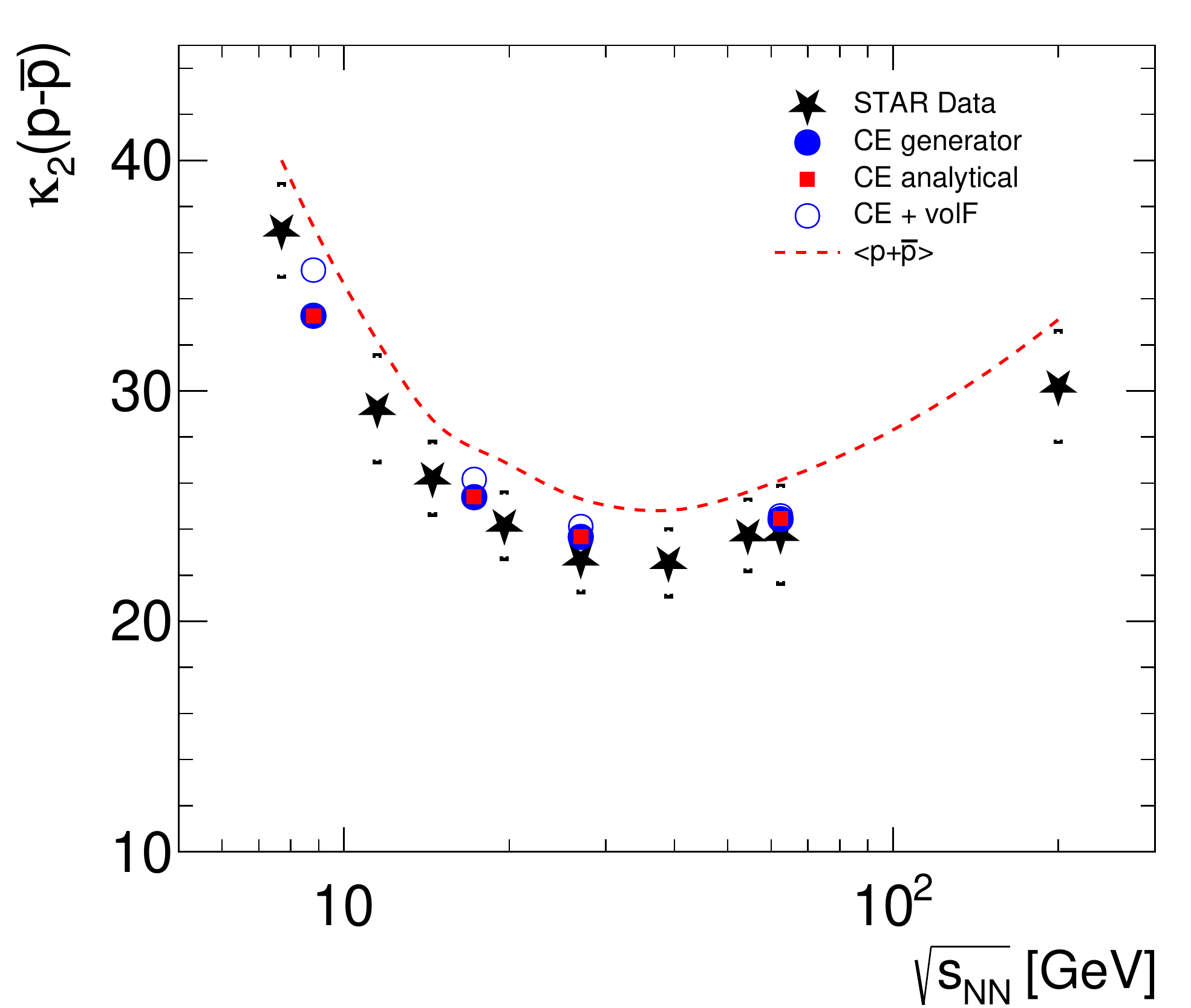}
    \includegraphics[width=.495\linewidth,clip=true]{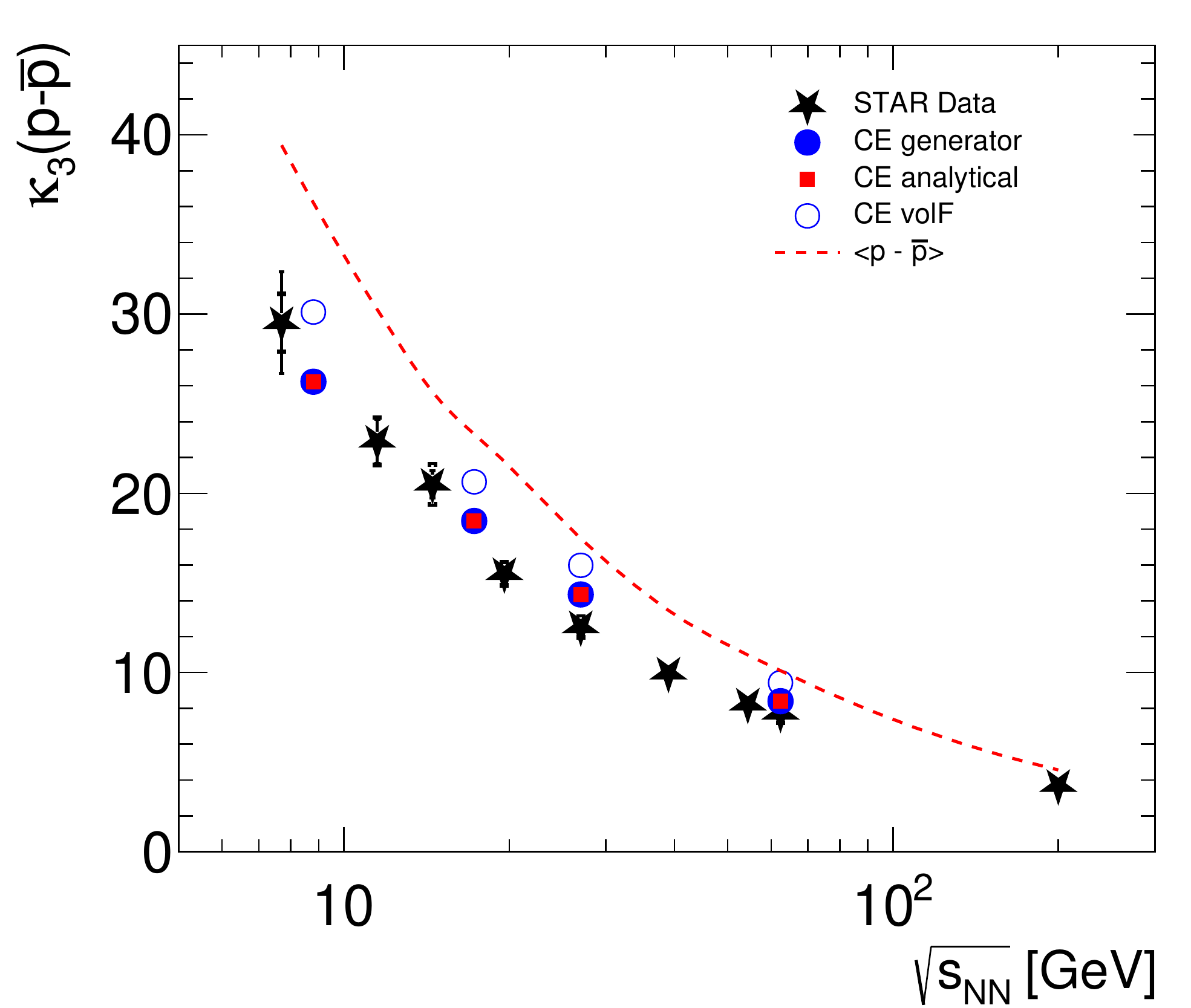}
    \includegraphics[width=.495\linewidth,clip=true]{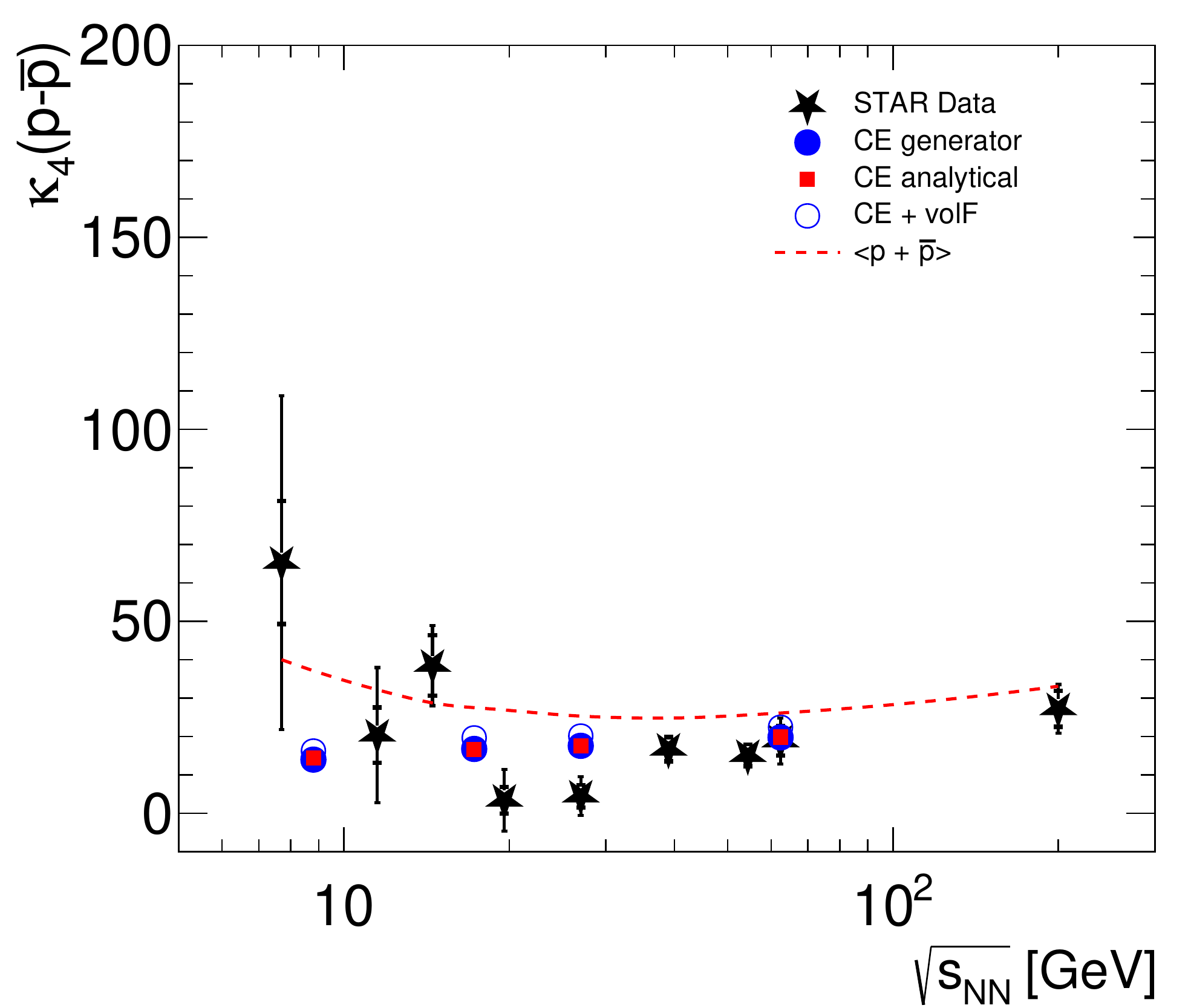}
    \caption{Cumulants of the net proton distributions. The red dashed line shows the HRG~\cite{Andronic:2017pug} baseline. The blue circles and red squares correspond to simulations and analytical calculations, respectively, accounting for baryon number conservation as presented in this work. They should be compared to the black symbols representing experimental results from the STAR collaboration \cite{Adam:2020unf}. Open blue circles include the additional contribution from reaction volume fluctuations. The STAR data are corrected with the Centrality Bin Width Correction method~\cite{Luo:2013bmi, Braun-Munzinger:2016yjz}, hence contributions from volume fluctuations in data, compared to the open circles, are suppressed.}
\label{fig:kn}
\end{figure} 

Using these acceptance values and baryon as well as antibaryon multiplicities, the net proton cumulants of arbitrary order can be computed  within the canonical formalism developed in section~\ref{sec:FlucCE}. They are presented in Fig.~\ref{fig:kn} both for the analytical calculations and the simulations. The results using the two methods are in excellent agreement with each other. Also shown are the experimental results of the STAR collaboration \cite{Adam:2020unf}. As can be seen in Fig.~\ref{fig:kn} the energy dependence of the net proton cumulants $\kappa_{n}$ are in remarkable agreement with the STAR data. The agreement for $\kappa_{1}$ is shown only as a consistency check, since this experimental information is used as input for our calculations. To demonstrate the importance of the canonical corrections for baryon number conservation, also the grand-canonical HRG baseline is shown. For even-order cumulants it is given by the sum of the measured protons and antiprotons in the acceptance $\langle p + \bar{p}\rangle$ and for the odd ones by $\langle p - \bar{p}\rangle$. Since the $\kappa_{1}$ values are used as input to the model, instead in Fig.~\ref{fig:kn} the HRG baseline is illustrated as tanh$(\mu_{B}/T_{ch})\times \langle p + \bar{p}\rangle$, where $\mu_{B}$ and $T_{ch}$ are the baryon chemical potential and temperature at chemical freeze-out, obtained by fitting the STAR hadron multiplicities with the  HRG model in the grand-canonical ensemble. As expected, all $\kappa_{n}$ ($n = 2, 3, 4$) values are suppressed with respect to the corresponding HRG baselines. For  $\kappa_{4}$, the data points fluctuate around the canonical baseline. The biggest differences are about 2 standard deviations. The 27 GeV point is low by this amount, while the point at the lowest energy is above by a similar amount (for a significance test, see below).

\begin{figure}[!htb]
\centering
    \includegraphics[width=1.\linewidth,clip=true]{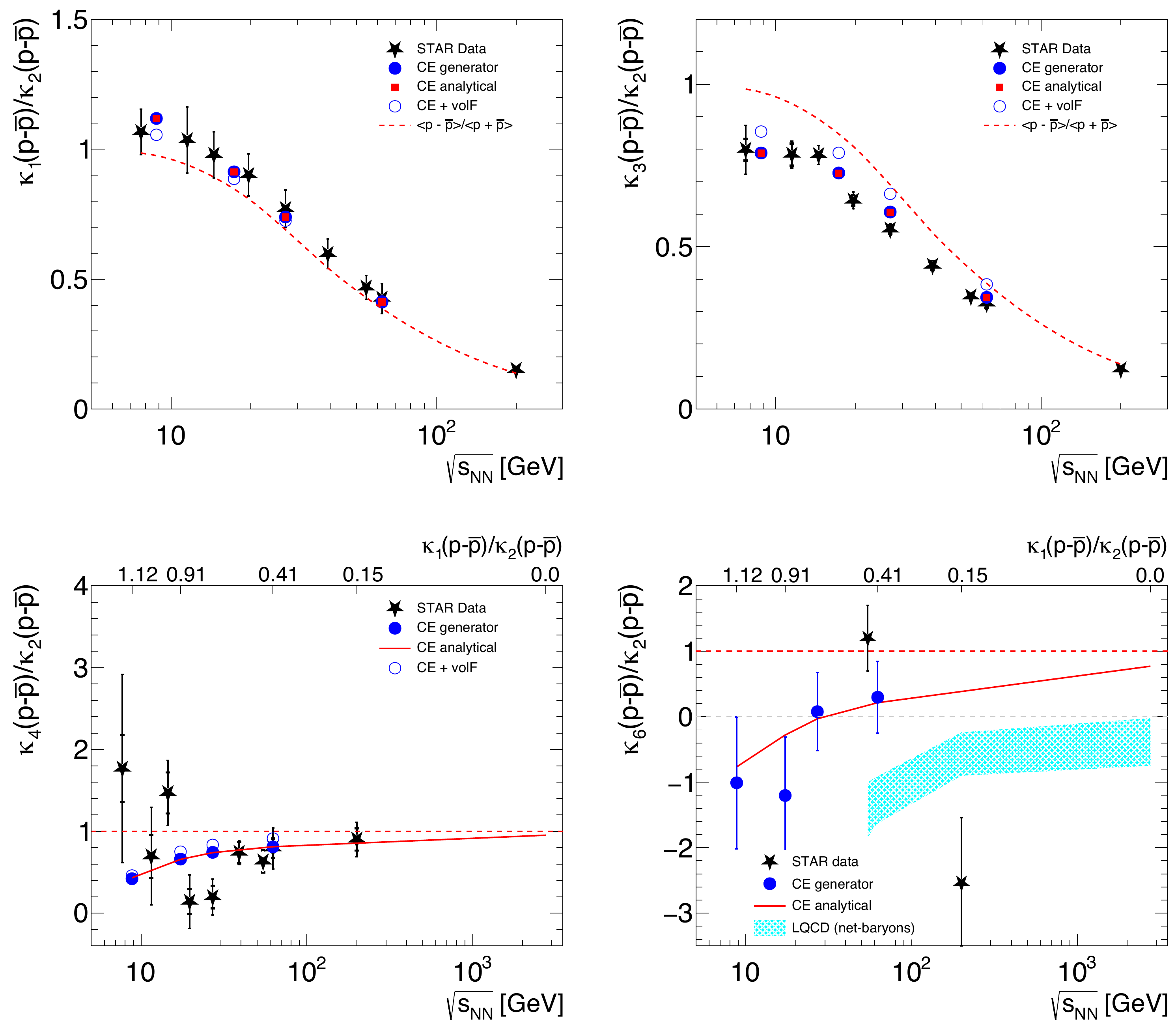}
   \caption{Cumulant ratios of the net proton distributions. The red dashed line shows the HRG value~\cite{Andronic:2017pug}. The blue circles show the results of the simulation, while the red squares and full lines indicate those of the analytical calculations. In both, baryon number conservation, as presented in this work, is accounted for. They should be compared to the black stars representing experimental results of the STAR collaboration \cite{Adam:2020unf}. Open blue circles include the additional contributions from reaction volume fluctuations~\cite{Braun-Munzinger:2016yjz}. The STAR data are corrected with the Centrality Bin Width Correction method~\cite{Luo:2013bmi, Braun-Munzinger:2016yjz}, hence contributions from volume fluctuations in data, compared to the open circles, are suppressed. Also shown as the cyan band in the lower right figure is the result from LQCD \cite{Bazavov:2020bjn} for net baryon cumulants. The LQCD results hence indicate qualitative trends but should not be used for quantitative comparison to data.}
\label{fig:kn-to-k2}
\end{figure} 

In Fig.~\ref{fig:kn-to-k2} we present the energy dependence of the normalized cumulants. Also shown for reference is the HRG baseline, i.e., the baseline for independent Poissonian fluctuations of protons and antiprotons. For $\kappa_{1}/\kappa_{2}$ and $\kappa_{3}/\kappa_{2}$ this corresponds to $\langle p - \bar{p}\rangle/\langle p + \bar{p}\rangle$, while for the ratio of even cumulants such as  $\kappa_{4}/\kappa_{2}$ and  $\kappa_{6}/\kappa_{2}$ it is equal to unity. As already shown in Fig.\ref{fig:kn}, in general there is good agreement over the full energy range covered experimentally between the results  developed here and the STAR data. In future studies the agreement can be improved further by accounting for simultaneous baryon number and electric charge conservation laws~\cite{Vovchenko:2020gne}. The latter however requires precise experimental measurements of cross-cumulants, which are not available yet. For $\kappa_{1}/\kappa_{2}$ the effect of baryon number conservation is small in the energy range shown here and our results are in excellent agreement with the STAR data. Within the experimental uncertainties the corresponding HRG baseline for the ratio $\kappa_{1}/\kappa_{2}$ is also in (marginal) agreement with the STAR data. For all cumulant ratios the STAR data, as well as our results, approach the HRG baseline for higher energies. This behaviour is a consequence of the decreasing acceptance parameters corresponding to the experimental setup or selected in a specific analysis. A fixed acceptance in rapidity, independent of the collision energy, effectively leads to a decreasing fraction of accepted protons with increasing energy, thereby reducing the effect of baryon number conservation. 

In general, baryon number conservation reduces the amount of fluctuations, at least for small acceptance probabilities, while other non-critical effects, such as fluctuations of the reaction volume~\cite{Braun-Munzinger:2016yjz}, introduce additional contributions. The latter, at least for cumulants up to third order, increase the measured fluctuations, as seen in Fig.\ref{fig:kn}. For higher order cumulants the volume fluctuations may eventually contribute with a negative sign. As the measured STAR data for $\kappa_{3}/\kappa_{2}$ are always below the corresponding HRG line, we conclude that baryon number conservation dominates over volume fluctuation effects. 

\begin{figure}[!htb]
\centering
    \includegraphics[width=.5\linewidth,clip=true]{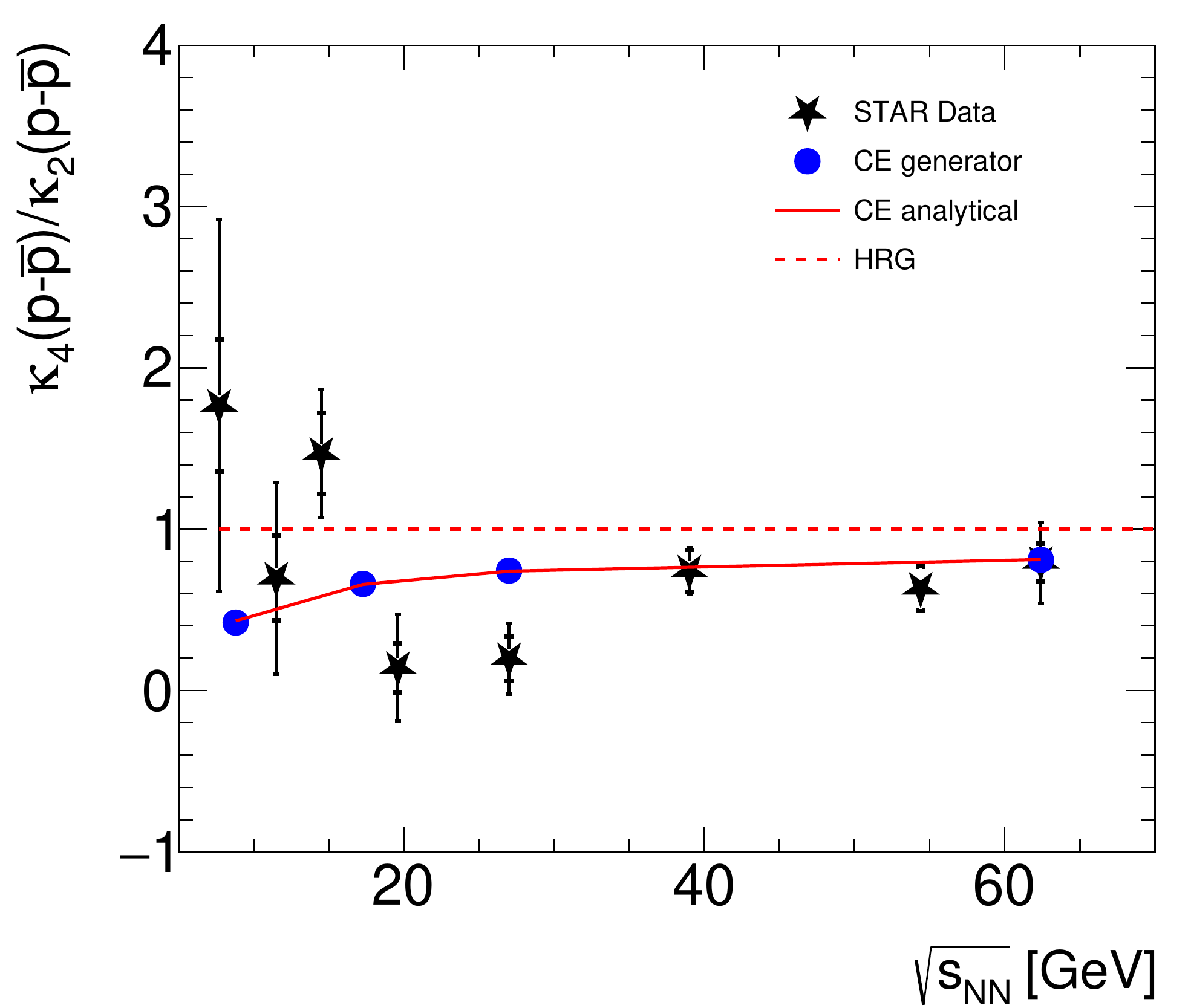}
    \caption{Enlarged view of he results shown in Fig.~\ref{fig:kn-to-k2} for the normalized fourth order cumulants of net protons in the collision energy range from $\sqrt{s_{NN}}=$ 7.7 to 62.4 GeV. A Kolmogorov-Smirnov test indicates that the deviations between the STAR data and our results, represented by the solid red line, are not statistically significant. The one-sided p-value is about 0.25.
    }
\label{fig:cum-KS}
\end{figure}

In Fig.~\ref{fig:kn-to-k2} we also show $\kappa_{6}/\kappa_{2}$ as function of energy, together with the STAR data and the corresponding LQCD results. In order to put the LQCD results for net-baryons on this plot, the ratio $\kappa_{1}/\kappa_{2}$ and it's empirical dependence on $\sqrt{s_{NN}}$ is used~\footnote{For a quantitative comparison between experimental data and LQCD results, a further step is needed to obtain empirical net-baryon cumulants, given the experimentally determined net-proton ones~\cite{Kitazawa:2012at}. However, such a calculation is beyond the scope of the present paper. Hence LQCD results are not used for making quantitative physics statements.}. This ratio is shown as an additional scale at the top of  the Figure. Here, baryon number conservation leads to a very significant effect. Starting at the HRG value of 1 at high collision energy, there is a significant reduction and even a sign change to negative values for this cumulant ratio for $\sqrt{s_{NN}} \leq$ 40 GeV. For this high order cumulant ratio, indeed even at the highest energy, a significant deviation from the HRG baseline is observed in the LQCD results \cite{Bazavov:2020bjn,Ratti:2010kj}. This is expected, since  the 6th order moment may receive a critical contribution due to the closeness of the pseudo-critical line of the chiral phase transition to the 2nd order O(4) phase transition obtained for vanishing light quark masses, even if the energy is far from that for a critical end point~\cite{Friman:2011pf}. In LQCD this cumulant ratio is negative for all energies shown. 
Moreover, while the STAR data at $\sqrt{s_{NN}}$ = 200 GeV could be in line with the combined suppression due to baryon number conservation and LQCD dynamics, the value at 54.4 GeV is significantly higher.

\begin{figure}[!htb]
\centering
    \includegraphics[width=.495\linewidth,clip=true]{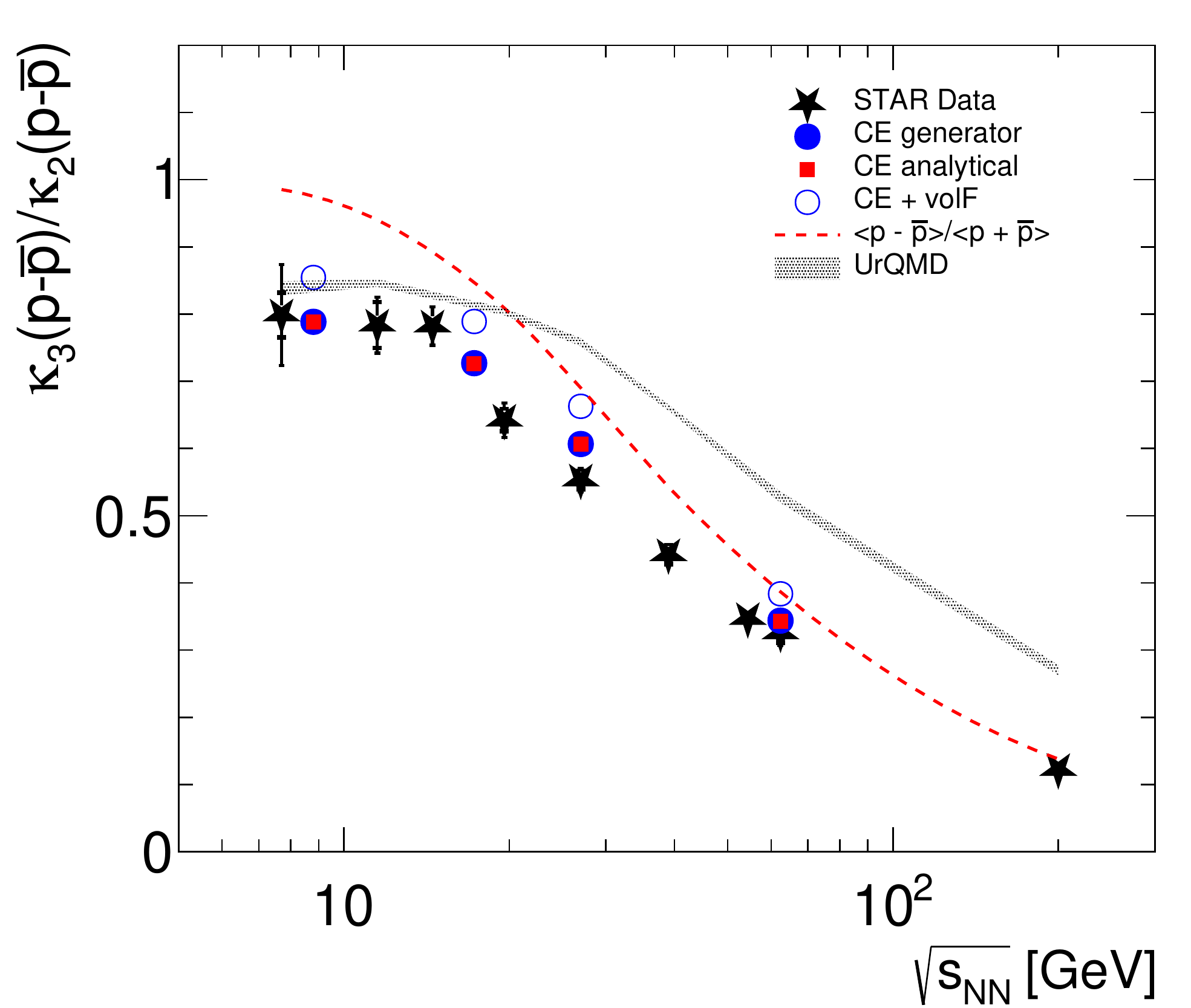}
    \includegraphics[width=.495\linewidth,clip=true]{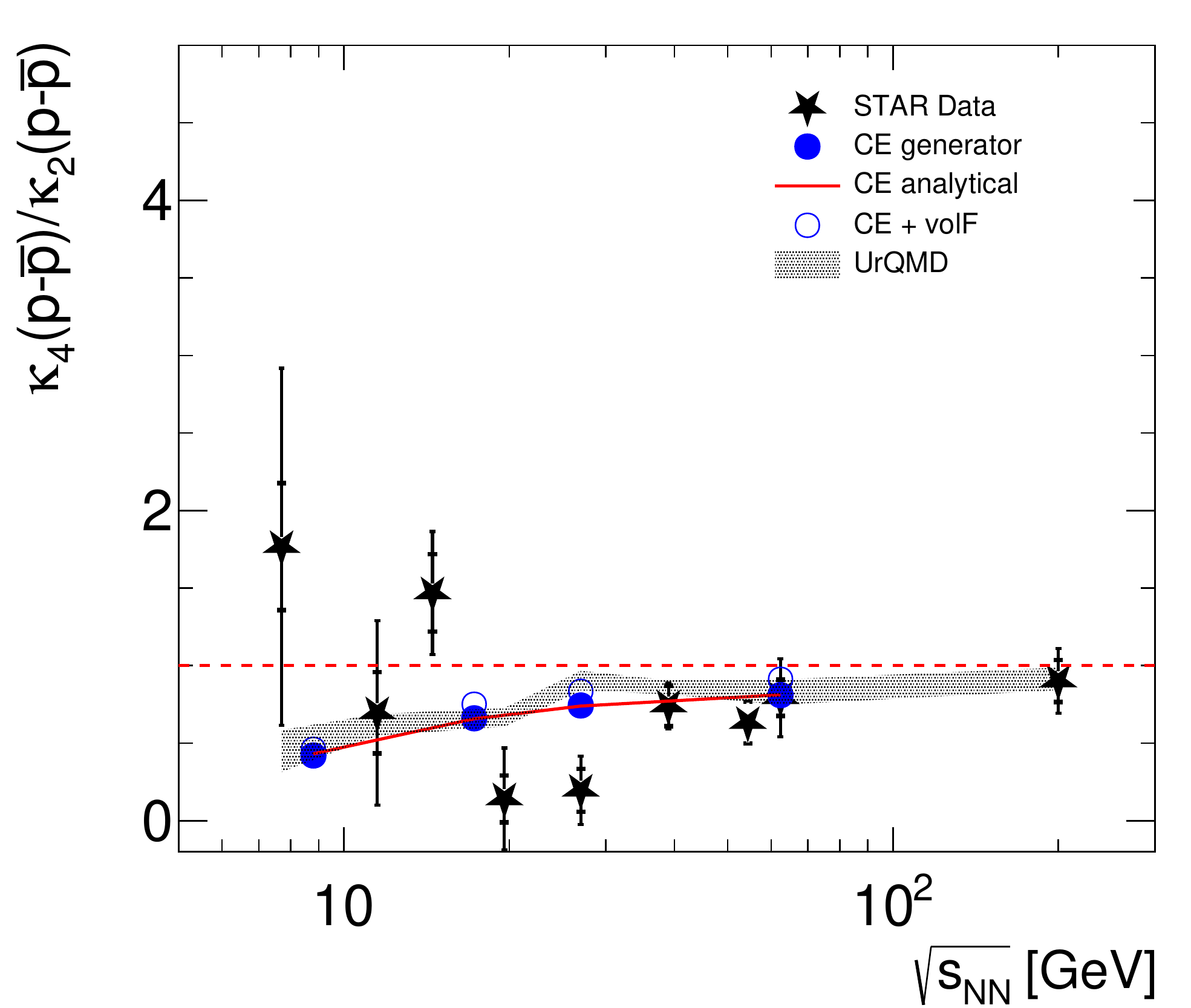}
    \caption{The third (left panel) and fourth (right panel) order cumulants normalized to the second order cumulant of net protons measured in central Au--Au collisions by the STAR experiment \cite{Adam:2020unf}. The red dashed line is the  baseline calculated with the HRG model~\cite{Andronic:2017pug} within the grand-canonical ensemble. The blue circles and red squares (red line in the right panel) correspond to simulations and analytical calculations, respectively, accounting for baryon number conservation. The open blue circles include the additional contributions from reaction volume fluctuations~\cite{Braun-Munzinger:2016yjz}. Also shown as the gray band are results obtained by STAR using the event generator UrQMD\cite{Adam:2020unf,urqmd}.}
\label{fig:cum-acc-dep}
\end{figure} 

Returning to the question whether the STAR data for the normalized fourth order cumulant are consistent with the non-critical baseline, including the effects of baryon number conservation, we display the results for $\kappa_4/\kappa_2$ in Fig.~\ref{fig:cum-KS} in the energy range from $\sqrt{s_{NN}}=$ 7.7 to 62.4 GeV. To quantify the significance of the deviations of the six STAR data points from the baseline we have performed a Kolmogorov-Smirnov test~\cite{KS-Test}.
The results of this test indicate that the observed deviations between the STAR data and our results, represented by the solid red line, are not statistically significant. The corresponding one-sided p-value is about 0.25, corresponding to a difference of $1.2\sigma$. We also performed a $\chi^{2}$ test. The resulting $\chi^2$ value is 12.6 for 8 degrees of freedom. This corresponds to a $1.5\sigma$ deviation from the canonical baseline. Consequently, both statistical tests show that there is no statistically significant difference between the canonical baseline and the STAR data.  We note that this result does not lend support to the indication for a non-monotonic energy dependence of $\kappa_{4}/\kappa_{2}$, reported in~\cite{Adam:2020unf}.

\begin{figure}[!htb]
\centering
    \includegraphics[width=.495\linewidth,clip=true]{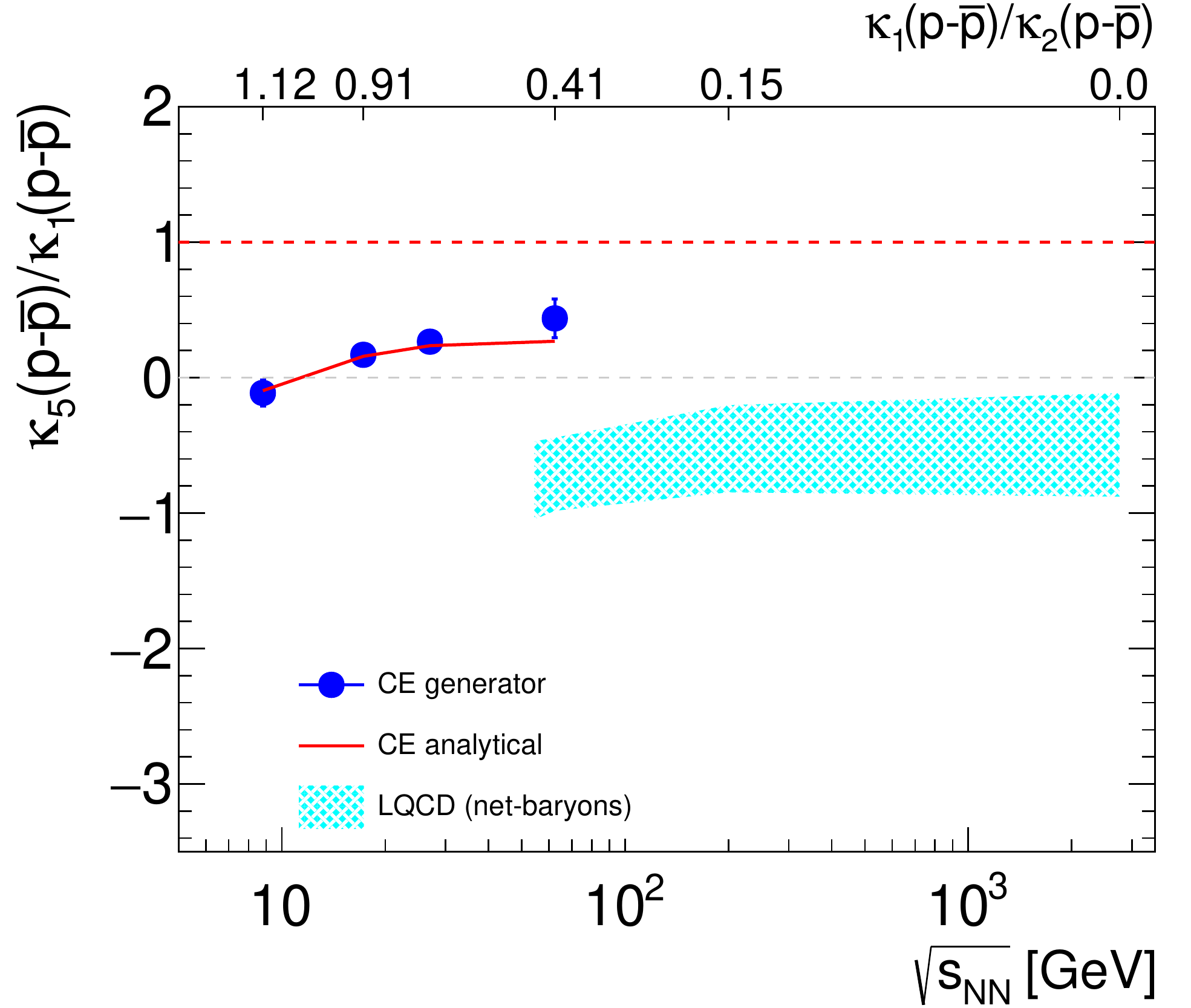}
     \includegraphics[width=.495\linewidth,clip=true]{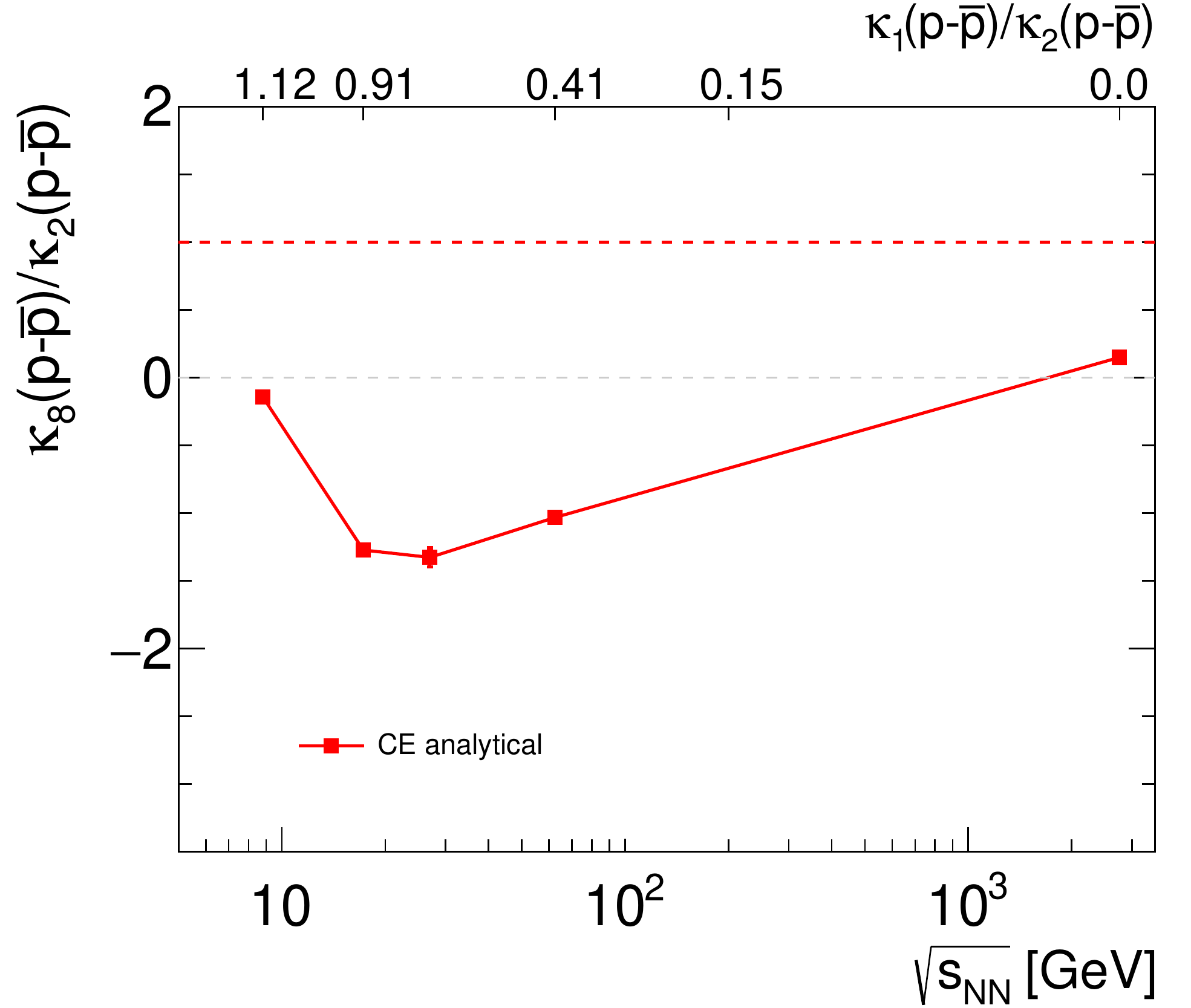}
    \caption{The normalized fifth (left panel) and eighth (right panel) order cumulants of net protons. For the fifth order cumulants also the result from LQCD is shown  \cite{Bazavov:2020bjn}. The LQCD results are for net-baryon cumulants and hence are shown here only to indicate qualitative physics trends. The non-monotonous dependence of $\kappa_8/\kappa_2$ is 
a reflection of the rather strong oscillatory dependence of this ratio on
the acceptance parameters $\alpha_B$ and $\alpha_{\bar{B}}$, which 
grow with decreasing energy (cf. Fig.~\ref{fig:acc-27}).}
\label{fig:cum-K5}
\end{figure} 

\begin{figure}[!htb]
\centering
    \includegraphics[width=.495\linewidth,clip=true]{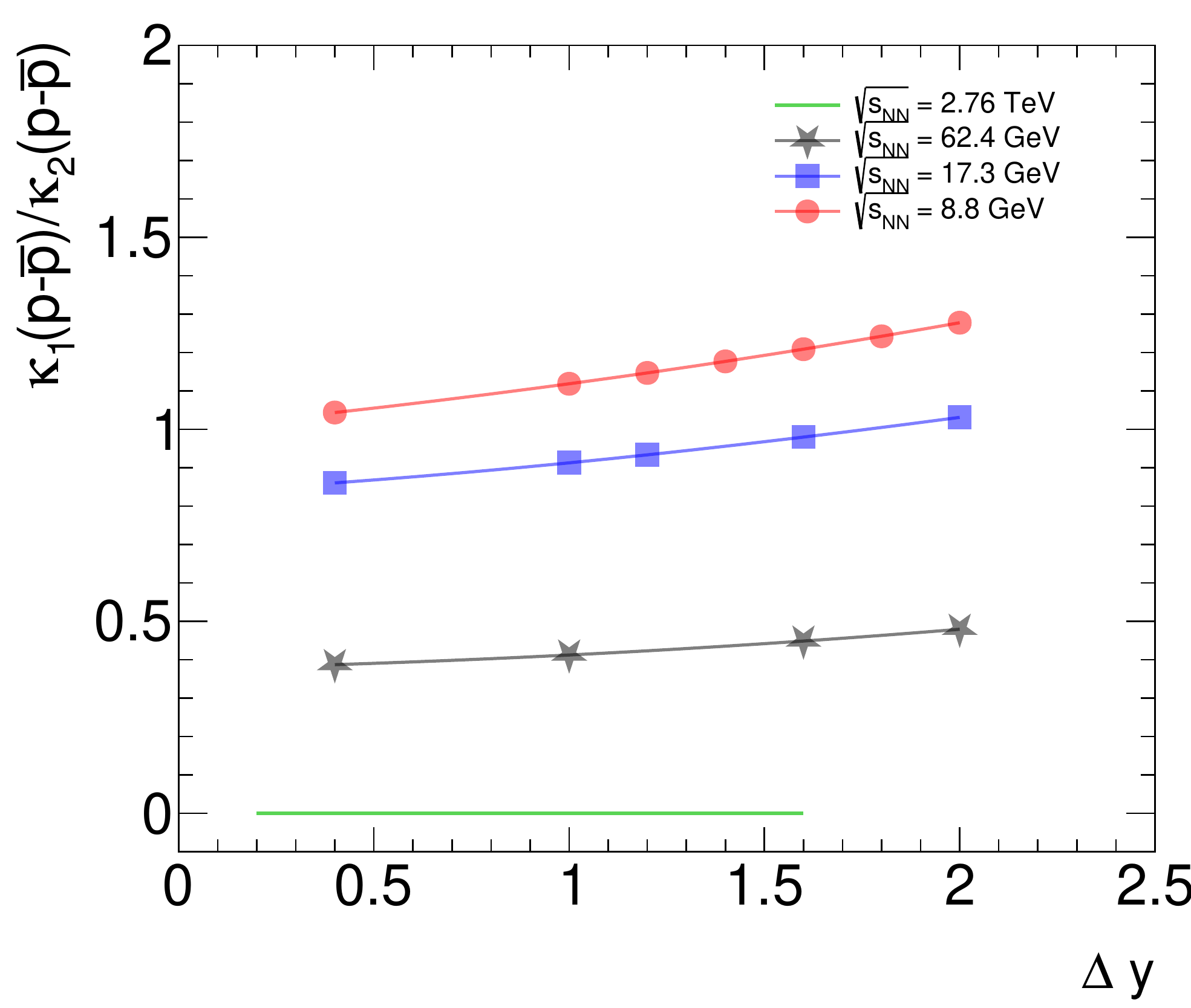}
    \includegraphics[width=.495\linewidth,clip=true]{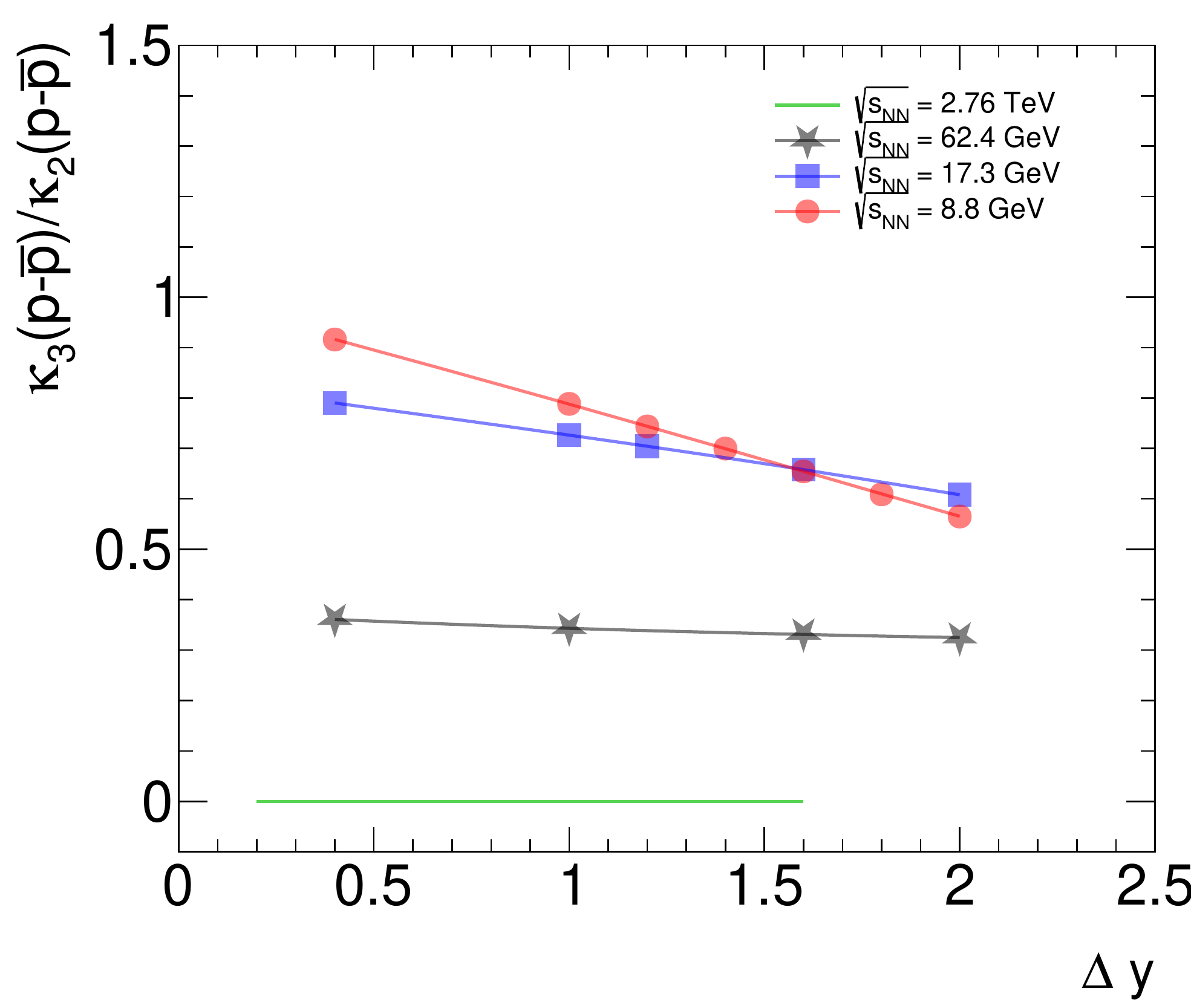}
    \includegraphics[width=.495\linewidth,clip=true]{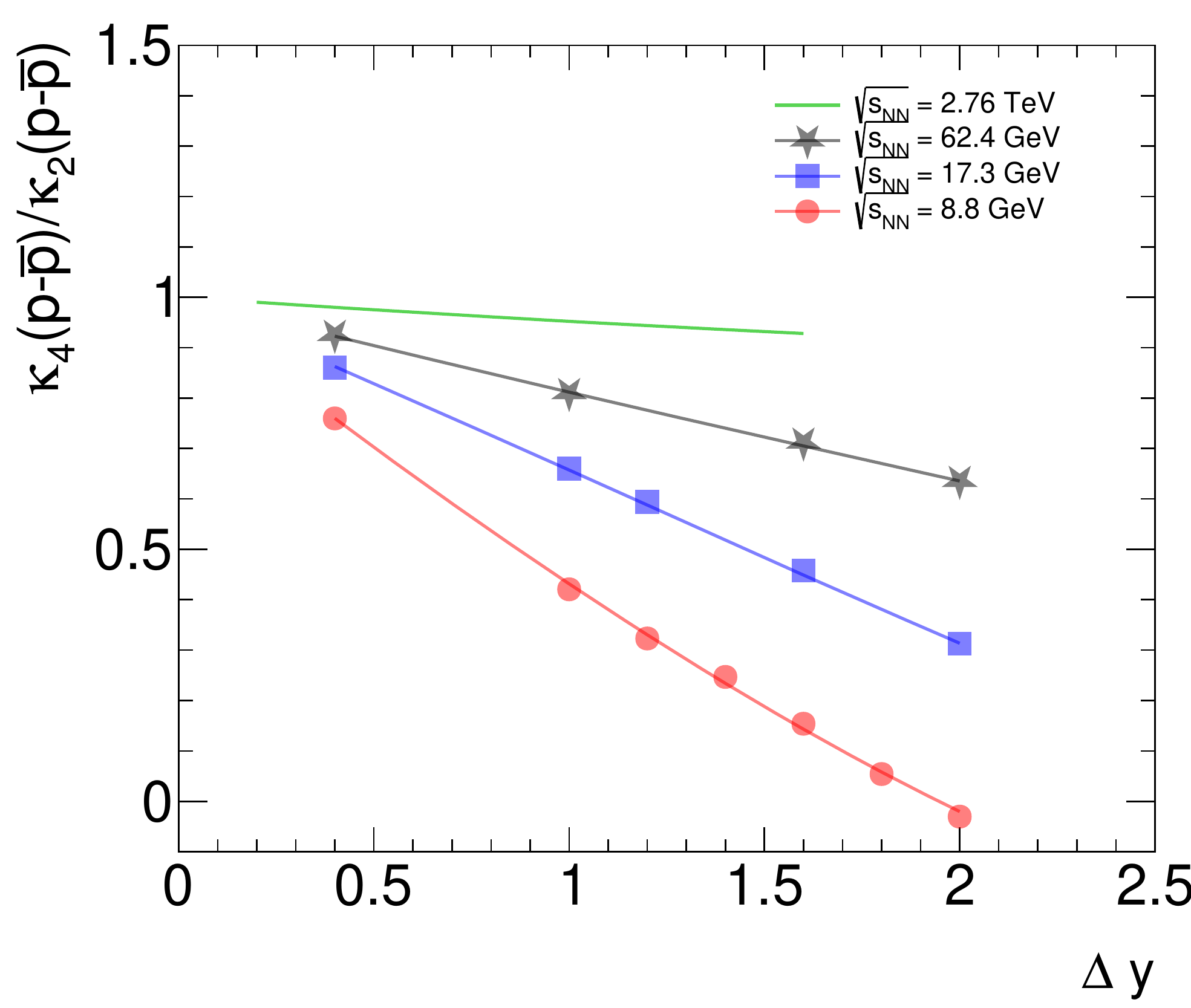}
    \includegraphics[width=.495\linewidth,clip=true]{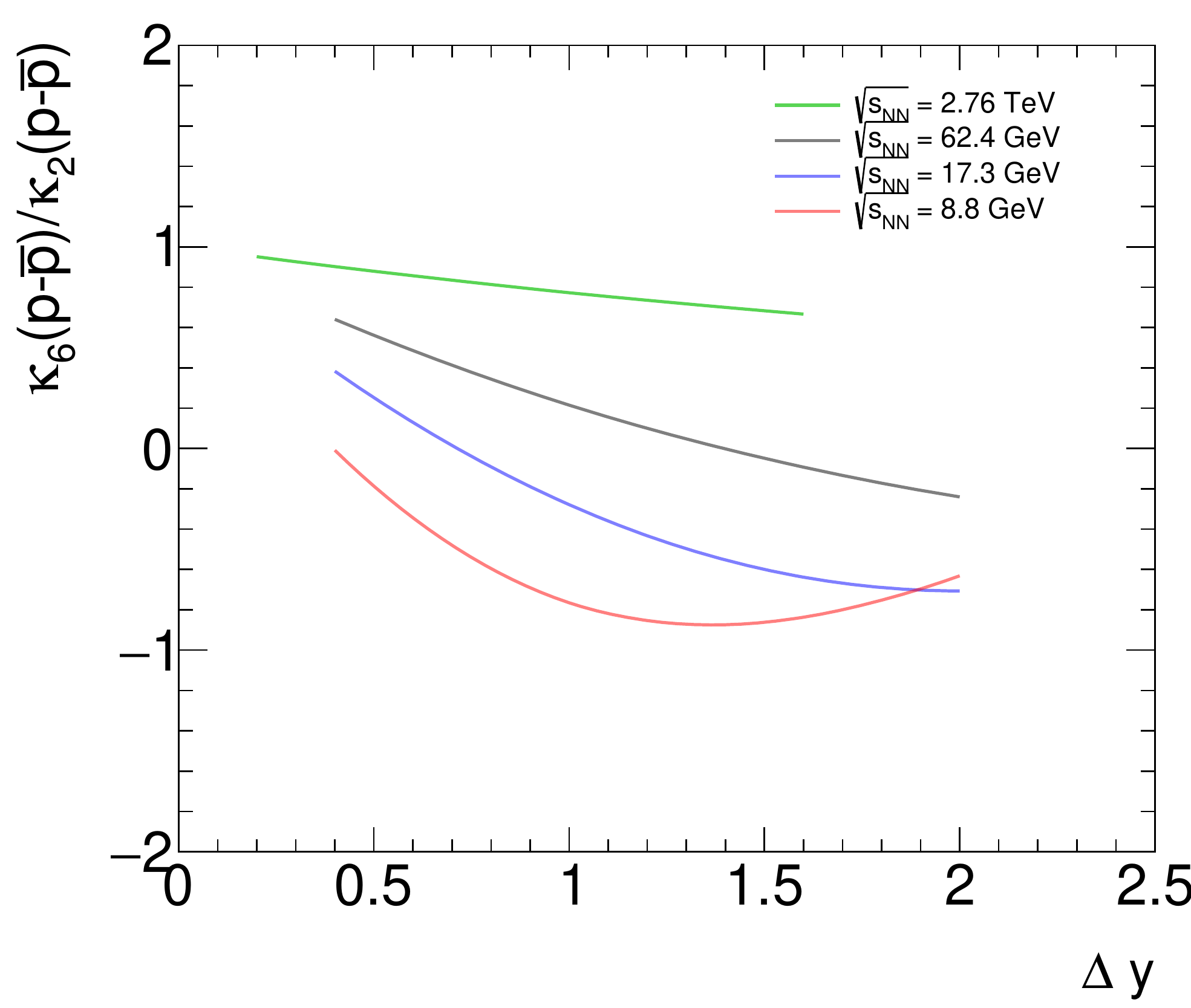}
    \caption{Acceptance dependence of normalized cumulants. The oscillatory dependence of higher cumulants on the acceptance parameters is reflected in a non-monotonous dependence of $\kappa_6/\kappa_2$ on $\Delta y$ at the lowest energy, which corresponds to the largest values of $\alpha_B$ (cf. Fig.~\ref{fig:acc-27}).}
\label{fig:cum-norm-acc-dep}
\end{figure}

In Fig.~\ref{fig:cum-acc-dep} we show, for the normalized 3rd and 4th order cumulants, in addition to the information shown in Fig.~\ref{fig:kn-to-k2} predictions obtained by STAR using the UrQMD event generator \cite{Adam:2020unf,urqmd}. While our results including baryon number conservation and volume fluctuations are in good agreement with the STAR data, for $\kappa_{3}/\kappa_{2}$, interestingly, for energies above 20 GeV the UrQMD results on $\kappa_{3}/\kappa_{2}$ are significantly above the measured STAR data and even above the HRG baseline. For the $\kappa_{4}/\kappa_{2}$ ratio we observe that the UrQMD results are close to our current results with the canonical suppression due to baryon number conservation.  In the past, the fact that $\kappa_{4}/\kappa_{2}$ for the UrQMD results is below the HRG baseline, was interpreted as due to baryon number conservation. The observation that for $\kappa_{3}/\kappa_{2}$, where the UrQMD prediction lies above the HGR baseline at higher energies and does not at all follow  the canonical suppression, casts doubt on that interpretation and makes it questionable whether the UrQMD results should be considered as a baseline for  $\kappa_{4}/\kappa_{2}$. 

Finally, event generators such as UrQMD and HIJING make use of the string fragmentation mechanism to describe baryon production. As demonstrated in ~\cite{Arslandok:2020mda,Acharya:2019izy} string fragmentation leads to rather short range correlations between produced baryons, at variance with the experimental observations at LHC energy. It would be interesting to investigate whether the anomaly in the UrQMD results on $\kappa_{3}/\kappa_{2}$ is connected to the particular mechanism of baryon production implemented in UrQMD.

Results for the effect of baryon number conservation are given in Fig.~\ref{fig:cum-K5} for the so far unmeasured normalized fifth and eighth order net proton cumulants. For the fifth order cumulant also results from LQCD are available
and shown in the Figure. It can be seen that for the fifth order cumulant the effects of baryon number conservation are sizable.
Combining this result with the predictions from LQCD, we  expect negative values for the fifth order cumulant for the entire RHIC and LHC collision energy range. For the eighth order cumulant the effect of baryon number conservation is already a nearly 100 \% correction at LHC energy.

For future reference for measurements and analyses performed with a different rapidity coverage or possibly also as a function of $\Delta y$ we show in Fig.~\ref{fig:cum-norm-acc-dep} the acceptance dependence of cumulant ratios at four different collision energies.

\section{Extraction of freeze-out parameters}

The determination of chemical freeze-out parameters by statistical hadro\-ni\-zation analysis of measured hadron abundances is well established. 

\begin{figure}[!htb]
\centering
    \includegraphics[width=0.6\linewidth,clip=true]{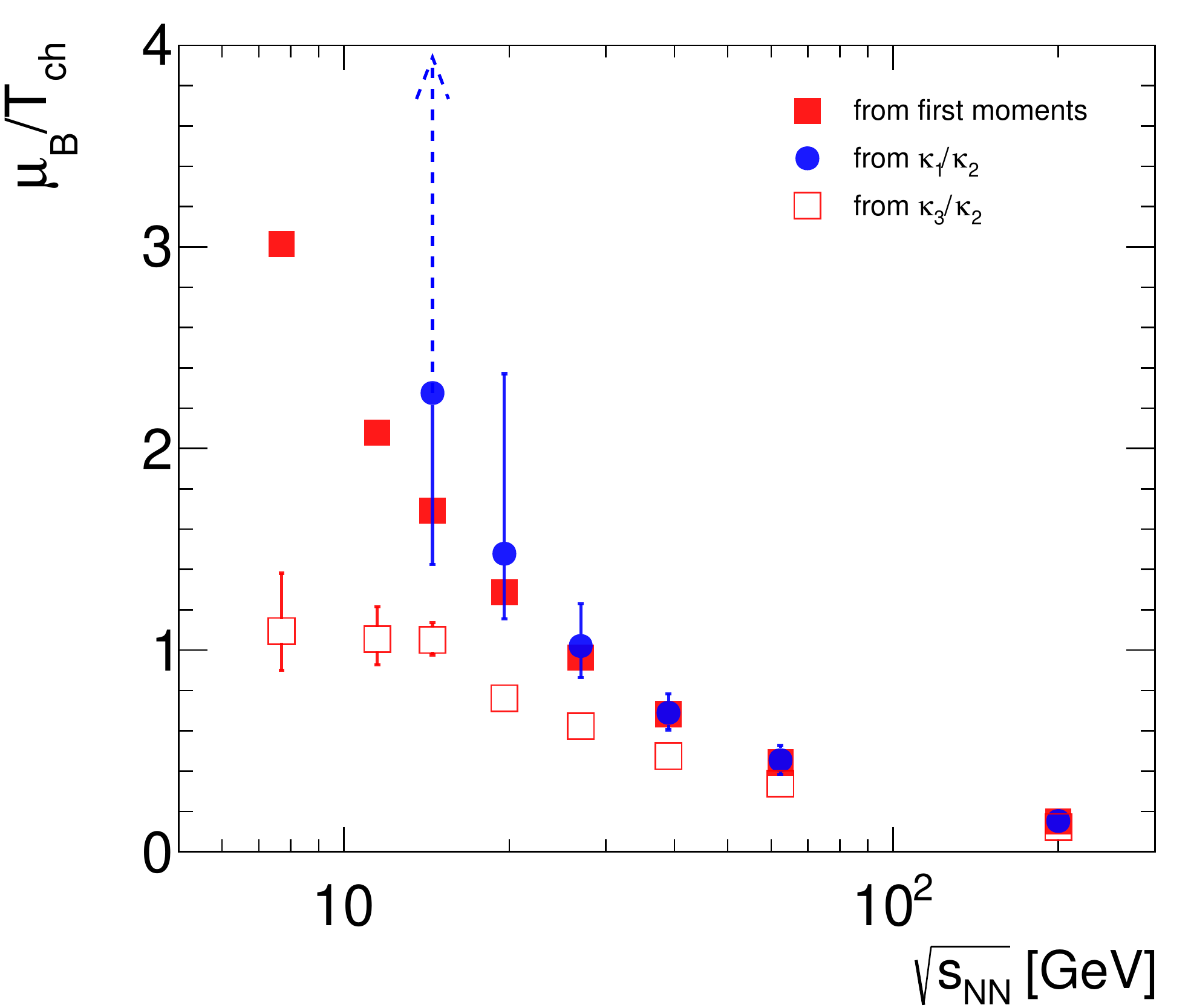}
    \caption{Determination of freeze-out parameters $\mu_B/T_{ch}$ from mean particle multiplicities and cumulant ratios $\kappa_1/\kappa_2$ and $\kappa_3/\kappa_2$.}
\label{fig:FO}
\end{figure}

For example, the $\mu_{B}/T_{ch}$ values presented in Fig.~\ref{fig:FO} are obtained by fitting the HRG model predictions to the measured STAR multiplicities for Au--Au collisions at $\sqrt{s_{NN}}$ = 7.7, 11.5, 14.5, 19.6, 27, 39, 62.4 and 200 GeV. In principle, the freeze-out parameters could also be extracted by analysis of data for higher cumulants.  In practice, this is complicated since  non-critical contributions such as those stemming from conservation laws and volume fluctuations influence the values of higher cumulants in a complex pattern, as discussed above.

To demonstrate this explicitly we show in Fig.~\ref{fig:FO} the $\mu_{B}/T_{ch}$ values calculated as  atanh${(\kappa_{1}/\kappa_2)}$ (blue circles)  and atanh${(\kappa_{3}/\kappa_{2})}$ (open squares), i.e., completely ignoring possible non-critical contributions. The resulting $\mu_{B}/T_{ch}$ values are systematically shifted with respect to values extracted by analysis of 1st moments  (shown as red solid squares) where such non-critical contributions are absent. Great care must hence be exercised if one wants to extract thermal parameters from an analysis of higher cumulants. We further note that the inverse hyperbolic tangent atanh$(x)$  yields real values only for values of $x$ in the domain $-1<x<1$. As seen in Fig.~\ref{fig:kn} for energies $\sqrt{s_{NN}}$ = 7.7 and 11.5 GeV the measured ${\kappa_{1}/\kappa_2}$ values exceed unity. This is the reason for missing freeze-out parameters in Fig.~\ref{fig:FO} extracted from the ${\kappa_{1}/\kappa_2}$ ratios. For the same reason the uncertainty of $\mu_{B}/T_{ch}$ as extracted from ${\kappa_{1}/\kappa_2}$ at $\sqrt{s_{NN}}$ = 14.5 GeV is infinitely large (cf. dashed arrow in Fig.~\ref{fig:FO}). The steepness of atanh$(x)$ makes it difficult to extract freeze-out parameters from second cumulants. 

The general conclusion is that, as soon as the effects of baryon number conservation significantly alter the cumulant ratios (as visible in Fig.~\ref{fig:kn-to-k2}), the extraction of freeze-out parameters using the grand-canonical ensemble and these cumulant ratios will give spurious results, as demonstrated in Fig.~\ref{fig:FO}. Since the effects of baryon number conservation grow with the order of the cumulant, the deviations of the extracted freeze-out parameters are expected to also grow with the order. For the second and third order cumulants this is what is seen in the Figure. Also, the deviation will be strongly dependent on the acceptance covered by the measurement/analysis, while the freeze-out parameters determined from particle abundances show a weak dependence on rapidity coverage.

\section{Software package}
\label{sec:software}

A Python package for calculating both analytical formulas and numerical values for net baryon cumulants of any order in a finite acceptance is available for download. 

\begin{figure}[!htb]
\centering
    \includegraphics[width=0.6\linewidth,clip=true]{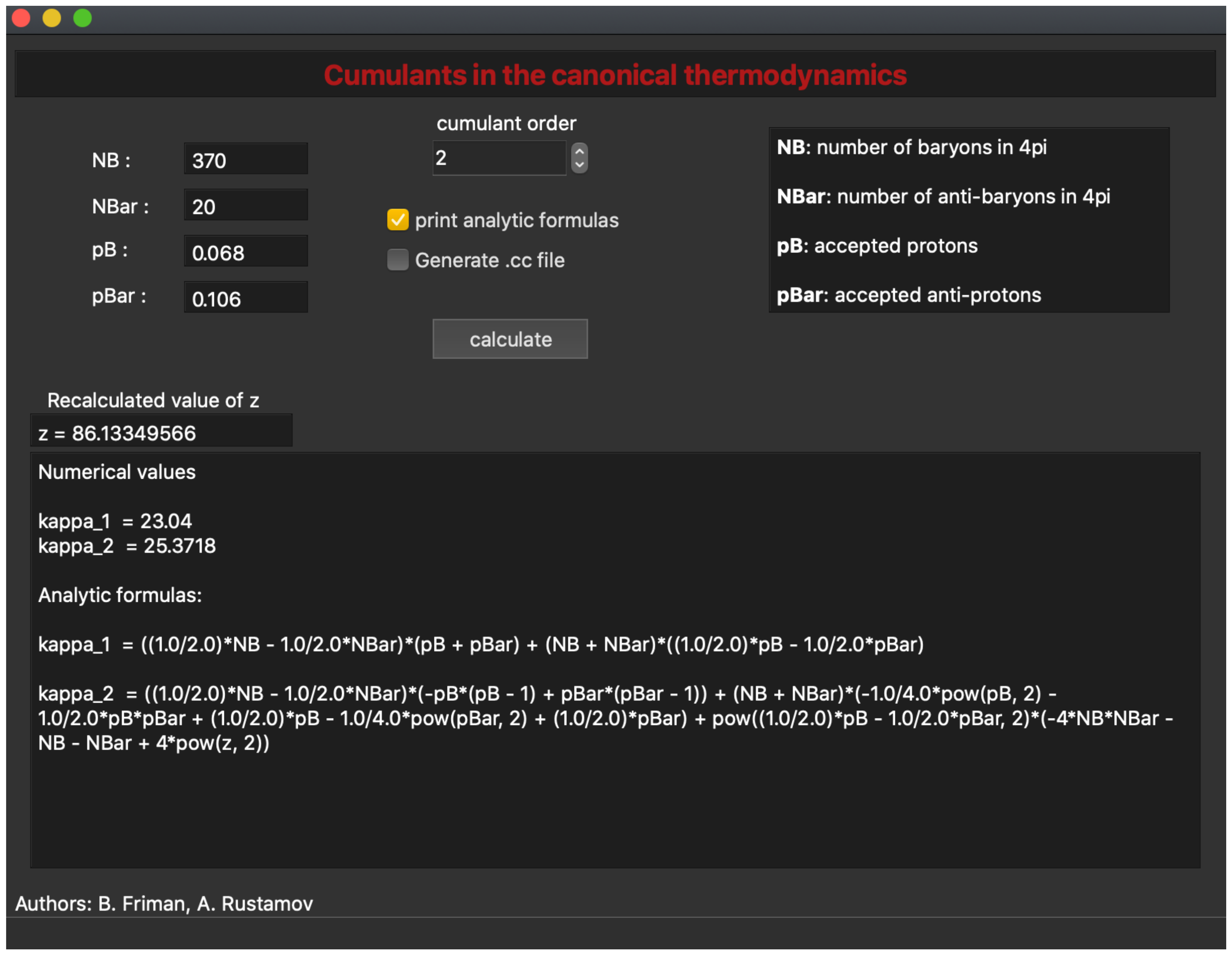}
    \caption{The graphical user interface of the software package that computes the cumulants of net-baryons (net-protons) of any order.}
\label{fig:CE-GUI}
\end{figure}

In the dedicated graphical user interface, presented in Fig.~\ref{fig:CE-GUI}, one enters the number of baryons and antibaryons in full phase space and the corresponding acceptance values. After pressing the "calculate" button the program computes the values of the net baryon cumulants up to the order selected by the user. Analytical formulas for cumulants are given if the checkbox "print also formulas" is activated. The package, including  further details on usage, can be downloaded via Ref.~\cite{refCE_GUI}.

\section{Summary and conclusions}
\label{sec:Conclusions}

We have provided a non-critical baseline for  net baryon  cumulants in relativistic nuclear  collisions, by explicitly taking into account effects of baryon number conservation. To this end the net-baryon cumulants are evaluated analytically in the framework of the canonical formulation of the HRG. Implementing finite experimental acceptance and allowing for a varying asymmetry between the mean numbers of baryons  and antibaryons  as a function of center-of-mass energy, we arrive at a compact description for the modification of cumulants of net baryons due to baryon number conservation. Together with data-based simulations of volume fluctuations this framework provides a quantitative basis for a search for critical behavior in experimental results on net baryon cumulants over the full energy range where data are available. For completeness, we note that deviations from the non-critical baseline considered here may also be due to non-critical contributions of higher order in the fugacity expansion, not considered in our approach. However, as mentioned above these are expected to be subleading, at least for cumulants of order $n<6$. 

In the present paper we have applied our approach to quantify the non-critical baseline for  cumulants of net-proton number fluctuations obtained by the STAR collaboration at different RHIC energies and by the ALICE collaboration at the LHC. The results demonstrate that, overall, the experimental data follow the non-critical baseline predictions well without statistically significant differences even at the lowest energies and for cumulant order larger than three.

We have prepared a dedicated software package with a graphical user interface which allows, using the framework developed in this paper,  to derive analytical formulas for net baryon cumulants of any order. 

With our approach we have made quantitative predictions for the energy dependence of cumulants and cumulant ratios of net baryons up to and including the eighth order which can be tested in the next generation of experiments at RHIC and LHC as well as at future facilities such as NICA and GSI/FAIR. 

\section{Acknowledgments}
\label{sec:Acknowledgments}
The authors acknowledge important and instructive discussions with Drs. Xiaofeng Luo and Nu Xu on the analysis of higher cumulants of net-baryon fluctuations in general and on the corresponding data from the STAR collaboration in particular.
This work is part of and supported by the DFG (German Research Foundation) -- Project-ID 273811115 -- SFB  1225 ISOQUANT.
K.R. acknowledges the support by the Polish National Science Center (NCN) under the Opus grant no. 2018/31/B\-/ST2/01663. K.R. also acknowledges partial support  of  the  Polish  Ministry  of  Science  and  Higher Education. B.F. acknowledges support by the Deutsche Forschungsgemeinschaft (DFG) through the CRC-TR 211 ``Strong-interaction matter under extreme conditions''– project number 315477589–TRR 211.

\appendix
\section{Derivation of canonical cumulants}
\label{sec:DerivationCE}
In this appendix we derive general expressions for the cumulants in a finite system taking the conservation of the net baryon number into account. The generating function for cumulants in a given acceptance, is given by~\cite{Bzdak:2012an}
\begin{eqnarray}\label{eq:generating-function}
g(t)&=&\ln\left(\sum_{B_A} P_A(B_A)e^{B_A t}\right)\nonumber\\
&=&\frac{B}{2}\left[\ln(q_1(t))-\ln(q_2(t))\right]+\ln\left\{I_B[2\, z\, \sqrt{q_1(t)\, q_2(t)}]\right\},
\end{eqnarray}
where $B$ is the total net baryon number, $q_1(t)=1-\pb+\pb\, e^t$ and $q_2(t)=1-\pbb+\pbb\,e^{-t}$.
Here $P_A(B_A)$ is the probability distribution for observing the net baryon number $B_A$ in the acceptance (\ref{eq:probability-acceptance-2}), while $\pb$ is the probability for observing a baryon and $\pbb$ that for an antibaryon. Moreover, $z=\sqrt{z_B\,z_{\bar{B}}}$ is the geometric mean of the single-particle partition functions $z_B$ and $z_{\bar{B}}$.

The $n$:th cumulant of the probability distribution is given by
\begin{equation}\label{eq:derivatives}
\kappa_n=\left.\frac{d^n\,g(t)}{d\,t^n}\right|_{t=0}.
\end{equation}
Thus, to compute the cumulants, we need the derivatives of composite functions. For a function of the form $f(h(x))$, the $n$th derivative is given by the formula of Fa\`{a} di Bruno~\cite{Riordan:1946,Weber:2002}
\begin{equation}\label{eq:faa-di-bruno}
\frac{d^n}{dx^n}f(h(x))=\sum_{k=1}^n f^{(k)}(h(x))\,B_{n,k}\left(h^{(1)}(x),h^{(2)}(x),\dots,h^{(n-k+1)}(x)\right),
\end{equation}
where $f^{(k)}$ and $h^{(k)}$ denote the $k$th derivatives and $B_{n,k}(y_1,y_2,\dots)$ are partial Bell polynomials~\cite{Bell:1927}.
For more complicated functions~\cite{Riordan:1946}, e.g., $f(h(k(x)))$, the higher derivatives can be obtained by repeated use of (\ref{eq:faa-di-bruno}). This will be needed for evaluating the derivatives of the last term in (\ref{eq:generating-function}). 

We shall first use (\ref{eq:faa-di-bruno}) to compute the contributions to the cumulants of the first two terms. The first one is of the same form as the cumulant generating function for the binomial distribution  
\begin{equation}\label{eq:gen-func-binomial}
g^{a}(t)=\ln(q_1(t))=\ln(1-\pb+\pb\,e^t),
\end{equation}
where we for the moment omit the trivial prefactor $B/2$.
We identify $f(x)=\ln(x)$ and $h(t)=1-\pb+\pb\,e^t$ and find the derivatives
\begin{equation}
f^{(n)}(x)|_{x=1}=(-1)^{(n-1)}\,(n-1)!,
\end{equation}
and 
\begin{equation}
h^{(n)}(t)|_{t=0}=\pb.
\end{equation}
Consequently, the explicit expression for the cumulants 
\begin{equation}\label{eq:cumulant-deriv-a}
c^{(n)}_B=\frac{d^n g^{a}(t)}{d\,t^n}|_{t=0}
\end{equation}
is given by
\begin{equation}\label{eq:Bell-poly}
c^{(n)}_B= \sum_{k=1}^n\,(-1)^{(k-1)}\,(k-1)!\,B_{n,k}(\pb,\pb,\dots,\pb),
\end{equation}
where there are $n-k+1$ arguments of the Bell polynomial. By noting that \cite{Comtet:1974}
\begin{equation}
B_{n,k}(\pb,\pb,\dots,\pb)=S(n,k)\, (\pb)^k,
\end{equation}
where $S(n,k)$ is the Stirling number of the second kind, we find
\begin{equation}\label{eq:stirling}
c^{(n)}_B= \sum_{k=1}^n\,(-1)^{(k-1)}\,(k-1)!\,S(n,k)\,(\pb)^k.
\end{equation}

In terms of the polylogarithm ${\rm Li}_n(z)=\sum_{k=1}^\infty\frac{z^k}{k^n}$, the cumulants can be expressed in closed form,~\cite{Truesdell:1945}
\begin{equation}\label{eq:polylog}
c^{(n)}_B= \delta_{n,1}+(-1)^{1+n}\, {\rm Li}_{1-n}(1-1/\pb).
\end{equation}
Using the following property of the polylogarithm,
\begin{equation}
z\frac{\partial}{\partial z}\,{\rm Li}_{-n}(z)={\rm Li}_{-n-1}(z),
\end{equation}
one obtains the recurrence relation
\begin{equation}\label{eq:rec-binomial-cumulants}
c^{(n+1)}_{B}=\pb\,(1-\pb)\,\frac{d}{d\,\pb}c^{(n)}_{B},
\end{equation}
which is identical to that for cumulants of the binomial distribution.

Similarly, to account  for the contribution of the second term in (\ref{eq:generating-function}), we define 
\begin{equation}
g^{b}(t)=-\ln(q_2(t))=-\ln(1-\pbb+\pbb\,e^{-t}),
\end{equation}
and note that 
\begin{equation}
q_2^{(n)}(t)|_{t=0}=(-1)^{n}\,\pbb.
\end{equation}
Thus, for
\begin{equation}\label{eq:cumulant-deriv-b}
c^{(n)}_{\bar{B}}=\frac{\partial^n g^{b}(t)}{\partial t^n}|_{t=0}
\end{equation}
Fa\`{a} di Bruno's formula yields
\begin{equation}\label{eq:Bell-poly-b}
c^{(n)}_{\bar{B}}=- \sum_{k=1}^n\,(-1)^{(k-1)}\,(k-1)!\,B_{n,k}(-\pbb,\pbb,-\pbb,\pbb,\dots),
\end{equation}
Then, using the relation \cite{Comtet:1974}
\begin{equation}
B_{n,k}(-\pbb,\pbb,-\pbb,\pbb,\dots)=(-1)^n\,S(n,k)\, (\pbb)^k,
\end{equation}
we find
\begin{equation}\label{eq:stirling-b}
c^{(n)}_{\bar{B}}=(-1)^{(n+1)}\, \sum_{k=1}^n\,(-1)^{(k-1)}\,(k-1)!\,S(n,k)\,(\pbb)^k.
\end{equation}
Again, this contribution to the cumulants can be expressed in closed form in terms of polylogarithms,
\begin{equation}\label{eq:polylog-b}
c^{(n)}_{\bar{B}}=\delta_{n,1}+{\rm Li}_{1-n}(1-1/\pbb),
\end{equation}
and one finds a recurrence relation, similar to (\ref{eq:rec-binomial-cumulants}),
\begin{equation}\label{eq:rec-antibaryon-cumulants}
c^{(n+1)}_{\bar{B}}=-\pbb\,(1-\pbb)\,\frac{d}{d\,\pbb}c^{(n)}_{\bar{B}}.
\end{equation}

The contribution of the first two terms in (\ref{eq:generating-function})  to the cumulants  can be summarized as follows
\begin{eqnarray}\label{eq:cumulants-a-b}
\kappa^{(a+b)}_n&=&\frac{B}{2}\left(c^{(n)}_B+c^{(n)}_{\bar{B}}\right)\\
&=&\frac{B}{2}\left(2\, \delta_{n,1}+(-1)^{(n+1)}{\rm Li}_{1-n}(1-1/\pb)\,+\,{\rm Li}_{1-n}(1-1/\pbb)\right).\nonumber
\end{eqnarray}
For later use, we introduce the notation
\begin{equation}\label{eq:k-plus}
k_+^{(n)}=\frac12\,(c^{(n)}_B+c^{(n)}_{\bar{B}}).    
\end{equation} 
The contributions to the first six cumulants are then given by
\begin{eqnarray}\label{A22}
\kappa_1^{(a+b)}&=&\frac{B}{2}\big[\pb+\pbb\big],\nonumber\\
\kappa_2^{(a+b)}&=&\frac{B}{2}\big[\pb(1-\pb)-\pbb(1-\pbb)\big],\nonumber\\
\kappa_3^{(a+b)}&=&\frac{B}{2}\big[\pb(1-\pb)(1-2\,\pb)+(\pb\to \pbb)\big],\\
\kappa_4^{(a+b)}&=&\frac{B}{2}\big[\pb(1-\pb)(1-6\,\pb+6\,\pb^2)-(\pb\to \pbb)\big],\nonumber\\
\kappa_5^{(a+b)}&=&\frac{B}{2}\,\big[ \pb\,(1-\pb)(1-2 \pb)(1-12 \pb+12 \pb^2)\nonumber\\
&+&(\pb\to \pbb)\big],\nonumber\\
\kappa_6^{(a+b)}&=&\frac{B}{2}\, \big[\pb\,(1-\pb)(1-30\,\pb+150\,\pb^2-240 \pb^3+120\,\pb^4)\nonumber\\
&-&(\pb\to \pbb)\big].\nonumber
\end{eqnarray} 

We now turn to the last term in (\ref{eq:generating-function})
\begin{equation}
g^{c}(t)=\log\left\{I_B[2\, z\, \sqrt{q_1(t)\, q_2(t)}]\right\}.
\end{equation}
In order to evaluate the derivatives of this term, we consider it to be of the form $f(z\,h(k(t)))$, with~\footnote{Clearly this identification is not unique. Our choice is motivated by the appearance of terms similar to those in $\kappa^{(a+b)}_n$.} 
\begin{eqnarray}
f(z\,x)&=&\ln\left\{I_B[2\,z\,x]\right\},\nonumber\\
h(y)&=&e^y,\\
k(t)&=&\frac12\left(\ln(q_1(t))+\ln(q_2(t))\right),\nonumber
\end{eqnarray}
and use the Fa\`{a} di Bruno formula twice. Since $k(0)=0$ and $h(k(0))=1$, the derivatives are evaluated at $x=1$ and $y=0$, respectively. 

The form of the derivatives of $k(t)=\frac12(g^a(t)-g^b(t))$,  
\begin{equation}\label{eq:k-deriv}
k_-^{(n)}=\frac12(c^{(n)}_B-c^{(n)}_{\bar{B}}),
\end{equation}
follows from (\ref{eq:cumulant-deriv-a}, \ref{eq:polylog}, \ref{eq:cumulant-deriv-b}) and (\ref{eq:polylog-b}). The explicit form of the first six derivatives of $k(t)$ are given below in (\ref{eq:k-p}).
Now, since the derivatives of $h(y)$ at $y=0$ are all equal to unity, the derivatives of $\hk(t)=h(k(t))$ are given by
\begin{eqnarray}\label{eq:deriv-binomial}
\hk^{(n)}= \frac{d^n}{d\,t^n}h(k(t))|_{t=0}
&=&\sum_{k=1}^n\,B_{n,k}\left(k_-^{(1)},k_-^{(2)},\dots,k_-^{(n-k+1)}\right)\\
&=&B_n\left(k_-^{(1)},k_-^{(2)},\dots,k_-^{(n)}\right),\nonumber
\end{eqnarray}
where $B_n\left(x_1,x_2\dots,x_n\right)$ is a complete Bell polynomial.
We provide the explicit form of the first six derivatives,
\begin{eqnarray}\label{eq:deriv-binomial-explicit}
\hk^{(1)}&=&k_-^{(1)}=\frac12\,\big[\pb-\pbb\big],\nonumber\\
\hk^{(2)}&=&k_-^{(2)}+\big[k_-^{(1)}\big]^2\nonumber\\
&=&\frac12\big[\pb(1-\pb)+\pbb(1-\pbb)\big]+\frac14\big[\pb-\pbb\big]^2,\nonumber\\
\hk^{(3)}&=&k_-^{(3)}+3\,k_-^{(1)}\,k_-^{(2)}+\big[k_-^{(1)}\big]^3,\\
\hk^{(4)}&=&k_-^{(4)}+4\,k_-^{(1)}\,k_-^{(3)}+3\big[k_-^{(2)}\big]^2+6\big[k_-^{(1)}\big]^2\,k_-^{(2)}+\big[k_-^{(1)}\big]^4, \nonumber\\
\hk^{(5)}&=&k_-^{(5)}+5\,k_-^{(1)}\,k_-^{(4)}+10\,k_-^{(2)}\,k_-^{(3)}+10\,\big[k_-^{(1)}\big]^2\,k_-^{(3)}\nonumber\\
&+&15\,k_-^{(1)}\,\big[k_-^{(2)}\big]^2+10\big[k_-^{(1)}\big]^3\,k_-{^{(2)}}+\big[k_-^{(1)}\big]^5,\nonumber\\
\hk^{(6)}&=&k_-^{(6)}+6\,k_-^{(1)}\,k_-^{(5)}+15\,k_-^{(2)}\,k_-^{(4)}+15\,\big[k_-^{(1)}\big]^2\,k_-^{(4)}+10\,\big[k_-^{(3)}\big]^2\nonumber\\
&+&60\,k_-^{(1)}\,k_-^{(2)}\,k_-^{(3)}+20\,\big[k_-^{(1)}\big]^3\,k_-^{(3)}+15\,\big[k_-^{(2)}\big]^3\nonumber\\
&+&45\big[k_-^{(1)}\big]^2\,\big[k_-^{(2)}\big]^2+15\,\big[k_-^{(1)}\big]^4\,k_-^{(2)}+\big[k_-^{(1)}\big]^6.\nonumber
\end{eqnarray}
The coefficients $r^{(n)}$ satisfy the recurrence relation
\begin{equation}
r^{(n+1)}=\pb\,(1-\pb)\,\frac{d\,r^{(n)}}{d \pb}-\pbb\,(1-\pbb)\,\frac{d\,r^{(n)}}{d \pbb}+r^{(n)}\,r^{(1)},
\end{equation}
which follows from (\ref{eq:rec-binomial-cumulants}), (\ref{eq:rec-antibaryon-cumulants}), (\ref{eq:k-deriv}) and the recurrence relation for the complete Bell polynomials~\footnote{$B_{n+1}(x_1,\dots,x_{n+1})=\left(x_1+\sum_{i=1}^n\,x_{i+1}\,\frac{\partial}{\partial\,x_i}\right)\,B_n(x_1,x_2,\dots,x_n)$ (cf. \cite{Riordan:1946}).}.

By applying Fa\`{a} di Bruno's formula to $\frac12(g^a(t)-g^b(t))=\log[h(k(t))]$, one finds the inverse of (\ref{eq:deriv-binomial})
\begin{equation}\label{eq:inverse-relation}
k_-^{(n)}=\sum_{k=1}^n\,(-1)^{k-1}\,(k-1)!\,B_{n,k}(\hk^{(1)},\hk^{(2)},\dots,\hk^{(n-k+1)}).
\end{equation}
We will make use of this relation below.

Finally, we compute the derivatives of $f(z\,x)=\ln\left\{I_B[2\, z\, x]\right\}$. Since $f(z\,x)$ is a function of the product $z\,x$, the derivatives with respect to $x$, evaluated at $x=1$, can be expressed in terms of derivatives with respect to $z$ for $x=1$, 
\begin{equation}
f^{(n)}(z)=\frac{d^n\,f(z\,x)}{d\, x^n}|_{x=1}=z^n\,\frac{d^n\,f(z)}{d\, z^n}%=z^n\,f^{(n)}.
\end{equation}
This implies that the derivatives $f^{(n)}(z)$ can be obtained using the recurrence relation
\begin{equation}
f^{(n+1)}(z)=z^{n+1}\frac{d}{d\,z}\left(f^{(n)}(z)/z^n\right).
\end{equation}
where $f^{(1)}(z)=\fti$ is the sum of the mean number of baryons and antibaryons  in the canonical ensemble. For the recurrence relation to be useful, we need the derivatives of $\langle N_B\rangle$ and $\langle N_{\bar{B}}\rangle$. They are easily evaluated using
Eqs. \ref{NCM} and \ref{NCP} 
with the result
\begin{equation}\label{eq:z-deriv-NB}
\frac{d}{d z}\,\langle N_B\rangle_C=\frac{d}{d z}\,\langle N_{\bar{B}}\rangle_C=\frac{2}{z}\,\left(\zsqminus\right).
\end{equation}

It is convenient to introduce the function $\tilde{f}(z)=f^{(1)}(z)$ and consider the derivatives thereof,
\begin{equation}
\tilde{f}^{(n)}(z)=\frac{d^n\,\tilde{f}(z)}{d\,z^n}.
\end{equation}
The relation between functions $f^{(n)}(z)$ and $\tilde{f}^{(n)}(z)$
is
\begin{eqnarray}\label{eq:relation-f-ftilde}
f^{(n)}(z)&=&(-1)^n\,(n-1)!\,\sum_{j=1}^n\,\frac{(-1)^j}{(j-1)!}\,z^{j-1}\,\tilde{f}^{(j-1)}(z)\\
&=&z^{n-1}\,\tilde{f}^{(n-1)}-(n-1)f^{(n-1)},\nonumber
\end{eqnarray}
with $\tilde{f}^{(0)}(z)=\tilde{f}(z)$.
Now, by differentiating $\tilde{f}(z)$ and making repeated use of (\ref{eq:z-deriv-NB}) we find
\begin{eqnarray}
\tilde{f}(z)&=&\ftis,\nonumber\\
\tilde{f}^{(1)}(z)&=&\frac{4}{z}\,\zsqminuss,\nonumber\\
\tilde{f}^{(2)}(z)&=&\frac{4}{z^2}\Big\{\zsqminuss-2\,W\Big\},\\
\tilde{f}^{(3)}(z)&=&\frac{8}{z^3}\Big\{W\,\big(2\,\ftis+1\big)-4\,\zsqminuss^2\Big\},\nonumber\\
\tilde{f}^{(4)}(z)&=&\frac{8}{z^4}\Big\{W\,\big(24\,\zsqminuss-4\ftis^2-8\ftis-3\big)
+8\,\zsqminuss^2\,\nplusns\Big\},\nonumber\\
\tilde{f}^{(5)}(z)&=&\frac{16}{z^5}\Big\{64\,\zsqminuss^3-2\,\zsqminuss^2
\big(4\nplusns^2+8\nplusns+3\big)\nonumber\\
&+&W\big(4\nplusns^3
+16\nplusns^2+19\nplusns-8\zsqminuss\big(7\,\nplusns+5\big)+6\big)\nonumber %\\
%&-&
-24\,W^2\Big\},\nonumber
\end{eqnarray}
where we have used the short-hand notation defined in (\ref{eq:SPQW})
The derivatives of $\tilde{f}(z)$ are also easily generated by using the derivatives
\begin{equation}
S'=\frac{4}{z}\,Q,\qquad  P'=\frac{2}{z}\,Q\,S,\qquad Q'=\frac{2}{z}\,\big(Q-W\big).  
\end{equation}
As described in the main text, the value for $z$ is determined by equating the canonical multiplicities (\ref{NCM}) and (\ref{NCP}) with the empirical particle numbers given in Table~\ref{tab:inputValues}.

Now, collecting the different terms, we then find for the contribution to the cumulants from the final term in (\ref{eq:generating-function})
\begin{equation}\label{eq:kappa_c}
\kappa_n^{(c)}=\sum_{k=1}^n\,f^{(k)}(z)\,B_{n,k}\left(\hk^{(1)},\hk^{(2)},\hk^{(3)},\dots,\hk^{(n-k+1)}\right),
\end{equation}
where the derivatives $f^{(n)}(z)$ are easily obtained using (\ref{eq:relation-f-ftilde}).

We note that the contribution to $\kappa_n^{(c)}$ that is proportional to $\tilde{f}(z)$, simplifies with the help of (\ref{eq:inverse-relation}),
\begin{eqnarray}
\delta_1\,\kappa_n^{(c)}&=&\tilde{f}(z)\,\sum_{k=1}^n\,(-1)^{k-1}\,(k-1)!\,B_{n,k}(\hk^{(1)},\dots,\hk^{(n-k+1)})\\
&=&S\,k_-^{(n)}=\frac12\,S(c^{(n)}_B-c^{(n)}_{\bar{B}}).\nonumber
\end{eqnarray}
 
We provide the explicit form of the contribution to the first six cumulants,
\begin{eqnarray}\label{eq:cumulant-c}
\kappa_1^{(c)}&=&S\,k_-^{(1)},\nonumber\\
\kappa_2^{(c)}&=&S \,k_-^{(2)}
+4\,Q\,\big[k_-^{(1)}\big]^2,\nonumber\\
\kappa_3^{(c)}&=&S\,k_-^{(3)}
+12\,Q \,k_-^{(1)}\,k_-^{(2)}
+8\,\Big(Q-W\Big)\,\big[k_-^{(1)}\big]^3,\nonumber\\
\kappa_4^{(c)}&=& S\,k_-^{(4)}+4\,Q\,\Big[3\,\big[k_-^{(2)}\big]^2
+4\,k_-^{(1)}\,k_-^{(3)}\Big]
+48\, \Big(Q-W\Big)\,\big[k_-^{(1)}\big]^2\,k_-^{(2)}\nonumber\\
&+&16\,\Big(W\,\big(S-1\big)-2Q^2+Q\Big)
\big[k_-^{(1)}\big]^4,\nonumber\\
\kappa_5^{(c)}&=&S\,k_-^{(5)}+20\,Q\,\Big[k_-^{(1)}\,k_-^{(4)}+2\,k_-^{(2)}\,k_-^{(3)}\Big]\\ 
&+&40\,\Big(Q-W\Big)\,\Big[3\,k_-^{(1)}\,\big(k_-^{(2)}\big)^2+2\,\big(k_-^{(1)}\big)^2\,k_-^{(3)}\Big]\nonumber\\
&+&160\,\Big(W\,\big(S-1)-2\,Q^2+Q\Big)\,\big[k_-^{(1)}\big]^3\,k_-^{(2)}\nonumber\\
&+&32\,\Big(W\,\big(6\,Q-S^2+S-1\big)
+Q\,\big(2\,Q\,S-6\,Q+1\big)\Big)\big[k_-^{(1)}\big]^5,\nonumber\\
\kappa_6^{(c)}&=&S\,k_-^{(6)}+4\,Q\,\Big[10\,\big[k_-^{(3)}\big]^2+15\,k_-^{(2)}\,k_-^{(4)}+6\,k_-^{(1)}\,k_-^{(5)}\Big]\nonumber\\
&+&120\Big(Q-W\Big)\,\Big[4\,k_-^{(1)}\,k_-^{(2)}\,k_-^{(3)}+\big[k_-^{(1)}\big]^2\,k_-^{(4)}+\big[k_-^{(2)}\big]^3\Big]\nonumber\\
&+&80\,\Big(W\,\big(S-1\big)-2\,Q^2+Q\Big)\,\Big[4\,\big[k_-^{(1)}\big]^3\,k_-^{(3)}+9\,\big[k_-^{(1)}\big]^2\,\big[k_-^{(2)}\big]^2\Big]\nonumber\\
&+&480\,\Big(W\,\big(6\,Q-S^2+S-1\big)
+Q\,\big(2\,Q\,S-6\,Q+1\big)\Big)\,\big[k_-^{(1)}\big]^4\,k_-^{(2)}\nonumber\\
&+&64\,\Big(W\,\big(S^3-S^2+S-1-14\,Q\,S+20\,Q\big)-6\,W^2\nonumber\\
&-&2\,Q^2\big(S^2-3\,S+7\big)+Q\,\big(16\,Q^2+1\big)\Big)\,\big[k_-^{(1)}\big]^6,\nonumber
\end{eqnarray}
where, using (\ref{eq:polylog}), (\ref{eq:polylog-b}) and (\ref{eq:k-deriv}),
\begin{eqnarray}\label{eq:k-p}
k_-^{(1)}&=&\frac12 \big[\pb-\pbb\big]\nonumber\\
k_-^{(2)}&=&\frac12  \big[\pb(1-\pb)+\pbb(1-\pbb)\big]\nonumber\\
k_-^{(3)}&=&\frac12  \big[\pb(1-\pb)(1-2\,\pb)
-(\pb\to \pbb)
\big]\\
k_-^{(4)}&=&\frac12  \big[\pb(1-\pb)(1-6\pb+6\pb^2)
+ (\pb\to \pbb)\big]\nonumber\\
k_-^{(5)}&=&\frac12 \big[\pb\,(1-\pb)(1-2 \pb)(1-12 \pb+12 \pb^2)-(\pb\to \pbb)
\big]\nonumber\\
k_-^{(6)}&=&\frac12\big[\pb\,(1-\pb)(1-30\,\pb+150\,\pb^2-240 \pb^3+120\,\pb^4)+(\pb\to \pbb)
\big]\nonumber
\end{eqnarray}

Adding the contributions, $\kappa_n^{(a+b)}+\kappa_n^{(c)}$ from (\ref{eq:cumulants-a-b}) and (\ref{eq:cumulant-c}), yields the result obtained by taking explicit derivatives in (\ref{eq:derivatives}). 

With the tools provided in this appendix, one can readily compute cumulants of any order, taking the conservation of the net baryon number into account.
Explicit expressions for the net-baryon number cumulants  up to the   sixth-order  are obtained by summing the corresponding contributions from  Eqs.  (\ref{A22}) and (\ref{eq:cumulant-c}). Moreover, the experimentally accessible cumulants of the net proton number are obtained by simply replacing replacing the binomial probabilities $\alpha_B$ and $\alpha_{\bar B}$ by $\alpha_p$ and $\alpha_{\bar p}$, respectively,   as discussed in section \ref{sec:FlucCE}. 

We note that for $\pb=\pbb=\alpha$ the cumulants simplify significantly~\cite{Bzdak:2012an}. In particular, in this case, for odd $n$, $\kappa_{n}^{(c)}=0$ since $r^{(2 n-1)}=k_-^{(2 n-1)}=0$ and consequently 
\begin{equation}
\kappa^{(eq)}_{2 n-1}=B\,\Big[\delta_{2 n-1,1}+ {\rm Li}_{2(1-n)}(1-1/\alpha)\Big],
\end{equation}
which implies that
\begin{eqnarray}
\kappa^{(eq)}_1&=&B\,\alpha,\nonumber\\
\kappa^{(eq)}_3&=&\kappa^{(eq)}_1\,(1-\alpha)\,(1-2\,\alpha),\\
\kappa^{(eq)}_5&=&\kappa^{(eq)}_3\,(1-12\,\alpha+12\,\alpha^2).\nonumber
\end{eqnarray}
Conversely, for even $n$, $\kappa_{n}^{(a+b)}=0$, since $k_+^{(2 n)}=0$ and it follows that in this case, 
\begin{equation}
\kappa^{(eq)}_{2 n}=\kappa_{2 n}^{(c)}|_{\alpha_B=\alpha_{\bar{B}}=\alpha}.
\end{equation}
The first three even cumulants are then
\begin{eqnarray}
\kappa^{(eq)}_2&=&S\,\alpha\,(1-\alpha),\nonumber\\
\kappa^{(eq)}_4&=&\kappa^{(eq)}_2\,(1-6\,\alpha+6\,\alpha^2)+12\,Q\,\alpha^2\,(1-\alpha)^2,\\
\kappa^{(eq)}_6&=&\kappa^{(eq)}_2\,(1-30\,\alpha+150\,\alpha^2-240\,\alpha^3+120\,\alpha^4)\nonumber\\
&+&60\big[\,Q\,(1-6\,\alpha+6\,\alpha^2)+2\,\big(Q-W\big)\,\alpha\,(1-\alpha)\big]\,\alpha^2\,\big(1-\alpha\big)^2.\nonumber
\end{eqnarray}

\section{High-energy limit}
\label{sec:High-energyLimit}
For applications to nuclear collisions at high energies, where the net baryon number is small compared to the number of baryons or antibaryons, approximate expressions for the cumulants can be useful. In this appendix, we derive the leading and sub-leading terms in this limit. To this end we need the asymptotic form of the modified Bessel functions $I_\nu(x)$ for $x>>|\nu^2-\frac14|$,
\begin{eqnarray}\label{eq:bessel-expand}
I_\nu(x)&\simeq& \frac{e^x}{\sqrt{2\pi x}}\Big[1-\frac{4\nu^2-1}{8x}+\frac{(4\nu^2-1)(4\nu^2-9)}{128\,x^2}\nonumber\\
&-&\frac{(4\nu^2-1)(4\nu^2-9)(4\nu^2-25)}{3072\,x^3}\dots\Big].
\end{eqnarray} 
Applied to the canonical particle number, the asymptotic form is useful for $|B^2-\frac14|<<2\,z = 2\,\sqrt{z_B\,z_{\bar{B}}}$. For the canonical baryon and antibaryon numbers we find
\begin{eqnarray}\label{eq:canonical-baryon}
\langle N_B\rangle&=& z+\frac{2\,B-1}{4}+\frac{4\,B^2-1}{32\,z}+\frac{4\,B^2-1}{64\,z^2}+\mathcal{O}(1/z^3),\\
\langle N_{\bar{B}}\rangle&=&  z-\frac{2\,B+1}{4}+\frac{4\,B^2-1}{32\,z}+\frac{4\,B^2-1}{64\,z^2}+\mathcal{O}(1/z^3).\label{eq:canonical-baryon-2}
\end{eqnarray}  
Owing to cancellations between terms from the numerator and denominator of $\langle N_B\rangle$ and $\langle N_{\bar{B}}\rangle$ (cf. Eqs. (\ref{NCM}) and (\ref{NCP})), the expansions (\ref{eq:canonical-baryon}) and (\ref{eq:canonical-baryon-2}) converge for ${\rm max}(B,1/2)<<2\,z$.

When the asymptotic expressions are implemented in $f^{(n)}(z)$, we find
\begin{equation}\label{eq:series-exp-fn}
f^{(n)}(z)=2\,z\,\delta_{n,1}-(-1)^{n-1}\,(n-1)!\Big(\frac12 -n\frac{4\,B^2-1}{16\,z}\Big)+\mathcal{O}(1/z^2).
\end{equation}
Thus, $f^{(1)}(z)$ yields the only contribution to $\kappa_n^{(c)}$  (\ref{eq:kappa_c}), that grows with increasing $z$.  Consequently, the leading-order contribution in the high-energy limit is  
\begin{eqnarray}
\kappa_n^{(c,LO)}&=&2\,z\,B_{n,1}\left(\hk^{(1)},\hk^{(2)},\hk^{(3)},\dots,\hk^{(n)}\right)\\
&=&\Big(S+\frac12\Big)\,\hk^{(n)},\nonumber
\end{eqnarray}
where we used $B_{n,1}(x_1,\dots,x_n)=x_n$. Moreover, in the second line we eliminated $z$ using (\ref{eq:canonical-baryon}), (\ref{eq:canonical-baryon-2}) and (\ref{eq:SPQW}), neglecting terms of order $\mathcal{O}(1/z)$ and higher. At next-to-leading order there are two terms: $\kappa_n^{(a+b)}$, which is proportional to the net baryon number $B$ and the contribution of the $z$-independent term in (\ref{eq:series-exp-fn}). The latter can be resummed using (\ref{eq:inverse-relation}),
\begin{eqnarray}\label{eq:c-NLO}
\kappa_n^{(c,NLO)}&=&-\frac12\,\sum_{k=1}^n\,(-1)^{k-1}(k-1)!\,B_{n,k}(r^{(1)},\dots,r^{(n-k+1)})\\
&=&-\frac12\,k_-^{(n)}.
\end{eqnarray}
Hence, including the LO and NLO contributions, the cumulants in the high-energy limit are given by
\begin{equation}\label{eq:high-energy}
\kappa_n^{(he)}=S\,r^{(n)}+B\,k_+^{(n)}+\frac12\,\big[r^{(n)}-k_-^{(n)}\big],
\end{equation}
where $r^{(n)}$, $k_+^{(n)}$ and $k_-^{(n)}$ are given by (\ref{eq:deriv-binomial}), (\ref{eq:k-plus}) and (\ref{eq:k-deriv}), respectively. For the first three cumulants this yields the following explicit forms,
\begin{eqnarray}
\kappa_1^{(he)}&=&\frac12\,\nplusns\,\big[\pb-\pbb\big]+\frac12\,B\,\big[\pb+\pbb\big],\nonumber\\
\kappa_2^{(he)}&=&\frac12\,\nplusns\,\big[\pb(1-\pb)+\pbb(1-\pbb)+\frac12(\pb-\pbb)^2\big]\\
&+&\frac12\,B\,\big[\pb(1-\pb)-\pbb(1-\pbb)\big]+\frac18\big[(\pb-\pbb)^2\big],\nonumber\\
\kappa_3^{(he)}&=&\frac12\,\nplusns\,\big[\pb(1-\pb)(1-2\,\pb)-\pbb(1-\pbb)(1-2\,\pbb)\big]\nonumber\\
&+&\frac12\,B\,\big[\pb(1-\pb)(1-2\,\pb)+\pbb(1-\pbb)(1-2\,\pbb)\big]\nonumber\\
&+&\frac18\Big(S+\frac{1}{2}\Big) \Big[6\big[\pb-\pbb\big]\big[\pb(1-\pb)+\pbb(1-\pbb)\big]+\big[\pb-\pbb\big]^3\Big]\nonumber
\end{eqnarray}
Higher cumulants are readily computed using (\ref{eq:high-energy}).
As an example, for $\langle N_B\rangle=1132$, $\langle N_{\bar{B}}\rangle=749$, $\pb=0.0107$ and $\pbb=0.0162$, the accuracy of the high-energy approximation is, for all even cumulants up to sixth order, better than one percent. The high-energy approximation yields very small odd cumulants, in agreement with the full calculation, although the relative errors are not so small. We note that for the (unphysical) values $B=\pm \frac12$, {\em all} terms of order $\mathcal{O}(1/z)$ and higher in (\ref{eq:canonical-baryon})-(\ref{eq:series-exp-fn}) vanish and the high-energy approximation (\ref{eq:high-energy}) is exact.

\section{Low-energy limit}
\label{sec:Low-energyLimit}

In the low-energy limit, the number of antibaryons vanishes.   This is realized by letting $z\to 0$. Using the Taylor expansion of the Bessel functions for small arguments, $I_\nu(x)= (x/2)^\nu/\nu!\left(1+\mathcal{O}(x^2)\right)$, in (\ref{NCM}) and (\ref{NCP}), one finds for $B>0$, $\nbc=B+\mathcal{O}(z^2/B)$ and $\nbbc=\mathcal{O}(z^2/B)$.  Moreover, the canonical probability distribution (\ref{prob}) reduces to~\footnote{In the strict $z\to 0$ limit, classical Maxwell-Boltzmann statistics is no longer applicable to a system of baryons. However, for relevant conditions, the lower limit on $z$, $z_{\rm min}=e^{-m/T}\,B$, is so small that the $z\to 0$ limit of the classical partition function provides a useful approximation. For instance, in heavy-ion collisions at a beam energy of $\sqrt{s_{NN}}=7.7$ GeV, where the relevant baryon chemical potential is $\mu\approx 406$ MeV, the temperature $T\approx 138$ MeV  and the net baryon number $B\approx 350$, the corresponding $z\approx e^{-\mu/T}\,B\approx 18.5$, while $z_{\rm min}\approx 0.66$.}
\begin{equation}
P_B(N_{\bar{B}})=\frac{B!}{(N_{\bar{B}}+B)!\,N_{\bar{B}}!}\,z^{N_{\bar{B}}}\underset{z\to 0}{\longrightarrow}\,\delta_{N_{\bar{B}},0}.
\end{equation}

Since, in the low-energy limit $z\to 0$ and $\nbbc\to 0$, one finds $P=Q=W=0$, which in turn implies that $\tilde{f}=S=B$, that all derivatives of $\tilde{f}$ vanish and consequently that 
\begin{equation}
f^{(n)}=(-1)^{n-1}\,(n-1)!\,B.    
\end{equation}
This implies that in the low-energy limit the cumulants are given by
\begin{equation}\label{eq:low-energy-cumulants}
\kappa_n^{(le)}=B\,\big[k_-^{(n)}+k_+^{(n)}\big]=B\,c^{(n)}_B.
\end{equation}
In this limit one thus recovers the cumulants of the binomial distribution,
\begin{eqnarray}
\kappa_1^{(le)}&=&B\,\pb,\nonumber\\
\kappa_2^{(le)}&=&B\,\pb(1-\pb),\nonumber\\
\kappa_3^{(le)}&=&B\,\pb(1-\pb)(1-2\,\pb),\\
\kappa_4^{(le)}&=&B\,\pb(1-\pb)(1-6\,\pb+6\,\pb^2),\nonumber\\
\kappa_5^{(le)}&=&B\,\pb(1-\pb)(1-2\,\pb)(1-12\,\pb+12\,\pb^2),\nonumber\\
\kappa_6^{(le)}&=&B\,\pb(1-\pb)(1-30\,\pb+150\,\pb^2-240\,\pb^3+120\,\pb^4).\nonumber
\end{eqnarray}

\end{document}